\input{aipcheck}

\documentclass[
    ,final            
  ]
  {aipproc}

\layoutstyle{8x11double}

\def\comment#1{}

\usepackage{amsmath}
\usepackage{amsfonts}
\usepackage{amssymb}
\def\lfrac#1#2{{#1/#2}}
\def\lfrac#1#2{{#1/#2}}

\begin{document}

\title{Dyadosphere formed in gravitational collapse}


\author{Remo Ruffini and She-Sheng Xue}{
  address={ICRANeT Piazzale della Repubblica, 10 -65122, Pescara, and\\
Dipartimento di Fisica, University of Rome ``La Sapienza", P.le A. Moro 5, 00185 Rome, Italy}
}



\begin{abstract}
We first recall the concept of Dyadosphere (electron-positron-photon plasma around a formed black holes) and its motivation, and 
recall on (i) the Dirac process: annihilation of electron-positron pairs to photons; (ii) the Breit-Wheeler process: production of electron-positron pairs by photons with the energy larger than electron-positron mass threshold; the Sauter-Euler-Heisenberg effective Lagrangian and rate for the process of electron-positron production in a constant electric field. We present a general formula for the pair-production rate in the semi-classical treatment of quantum mechanical tunneling. 
We also present in the \emph{Quantum Electro-Dynamics} framework, the calculations of the Schwinger rate and effective Lagrangian for constant electromagnetic fields.  
We give a review on the electron-positron plasma oscillation in constant electric fields, and its interaction with photons leading to energy and number equipartition of photons, electrons and positrons. The possibility of creating an overcritical field 
in astrophysical condition is pointed out. We present the discussions and calculations on (i) energy extraction from gravitational collapse; (ii) the formation of Dyadosphere in gravitational collapsing process, and (iii) 
its hydrodynamical expansion in Reissner Nordstr\"om geometry.  We calculate the spectrum and flux of photon radiation at the point of transparency, and make predictions 
for short Gamma-Ray Bursts.
\end{abstract}

\maketitle


\section{Introduction}

\noindent{\it Motivations and Dyadosphere.}\hskip0.3cm
It is an one of most important issues in modern physics to understand how gravitational energy  
transforms to electromagnetic and rotational energies to during the process of gravitational
collapses to black holes, in connection with observations.  
The primal steps toward the understanding of this issue are studies of 
electromagnetic properties of spinning and non-spinning black holes: (i) reversible and irreversible 
transformations -- the Christodoulou-Ruffini formula, (ii) electron-positron pair-production in Kerr-Newmann geometry -- the 
Damour-Ruffini proposal for Gamma Ray Bursts (GRBs), (iii) formation of electron-positron-photon plasma -- 
Preparata-Ruffini-Xue Dyadopshere.

\noindent{\it Pair-production.}\hskip0.3cm 
The annihilation of electron-positron pair to two photons, and its inverse process -- the production of electron-positron pair by
the collision of two photons, as well as the electron-positron pair 
production from the vacuum in constant electromagnetic fields,
were studied in quantum mechanics by Dirac, Breit, Wheeler, Sauter, Euler, Heisenberg 
respectively in 1930's, and Schwinger in {\it Quantum Electro-Dynamics} (QED) in 1951. 

\noindent{\it Nonuniform fields.}\hskip0.3cm 
It has been a difficult task to obtain the rate of electron-positron pair production 
in varying electromagnetic fields in space and time. This issue has attracted 
much attention not only for its theoretical viewpoint, but also its possible applications in 
heavy-ion collisions and high-energy laser beams, as well as astrophysics. 

\noindent{\it Plasma oscillation.}\hskip0.3cm 
A naive expectation is that such external electric field rapidly vanishes for its source neutralized by electrons and positrons produced. 
However, the back reaction (screening effect) of electron-positron pairs on the external fields leads to the plasma oscillation 
phenomenon: electrons and positron oscillating back and forth in phase with alternating electric field. 
Beside electron-positron pairs oscillating together with the electric field, 
they interact with photons via the Dirac and Breit-Wheeler processes, and approach to a thermal configuration.  
  
\noindent{\it Critical fields on the surface of massive nuclear cores.}\hskip0.3cm
In ground based laboratories, it is rather difficult to built up
electromagnetic fields at the order of the critical field value $E_c$ in macroscopic space-time scales.
However, in the arena of astrophysics, supercritical electric fields are energetic-favorably developed on 
the surface of neutron star cores, due to strong, electroweak and gravitational interactions of 
degenerate nucleons and electrons. 

\noindent{\it Dyadosphere formed in gravitational collapse.}\hskip0.3cm
Initiating with supercritical electric fields on the surface, gravitational collapses of nuclear massive cores and processes of 
pair production, annihilation and oscillation lead to the formation of high energetic and dense plasma of 
electrons, positrons and photons, {\it Dyadosphere} that we proposed in 1998.

\noindent{\it Hydrodynamic expansion after gravitational collapse.}\hskip0.3cm
The adiabatic and hydrodynamic expansion of the electron-positron-photon plasma after gravitational collapse, up to 
the transparency to photons, account for daily observing 
phenomena of Gamma Ray Bursts (GRBs).    
  
\noindent{\it Predications in connection with short Gamma ray Bursts.}\hskip0.3cm  
Armed with a complete knowledge of all these fundamental 
processes, we present our understanding on the genuine origin of GRB-phenomenon, 
and make some predictions in connection with observations of short GRBs.

\section{Basic motivations and Dyadosphere}

\noindent{\it Energetics of Electromagnetic Black Holes.}\hskip0.3cm

The process of gravitational collapse of a massive core generally leads to a black hole
characterized by \textit{all} the three fundamental parameters: the
mass-energy $M$, the angular momentum $L$, and charge $Q$ \cite{rw}. 
The phenomenon of gravitational collapse is crucial for
the evolution of the system. Nonetheless in order not to involve its complex
dynamics at this stage, we assume that the collapse has
already occurred. Correspondingly a generally charged and rotating black
hole has been formed whose curved space-time is described by the stationary Kerr-Newmann
geometry in Boyer-Lindquist coordinates $(t,r,\theta,\phi)$
\begin{eqnarray}
ds^{2}&=&{\frac{\Sigma}{\Delta}}dr^{2}+\Sigma d\theta^{2}+{\frac{\Delta}{\Sigma
}}(dt-a\sin^{2}\theta d\phi)^{2}\nonumber\\
&+&{\frac{\sin^{2}\theta}{\Sigma}}\left[
(r^{2}+a^{2})d\phi-adt\right]  ^{2},\label{kerrnewmannBL}%
\end{eqnarray}
where $\Delta=r^{2}-2Mr+a^{2}+Q^{2}$ and $\Sigma=r^{2}+a^{2}\cos^{2}\theta$,
$a=L/M$ being the angular momentum per unit mass of the black hole. The
Reissner-Nordstr\"{o}m and Kerr geometries are particular cases for
non-rotating $a=0$, and uncharged $Q=0$, black holes respectively. 
The total energy in terms of the Coulomb and rotational energies is described by the 
Christodoulou--Ruffini mass formula \cite{dr2} 
\begin{eqnarray}
&&M^{2}c^{4}=\left(  M_{\mathrm{ir}}c^{2}+{\frac{c^{2}Q^{2}}{4GM_{\mathrm{ir}}}%
}\right)  ^{2}+{\frac{L^{2}c^{8}}{4G^{2}M_{\mathrm{ir}}^{2}}},\nonumber\\
&&\left(  {\frac{c^{2}}{16G^{2}M_{\mathrm{ir}}^{4}}}\right)  \left(
Q^{4}+4L^{2}c^{4}\right)  \leq 1,
\label{MassForm}%
\end{eqnarray}
where $M_{\mathrm{ir}}$ is the irreducible mass. The reversible (irreducible) 
process of the black hole, characterized by constant (increasing) irreducible mass, can (cannot) 
be inverted bringing the black hole to its original state. Energy can be extracted approaching arbitrarily 
close to reversible processes which are the most efficient
ones. Namely, from Eq.~(\ref{MassForm}) it follows that up to 29$\%$ of the
mass-energy of an extreme Kerr black hole ($M^{2}=a^{2}$) stored in its
rotational energy term ${\frac{Lc^{4}}{2GM_{\mathrm{ir}}}}$, whereas 
up to 50$\%$ of the mass energy of an extreme EMBH with $(Q=M)$ stored 
in the electromagnetic energy term ${\frac{c^{2}Q^{2}}{4GM_{\mathrm{ir}}}}$, 
can be in principle extracted. 

\noindent{\it Vacuum polarization around an Electromagnetic Black Hole.}\hskip0.3cm

It was pointed \cite{dr} that via Sauter, Heisenberg, Euler and Schwinger process,
the electron-positron pair production occurring around an superritical Electromagnetic Black Hole (EMBH) is
actually a very efficient almost reversible process of energy extraction, 
and extractable energy is up to $10^{54}$ergs that accounts for very energetic phenomenon of GRBs. 
In order to study the pair production in the Kerr-Newmann geometry, 
at each event $(t,r,\theta,\phi)$ a local Lorentz frame is introduced, associated with a
stationary observer ${\mathcal{O}}$ at the event $(t,r,\theta,\phi)$. A
convenient frame is defined by the following orthogonal tetrad 
\begin{eqnarray}
{ \bf\omega}^{(0)} &  =&(\Delta/\Sigma)^{1/2}(dt-a\sin^{2}\theta
d\phi),\label{tetrad1}\\
{\bf\omega}^{(1)} &  =&(\Sigma/\Delta)^{1/2}dr,\label{tetrad2}\\
{\bf\omega}^{(2)} &  =&\Sigma^{1/2}d\theta,\label{tetrad3}\\
{\bf\omega}^{(3)} &  =&\sin\theta\Sigma^{-1/2}((r^{2}+a^{2}%
)d\phi-adt).\label{tetrad4}%
\end{eqnarray}
In the so fixed Lorentz frame, the electric potential $A_{0}$, the electric
field ${\bf E}$ and the magnetic field ${\bf B}$ are given by the following
formulas,
\begin{equation}
A_{0}  =\boldsymbol{\omega}_{a}^{(0)}A^{a},\quad
{\bf E}^{\alpha}  =\boldsymbol{\omega}_{\beta}^{(0)}F^{\alpha\beta},\quad
{\bf B}^{\beta} ={\frac{1}{2}}\boldsymbol{\omega}_{\gamma}^{(0)}%
\epsilon^{\alpha\gamma\delta\beta}F_{\gamma\delta}.
\end{equation}
One then obtains $A_{0}=-Qr(\Sigma\Delta)^{-1/2}$,
while the electromagnetic fields ${\bf E}$ and ${\bf B}$ are parallel to the
direction of $\boldsymbol{\omega}^{(1)}$
\begin{eqnarray}
E_{(1)} &  = &Q\Sigma^{-2}(r^{2}-a^{2}\cos^{2}\theta),\label{e1}\\
B_{(1)} &  = &Q\Sigma^{-2}2ar\cos\theta,\label{b1}%
\end{eqnarray}
respectively. The spatial
variation scale $GM/c^{2}$ of these background fields is much larger than the
Compton wavelength $\hbar/m_ec$ of the quantum field, then, for what concern pair production, 
it is possible to consider the electric and magnetic fields defined by Eqs.~(\ref{e1},\ref{b1}) as constants
in a neighborhood of a few wavelengths around any events $(r,\theta,\phi,t)$.
Based on the equivalence principle, the rate of pair-production 
process in a constant field over a flat space-time can be locally applied to the case of the
curved Kerr-Newmann geometry:
\begin{eqnarray}
{\frac{dN}{\sqrt{-g}d^{4}x}}&=&{\frac{e^2E_{(1)}B_{(1)}}{4\pi^{2}}}%
\sum_{n=1}^{\infty}{\frac{1}{n}}%
\coth\left(  {\frac{n\pi B_{(1)}}{E_{(1)}}}\right)\nonumber\\
&\cdot &  \exp\left(
-{\frac{n\pi E_{\mathrm{c}}}{E_{(1)}}}\right)  ,\label{drw}%
\end{eqnarray}
where the critical field $E_c=m_e^2c^3/e\hbar$.
It was assumed that electron and positron produced fly apart from each other, one goes inward to neutralize EMBHs 
and another goes to infinity. This view was fundamentally modified in Refs.~\cite{prx98,rjapan,prxprl} 
by the novel concept of the Dyadosphere.

\noindent{\it Dyadosphere: electron-positron-photon plasma.}\hskip0.3cm

We start with the Reissner-Nordstr\"{o}m black holes 
and consider a spherical shell of proper
thickness $\delta={\frac{\hbar}{mc}}\ll{\frac{MG}{c^{2}}}$ centered on the EMBH, the electric
field is approximately constant in it. We can then at each value of the radius
$r$ model the electric field as created by a capacitor of width $\delta$ and
surface charge density
$\sigma(r)={\frac{Q}{4\pi r^{2}}}$,
and express Eq.~(\ref{drw}) as,
\begin{equation}
{\frac{dN}{\sqrt{-g}d^{4}x}}={\frac{1}{4\pi c}}\left(  {\frac{eE}{\pi\hbar}%
}\right)  ^{2}e^{-{\frac{\pi E_{\mathrm{c}}}{E}}}={\frac{1}{4\pi c}}\left(
{\frac{4e\sigma}{\hbar}}\right)  ^{2}e^{-{\frac{\pi\sigma_{\mathrm{c}}}%
{\sigma}}},\label{rate1}%
\end{equation}
where electric field $E=4\pi\sigma$, $\sigma_{\mathrm{c}}={\frac{1}{4\pi}}E_{\mathrm{c}}$ is
the critical surface charge density. 
The pair creation process in these shells will continue until a value of the surface charge density reaches the 
critical value $\sigma_{\mathrm{c}}$, and it takes 
\begin{eqnarray}
\Delta\tau &=&{\frac{\sigma-\sigma_{\mathrm{c}}}{{\frac{e}{4\pi c}}\left(
{\frac{4e\sigma}{\hbar}}\right)  ^{2}e^{-{\frac{\pi\sigma_{\mathrm{c}}}%
{\sigma}}}\left(  {\frac{\hbar}{mc}}\right)  }}\lesssim1.99\left(
{\frac{\hbar}{mc^{2}\alpha}}\right) \nonumber\\
& =&1.76\cdot 10^{-19}\mathrm{s}.\label{dis}%
\end{eqnarray}
This time is so short that the light travel time is
smaller or approximately equal to the width $\delta$. Under these circumstances
the correlation between shells can be approximately neglected, thus we can
justify the approximation of describing the pair creation process shell by shell.

\begin{figure}[ptb]
\def\fsz{\footnotesize}
\def\ssz{\scriptsize}
\def\tsz{\tiny}
\def\dst{\displaystyle}\unitlength1mm
\includegraphics[width=6cm,height=6cm]{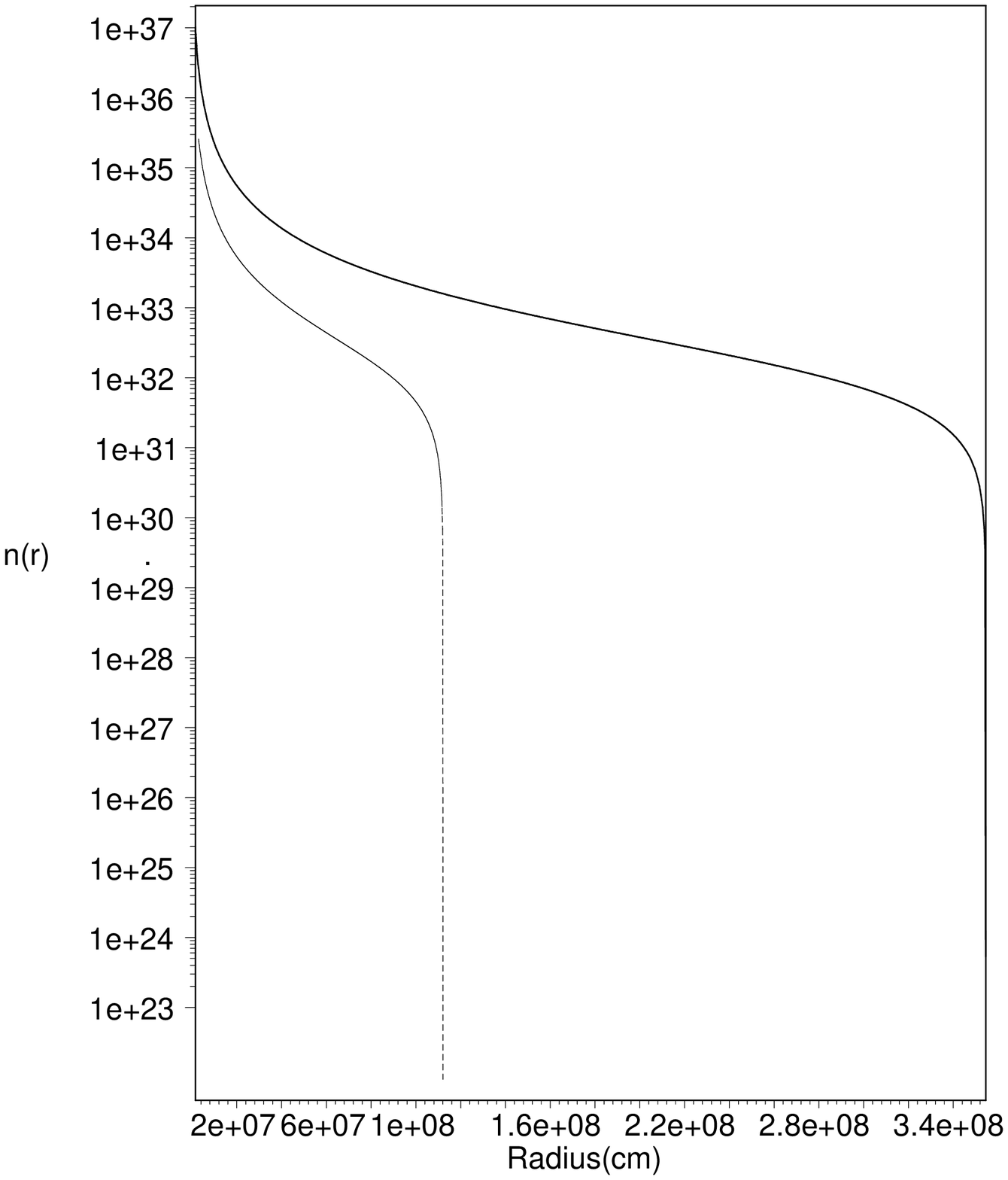}
\includegraphics[width=6cm,height=6cm]{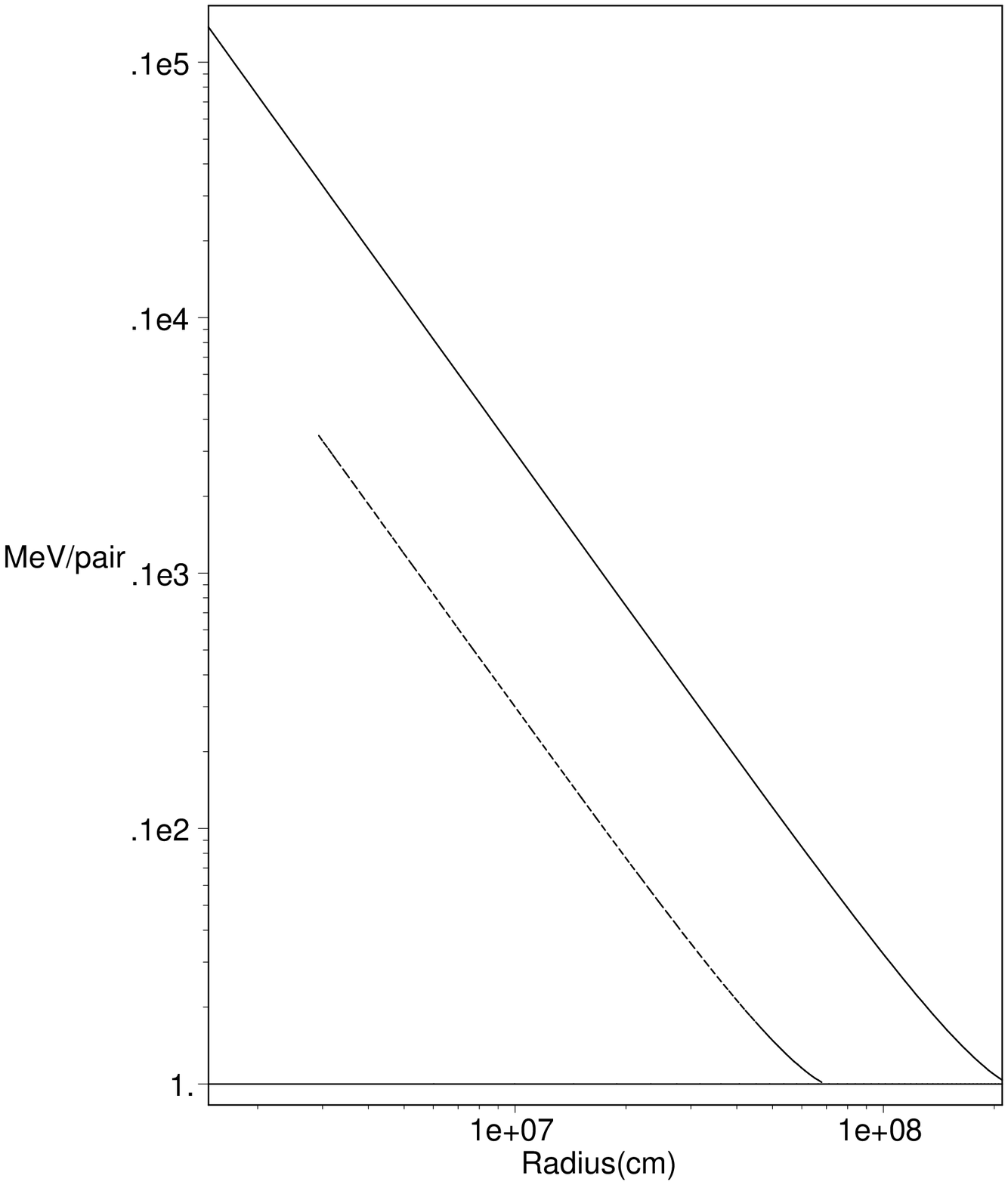}
\caption{The number-density $n_{e^{+}e^{-}}(r)$ (left) and average energy per pair in MeV (right) are plotted
as a function of the radial coordinate for $\mu=M/M_{\odot}=10$ and $\xi=Q/M=1$ (upper curve) 
and $\xi=0.1$ (lower curve).} 
\label{a&afig.2anda&afig.3}%
\end{figure}

\begin{figure}[ptb]
\def\fsz{\footnotesize}
\def\ssz{\scriptsize}
\def\tsz{\tiny}
\def\dst{\displaystyle}\unitlength1mm
\includegraphics[width=7cm,height=5cm]{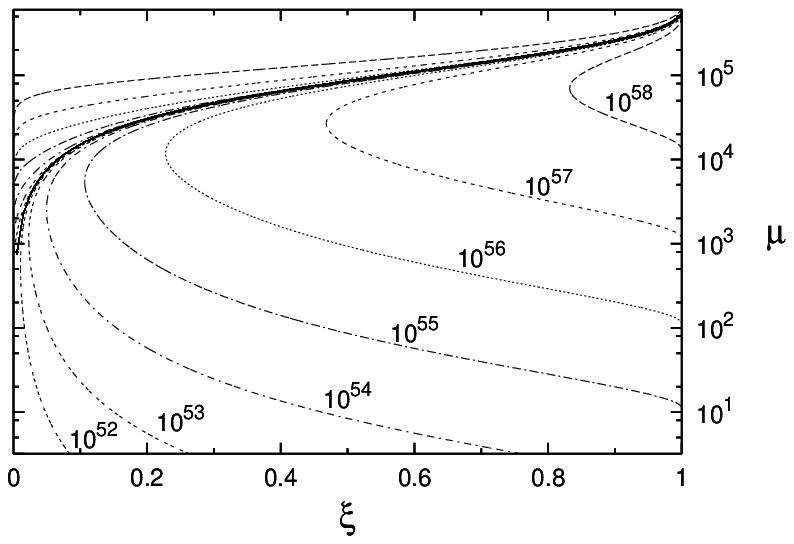}
\includegraphics[width=6cm,height=5.4cm]{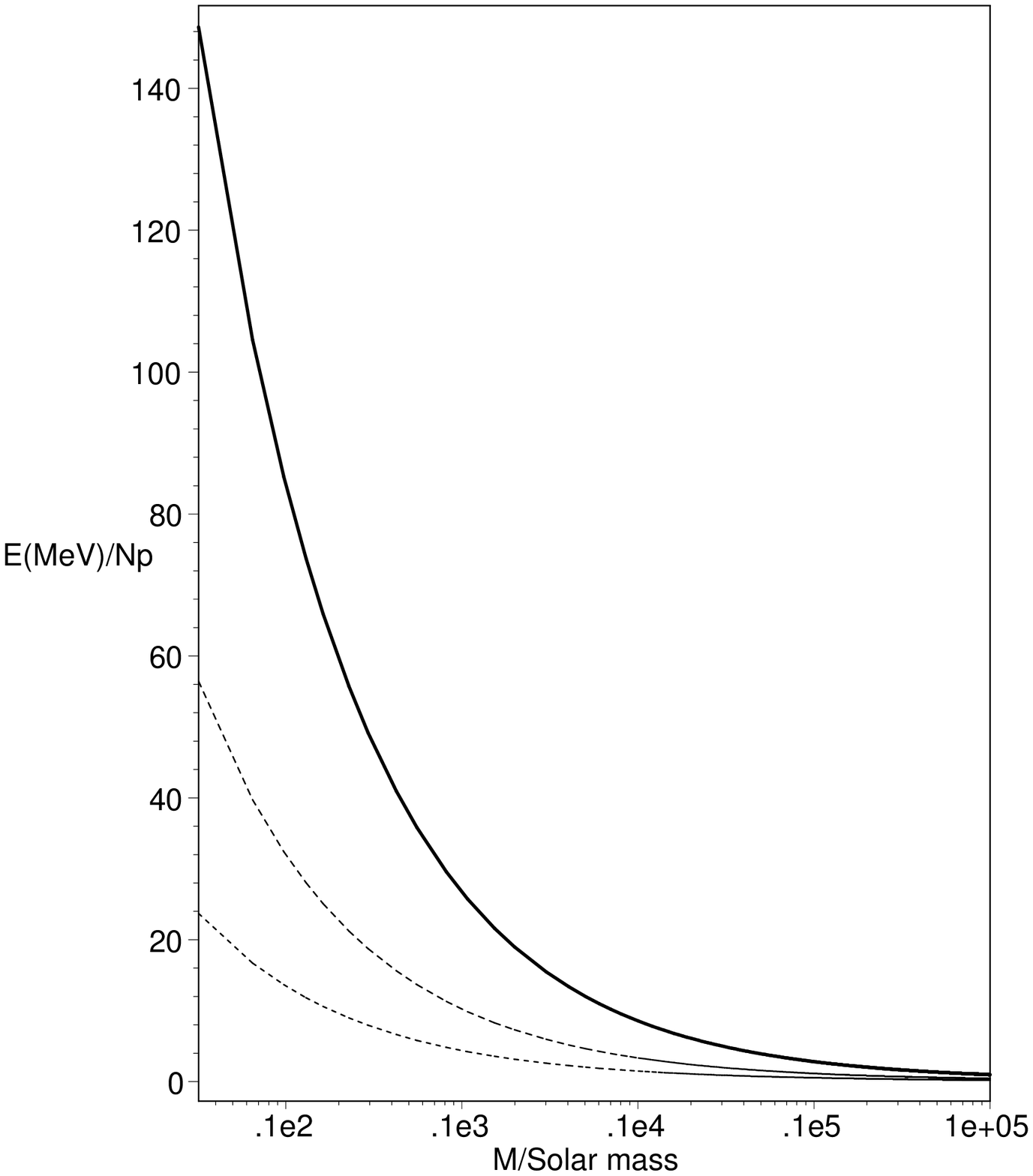}
\caption{Left: Total energy of Dyadosphere as a function of 
EMBHs' mass and charge parameters $\mu,\xi$. 
Right: The average energy per pair is shown here as a function of the EMBH
mass in solar mass units for $\xi=1$ (solid line), $\xi=0.5$ (dashed line) and
$\xi=0.1$ (dashed and dotted line).} 
\label{fig: totalenergyanda&afig.4}%
\end{figure}

The Dyadosphere is composed by these shells from the horizon $r_+$ to $r_{ds}$, which is given by $E(r_{ds})=E_c$ 
and can be expressed as,
\begin{eqnarray}
r_{\mathrm{ds}}&=&\left(  {\frac{\hbar}{mc}}\right)  ^{\frac{1}{2}}\left(
{\frac{GM}{c^{2}}}\right)  ^{\frac{1}{2}}\left(  {\frac{m_{\mathrm{p}}}{m}%
}\right)  ^{\frac{1}{2}}\left(  {\frac{e}{q_{\mathrm{p}}}}\right)  ^{\frac
{1}{2}}\nonumber\\
&\cdot&\left(  {\frac{Q}{\sqrt{G}M}}\right)  ^{\frac{1}{2}}
\gg {\frac{GM}{c^{2}}},\label{rc}%
\end{eqnarray}
using the Planck charge
$q_{\mathrm{p}}=(\hbar c)^{1/2}$ and the Planck mass $m_{\mathrm{p}%
}=(\hbar c/G)^{1/2}$, which clearly shows the hybrid gravitational and quantum nature of this
quantity. The total number of shells is about $(r_{ds}-r_{+})/{\frac{\hbar}{mc}}$ and the total
number of pairs,
\begin{equation}
N_{e^{+}e^{-}}\simeq{\frac{Q-Q_{\mathrm{c}}}{e}}\left[  1+{\frac{(r_{ds}%
-r_{+})}{{\frac{\hbar}{mc}}}}\right]  .\label{n}%
\end{equation}
We calculate the number and energy densities of pairs in the Dyadosphere
\begin{eqnarray}
\!\!\!\!n_{e^{+}e^{-}}(r)&=&{\frac{Q}{e4\pi r^{2}\left(  {\frac{\hbar}{mc}}\right)  }}\left[
1-\left(  {\frac{r}{r_{ds}}}\right)  ^{2}\right], \nonumber\\
\epsilon_{e^{+}e^{-}}(r)  
 &=&{\frac{Q^{2}}{8\pi r^{4}}}\left[  1-\left(  {\frac{r}{r_{ds}}}\right)
^{4}\right],\label{nd}%
\end{eqnarray}
as shown in Figs.~\ref{a&afig.2anda&afig.3}, 
and total energy is then
\begin{equation}
E_{e^{+}e^{-}}^{\mathrm{tot}}={\frac{1}{2}}{\frac{Q^{2}}{r_{+}}}\left(
1-{\frac{r_{+}}{r_{\mathrm{ds}}}}\right)  \left[  1-\left(  {\frac{r_{+}%
}{r_{\mathrm{ds}}}}\right)  ^{2}\right]  .\label{tee}%
\end{equation}
Due to the very large pair density given by Eq.~(\ref{nd}) and to the sizes of
the cross-sections for the process $e^{+}e^{-}\leftrightarrow\gamma+\gamma$,
the system is expected to thermalize to a plasma configuration for which
\begin{equation}
N_{e^{+}}=N_{e^{-}}=N_{\gamma}=N_{e^{+}e^{-}},\quad T_{\circ}={\frac{ E^{\mathrm{tot}}_{e^{+}e^{-}}}{3N_{e^{+}e^{-}}\cdot2.7}}.
\label{t}%
\end{equation}
In Fig.~\ref{fig: totalenergyanda&afig.4}, the total energy (\ref{tee}) and the average energy per pair ${\frac{E^{\mathrm{tot}}_{e^{+}e^{-}}}{N_{e^{+}e^{-}}}}$ are shown 
in terms of EMBH's mass $\mu$ and charge $\xi$.
Recently, the Dyadotorus: the plasma of electron-positron-photon created in Kerr-Newmann black holes is studied \cite{cgrr2007}.

\section{Production of electron-positron pairs }\label{pair}


\noindent{\it Early quantum electrodynamics.}\hskip0.3cm

\begin{figure}[ptb]
\includegraphics[height=4.8cm,width=8.8cm]{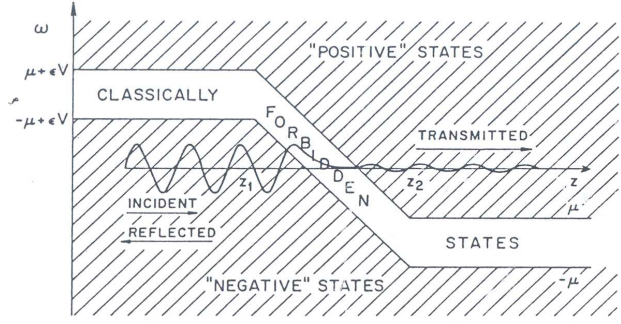}
\caption{In presence of a strong enough electric field the boundaries of the classically allowed states 
(``positive'' or ``negative'') can be so tilted that a ``negative'' is at the same level as a ``positive'' 
(level crossing). Therefore a ``negative'' wave-packet from the left will be partially transmitted,
after an exponential damping due to the tunneling through the classically forbidden states, as s ``positive''
wave-packet outgoing to the right. $\mu$ is particle's mass, $\epsilon V$ potential energy and $\omega$ energy.}%
\label{demourgape}%
\end{figure}

We recall three results, which played a crucial role in the development of the \emph{Quantum Electro-Dynamics} (QED).
The first is the Dirac process of an electron-positron pair annihilation into two photons,
\begin{equation}\label{ee2gamma}
e^{+}+e^{-}\rightarrow\gamma_{1}+\gamma_{2},
\end{equation}
and the cross-section in the rest frame of electron:
\begin{equation}
\sigma_{e^+ e^-}\simeq\frac{\pi}{\gamma}\left(\frac{\alpha\hbar}{m_e\,c}\right)^2\left[\ln\left(2\gamma\right)-1\right];
\label{Dirac section gg1}
\end{equation}
where $\gamma\equiv{\mathcal E_{+}}/m_e\,c^{2}\gg 1$ is the energy of the positron and $\alpha=e^2/(4\pi\hbar c)$ 
is the fine structure constant.
The second is the Breit-Wheeler process of electron-positron pair production by two photons collision, 
\begin{equation}\label{2gammaee}
\gamma_{1}+\gamma_{2}\rightarrow e^{+}+e^{-},
\end{equation}
which is the inverse Dirac-process (\ref{ee2gamma}) and the cross-section is related to (\ref{ee2gamma}) by the ${\mathcal CPT}$-theorem ,
\begin{equation}
\sigma_{\gamma\gamma}=2\beta^2\sigma_{e^+ e^-},
\label{Dirac section0}
\end{equation}
where $\beta$ is the relative velocity of electron and positron.
The third is the vacuum polarization in external uniform electromagnetic field, studied by Heisenberg and Euler, 
following Sauter's work on quantum tunneling probability 
from negative energy states [see Fig.~(\ref{demourgape})],  
\begin{eqnarray}
|T|^2&=&\frac{|{\rm transmission\hskip0.2cm flux}|}{|{\rm incident\hskip0.2cm flux}|}
\sim e^{-\pi\frac{m_e^2c^3}{\hbar e E}},
\label{transmission}\\
E_c &\equiv& \frac{m_e^2c^3}{e\hbar},\quad {\rm critical-field},\nonumber       
\end{eqnarray}
$E_c\simeq 1.3\cdot 10^{16}\, {\rm V/cm} $, $10^6$ larger than the value required to ionize a hydrogen atom. 
Heisenberg and Euler obtained nonlinear Lagrangian from the Dirac theory,     
\begin{eqnarray}
 \Delta {\mathcal L}_{\rm eff}&=&\frac{e^2}{16\pi^2\hbar c}\int^\infty_0
e^{-s}\frac{ds}{s^3}\nonumber\\
&\cdot&\Big[is^2\,\bar E \bar B
\frac{\cos(s[\bar E^2-\bar B^2+2i(\bar E\bar B)]^{1/2})
+{\rm c.c.}}{\cos(s[\bar E^2-\bar B^2+2i(\bar E\bar B)]^{1/2})-{\rm c.c.}}\nonumber\\
&+& \left(\frac{m_e^2c^3}{e\hbar}\right)^2
+\frac{s^2}{3}(|\bar B|^2-|\bar E|^2)\Big];\nonumber\\
\bar E &=& |{\bf E}|/E_c,\quad \bar B=|{\bf B}|/E_c,
\label{effectiveint}
\end{eqnarray}
and its series expansion in powers of $\alpha$,
\begin{align}
\Delta{\mathcal L}_{\rm eff}&=
\frac{2 \alpha ^2}{45 m_e^4}\left\{
({\bf E}^2\!-\!{\bf B}^2)^2+7 ({\bf E}\cdot {\bf B})^2 \right\}+\cdot\cdot\cdot .
\label{Kleinert1}
\end{align}
They found facts that $\Delta {\mathcal L}_{\rm eff}$ is
a complex function of ${\bf E}$ and ${\bf B}$, the imaginary part is associated with pair production 
when the electric field $E\gtrsim E_c$, and the vacuum behaves as a dielectric and permeable medium in which,
\begin{equation}
D_i=\sum_k\epsilon_{ik}E_k,\hskip0.5cm H_i=\sum_k\mu_{ik}B_k,
\label{dh1}
\end{equation}
where complex $\epsilon_{ik}$ and $\mu_{ik}$ are the field-dependent dielectric and permeability tensors of the vacuum.

\noindent{\it Quantum electrodynamics.}\hskip0.3cm 

The QED-Lagrangian describing the interacting system
of photons, electrons, and positrons reads
\begin{eqnarray}
{\mathcal L} &=& {\mathcal L}_0^{e^+e^-\gamma}(\bar\psi,\psi,A_\mu) + {\mathcal L}_{\rm int}(\bar\psi,\psi,A_\mu),
\label{qcdl}
\end{eqnarray}
where ${\mathcal L}_0^{e^+e^-\gamma}$ is for free electrons, positron and photons.
An external field $A^{\rm e}_\mu$
is incorporated by adding to the quantum field
$A_\mu$ in 
\begin{equation}
{\mathcal L}_{\rm int}+
{\mathcal L}^{\rm e}_{\rm int}
 = -e\bar\psi(x)\gamma^\mu
\psi(x)
\left[
A_\mu(x)+
A^{\rm e}_\mu(x)\right] .
\label{intc}
\end{equation}
The
amplitude
for the vacuum to vacuum transition in the presence of$A^{\rm e}$:
\begin{eqnarray}
\langle 0|0\rangle &\!\!\!\!\!\!=\!\!\!\!\!\!& \frac{Z[A^{\rm e}]}{ Z[0]},\nonumber\\
Z[A^{\rm e}] &\!\!\!\!\!\!=\!\!\!\!\!\!&\int [{\mathcal D}\psi {\mathcal D}\bar\psi {\mathcal D}A_\mu]
\exp \left[ i\int d^4x ({\mathcal L} + {\mathcal L}^{\rm e}_{\rm int}
) \right].
\label{vvamplitude}
\end{eqnarray}
The effective action as a functional of $A^{\rm e}$ is: 
\begin{equation}
\Delta{\mathcal A}_{\rm eff}[A^{\rm e}]\equiv -i\ln \langle  0|0\rangle.
\label{eaction}
\end{equation}
Under the assumption that $A^{\rm e}(x)$ varies smoothly over a finite spacetime region, 
we may define an approximately local 
effective Lagrangian $\Delta{\mathcal L}_{\rm eff}[A^{\rm e}(x)]$,
\begin{equation}
\Delta{\mathcal A}_{\rm eff}[A^{\rm e}]\simeq \int d^4x \Delta{\mathcal L}_{\rm eff}[A^{\rm e}(x)] 
\approx V\Delta t \Delta{\mathcal L}_{\rm eff}[A^{\rm e}],
\label{effl}
\end{equation}
where $V$ is the spatial volume and time interval $\Delta t$, over which 
the field is nonzero.  
The amplitude of the vacuum 
to vacuum transition (\ref{vvamplitude}) has the form,
\begin{equation}
\langle 0|0 \rangle = e^{-i(\Delta{\mathcal E}_0-i\Gamma/2)\Delta t},
\label{vvamplitude1}
\end{equation}
where vacuum-energy difference $\Delta{\mathcal E}_0={\mathcal E}_0(A^{\rm e})-{\mathcal E}_0(0)$, 
and $\Gamma$ is the vacuum decay rate. 
The probability that the vacuum remains as it is in the presence of the external field $A^{\rm e}$ is
\begin{equation}
|\langle 0|0\rangle|^2 = e^{-2{\rm Im}\Delta{\mathcal A}_{\rm eff}[A^{\rm e}]}.
\label{probability}
\end{equation}
This determines the decay rate of the vacuum caused by the production of electron and positron pairs:
\begin{equation}
\frac{ \Gamma}{V}= \frac{2{\rm \,Im}\Delta{\mathcal A}_{\rm eff}[A^{\rm e}]}{\Delta t V}
\approx 2{\rm \,Im}\Delta{\mathcal L}_{\rm eff}[A^{\rm e}];
\label{path21}
\end{equation}
and vacuum-energy variation  
\begin{equation}
\frac{ \Delta{\mathcal E}_0}{V}= -\frac{{\rm \,Re}\Delta{\mathcal A}_{\rm eff}[A^{\rm e}]}{\Delta t V}
\approx -{\rm \,Re}\Delta{\mathcal L}_{\rm eff}[A^{\rm e}].
\label{path21e}
\end{equation}
We calculate the imaginary part (\ref{path21}) and reproduce the Schwinger formula,
\begin{equation}
\frac{ \Gamma }{V}=\frac{  \alpha   \varepsilon^2}{ \pi^2 }
\sum_{n=1}  \frac{1}{n^2}
\frac{ n\pi\beta / \varepsilon }
{\tanh {n\pi \beta/ \varepsilon}}\exp\left(-\frac{n\pi E_c}{ \varepsilon}\right),
\label{probabilityeh}
\end{equation}
where $\varepsilon ^2- \beta ^2\equiv {\bf E}^2-{\bf B}^2$ and $\varepsilon \beta \equiv {\bf E}\,{\bf B}$.
In addition, we calculate \cite{hagen1,rvxreport} the real part (\ref{path21e}),
\begin{eqnarray}
{\rm Re}(\Delta{\mathcal L}_{\rm eff})  &=&
\frac{1}{2(2\pi)^2}\sum_{n,m=-\infty}^{\infty}{\!\!\!}' \frac{1}{ \tau^2_m+\tau^2_n}\nonumber\\
&\cdot&
\Big[\bar\delta_{m0}J(i\tau_m m^2_e)-\bar\delta_{n0}J(\tau_n m^2_e)\Big],
\label{pertur}
\end{eqnarray}
where $\tau_n=n\pi/e \varepsilon$, $\tau_m=m\pi/e \beta$, 
\begin{equation}
J(z)
= -\frac{1}{2}\Big[e^{-z}{\rm Ei}(z)
+ e^{z}{\rm Ei}(-z)\Big],
\label{J(z)1}
\end{equation}
and ${\rm Ei}(z)$ is the exponential-integral function. In the weak-field expansion, we obtain Eq.~(\ref{Kleinert1}).
In the strong-field expansion, we obtain,
\begin{eqnarray}
{\rm Re}(\Delta{\mathcal L}_{\rm eff})  
&=&\frac{1}{2(2\pi)^2}\sum_{n,m=-\infty}^{\infty '} \frac{1}{ \tau^2_m+\tau^2_n}\label{strongexp}\\
&\cdot&\Big[\bar\delta_{n0}\ln(\tau_n m^2_e)-\bar\delta_{m0}\ln(\tau_m m^2_e)\Big]+\cdot\cdot\cdot .
\nonumber
\end{eqnarray}
In the case $E\gg 1$, $B=0$ and $m=0$, we obtain,
\begin{eqnarray}
(\Delta{\mathcal L}_{\rm eff})  &=&\frac{1}{2(2\pi)^2}\sum_{n=1}^{\infty} \frac{1}{ \tau^2_n}
\ln(\tau_n m^2_e)+\cdot\cdot\cdot \nonumber\\
&=&\frac{e^2E^2}{8\pi^4}\sum_{n=1}^{\infty} \frac{1}{ n^2}
\ln\left(\frac{n\pi E_c}{E}\right)+\cdot\cdot\cdot .
\label{strongexpe}
\end{eqnarray}
In the case $B \gg 1$, $E=0$ and $n=0$, we obtain,
\begin{eqnarray}
(\Delta{\mathcal L}_{\rm eff})  &=& -\frac{1}{2(2\pi)^2}\sum_{m=1}^{\infty} \frac{1}{ \tau^2_m}
\ln(\tau_m m^2_e)+\cdot\cdot\cdot \nonumber\\
&=&-\frac{e^2B^2}{8\pi^4}\sum_{m=1}^{\infty} \frac{1}{ m^2}
\ln\left(\frac{m\pi E_c}{B}\right)+\cdot\cdot\cdot .
\label{strongexpb}
\end{eqnarray}


\noindent{\it Nonuniform electric fields.}\hskip0.3cm 

Let the field vector ${\bf E}(z)$
point in the $\hat {\bf z}$-direction.
The one-dimensional electric potential $A_0(z)=-\int^z dz'E(z')$ 
and the positive and negative continuum energy-spectra are
\begin{equation}
{\mathcal E}_\pm=\pm\sqrt{(cp_z)^2+c^2{\bf p}_\perp^2+(m_ec^2)^2}+V(z),
\label{energyl+-}
\end{equation}
where $p_z$ is the classical momentum in $\hat {\bf z}$-direction, ${\bf p}_\perp$ transverse momenta, and
$V(z)=eA_0(z)$ potential energy. The crossing energy-levels between two energy-spectra 
${\mathcal E}_\pm$ (\ref{energyl+-}) appear, $\epsilon\equiv {\mathcal E}_+={\mathcal E}_-$. The probability amplitude for quantum tunneling
process can be estimated by a semi-classical calculation using WKB method
(see e.g. \cite{krx2007}):
\begin{eqnarray}
W_{\rm WKB}(|{\bf p}_\perp |,\epsilon) &\equiv  & \exp\left\{-{\frac{2}{\hbar}}\int_{z_-}^{z_+} \kappa_zdz\right\},
\label{tprobability1}
\end{eqnarray}
where $\kappa=-ip_z$ and the turning points $z_\pm$ determined by setting $p_z=0$
\begin{equation}
V[z_\pm]=\mp\big[c^2{\bf p}_\perp ^2+m_e^2c^4\big]^{1/2}+\epsilon .
\label{crosspoint+-}
\end{equation}
Changing the variable of integration from  $z$ to $y(z)$,
\begin{equation}
y(z)= \frac{\epsilon-V(z)}{c{\sqrt{{\bf p}_\perp ^2+m_e^2c^2}}},
\label{y(x)}
\end{equation}
we obtain $
y_-(z_-) =-1,y_+(z_+) = +1
$
and
\begin{eqnarray}
W_{\rm WKB}(|{\bf p}_\perp |,\epsilon) &= & \exp\Big[ -\frac{2E_c}{E_0}  \left(1+\frac{{\bf p}_\perp ^2}{m_e^2c^2}\right)\nonumber\\
&\cdot&\int^{1}_{-1} dy\frac{\sqrt{1-y^2}}{\bar E(y)}\Big],
\label{wwkbp}
\end{eqnarray}
where $\bar E(y)=E[z(y)]/E_0$ and $z(y)$ is the inverse function of Eq.~(\ref{y(x)}).
The flux density of virtual particles attempt for tunneling at $z_-$ is
\begin{eqnarray}
d^3J_z=v_z D_s \frac{d^2{\bf p}_\perp dp_z}{(2\pi\hbar)^3} ,\quad v_z=\partial\epsilon/\partial p_z,
\label{nflux}
\end{eqnarray}
and the energy-variation $d\epsilon=|eE(z_-)|dz$. Using Eqs.~(\ref{wwkbp},\ref{nflux}) and 
expanding up to ${\bf p}_\perp ^2/(m_e^2c^2)$, we obtain the WKB-rate of pair production per unit volume at given 
crossing energy-level $\epsilon(z_-)$ is 
\begin{eqnarray} 
\frac{\Gamma _{\rm WKB}[\epsilon(z_-),z_-]}{V}&=&D_s|eE(z_-)|
\int\frac{d^2{\bf p}_\perp}{(2\pi\hbar)^3}\label{gwkbprobability}\\
&\cdot& e^{-{\pi c(G/2+g)} \lfrac{{\bf p}_\perp ^2}{|eE_0|\hbar }}e^{-{\pi G} \lfrac{E_c}{E_0 }}\!\nonumber\\
&\simeq & D_s\frac{\alpha E_0 E(z_-)}{2\pi^2\hbar (G/2+g)}
e^{-\lfrac{\pi G E_c}{E_0}},\nonumber
\end{eqnarray}
where $V\equiv V_\perp dz, V_\perp=\int dxdy$; $D_s=2$ for a spin-$1/2$ particle and $D_s=1$ for spin-$0$. The 
$G$ and $g$ functions are 
\begin{eqnarray}
G[\epsilon(z_-)]&=&\frac{2}{\pi}\int^{1}_{-1} dy\frac{\sqrt{1-y^2}}{\bar E(y)},\nonumber\\  
g[\epsilon(z_-)]&=&\frac{1}{\pi}\int^{1}_{-1} \frac{y^2}{\sqrt{1-y^2}}\frac{dy}{\bar E(y)}.
\label{gf}
\end{eqnarray}
Eq.~(\ref{gwkbprobability}) gives the semi-classical WKB-rate of pair-production per unit volume for any one-dimensional 
electric field ${\bf E}(z)$ and potential $V(z)$  \cite{krx2007}, 
provided crossing energy-levels $\epsilon$ between negative and positive energy-spectrum occur. 

We apply our formula (\ref{gwkbprobability},\ref{gf}) to a uniform field case, obtain $G=1, g=G/2$,
\begin{equation} 
\frac{\Gamma _{\rm WKB}}{V_\perp \Delta z}\simeq D_s\frac{\alpha E^2}{ 2\pi^2\hbar}
e^{-\lfrac{\pi E_c}{E}},
\label{wkbprobability}
\end{equation}
which is independent of crossing energy-levels $\epsilon$ and coordinate $z$.
gives the Sauter factor
(\ref{transmission}) and Heisenberg-Euler prefactor obtained from (\ref{effectiveint}).

\noindent{\it Sauter electric field.}\hskip0.3cm

\begin{figure}[ptb]
\includegraphics[height=3.8cm,width=7.8cm]{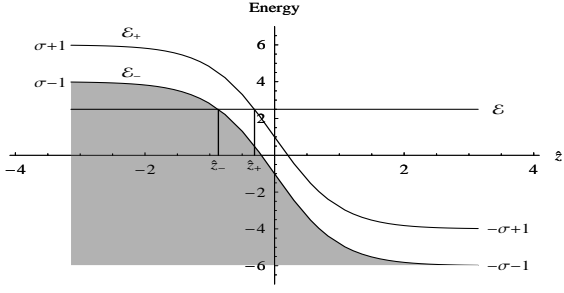}
\caption{Positive and negative energy-spectra ${\mathcal E}_\pm(z)$
of Eq.~(\ref{energyl+-})
in units of $m_ec^2$,
with  $p_z=p_\perp =0$
as a function of $x=z/\ell $
for the Sauter potential $V_\pm(z)$ (\ref{sauterv}) for $\sigma=5$. }%
\label{sauterf}%
\end{figure}

\begin{figure}[ptb]
\def\fsz{\footnotesize}
\def\ssz{\scriptsize}
\def\tsz{\tiny}
\def\dst{\displaystyle}\unitlength1mm
\includegraphics[width=6cm,clip]{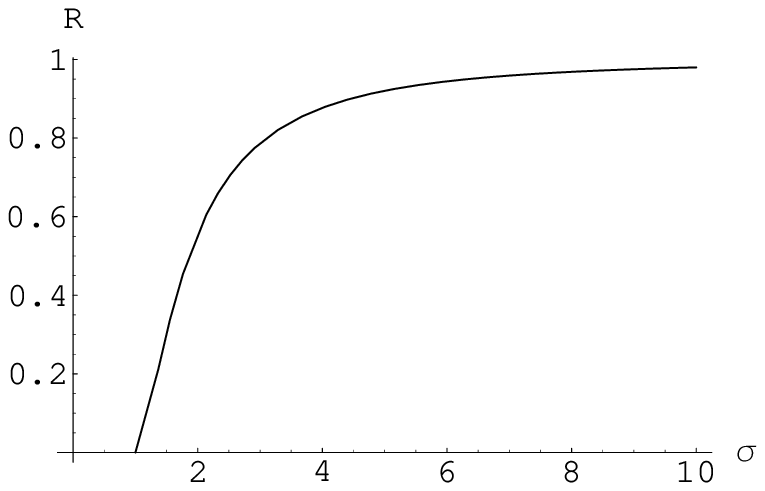}
\includegraphics[width=6cm,clip]{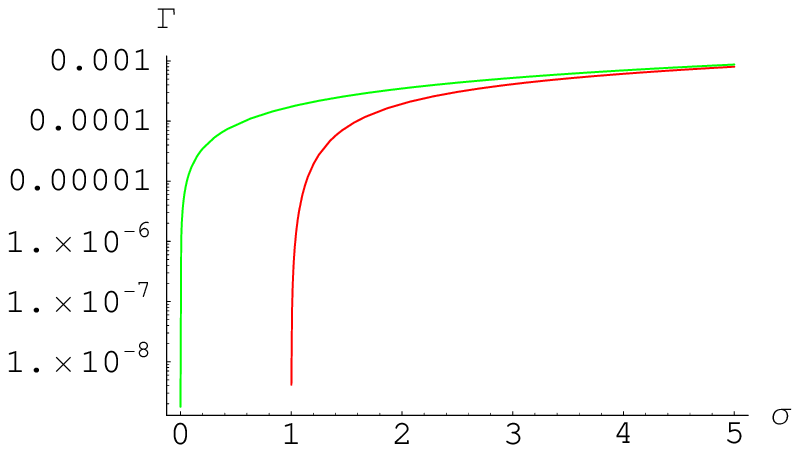}
\caption{For $E_0=E_c$, $\sigma=\ell /\lambda_C$ (\ref{@gamm}) is the spatial size where electric field $E\not=0$.
The ratio $R_{\rm rate}$ (\ref{compcs}) is plotted as function of $\sigma$ in the left figure. 
The number of pairs created in Compton area and time as functions of $\sigma$
(up curve for Schwinger constant field (\ref{wkbprobability}) and low one is nonuniform Sauter field (\ref{expgwkb1})) in the right figure.} 
\label{rratef}%
\end{figure}

As an example, we consider the
nonuniform Sauter electric field
localized within finite slice of space of width $\ell $
in the $xy$-plane \cite{krx2007}. 
Electric field ${\bf E}=E(z)\hat{\bf z}$ and
potential energy $V(z)$ are given by
\begin{eqnarray}
E(z)&=&E_0/{\rm cosh}^2\left({z}/{\ell }\right),
\label{sfield}\\
V(z)&=& \sigma\, m_ec^2\tanh\left({z}/{\ell }\right),
\label{sauterv}
\end{eqnarray}
where
\begin{equation}
\sigma\equiv |eE_0|\ell /m_ec^2=(\ell /\lambda_C)(E_0/E_c)=\ell |eE_0|/m_ec^2,
\label{@gamm}\end{equation}
and see Fig.~\ref{sauterf}.
Using our formula (\ref{gwkbprobability},\ref{gf}) and integrating over $\epsilon(z_-)$, we approximately obtain
\begin{eqnarray}
\frac{\Gamma _{\rm WKB}}{V_\perp \ell }
\simeq D_s\frac{e^2 E^{2}_0 }{8\pi^3\hbar}
 \sqrt{\frac{E_0}{E_c}}\frac{(\sigma^2-1)^{5/4}}{\sigma^{5/2}}
e^{-\lfrac{\pi G( 0 ) E_c}{E_0}}.
\label{expgwkb1}
\end{eqnarray}
The comparison between pair-production rates in unbound uniform and bound nonuniform fields is given by the ratio $R_{\rm rate}$ of 
Eq.~(\ref{wkbprobability}) and Eq.~(\ref{expgwkb1})
\begin{eqnarray}
R_{\rm rate}= \sqrt{\frac{E_0}{E_c}}e^{\lfrac{\pi E_c}{E_0}}\frac{(\sigma^2-1)^{5/4}}{\sigma^{5/2}}
e^{-\lfrac{\pi G( 0 ) E_c}{E_0}}.
\label{compcs}
\end{eqnarray}
In Fig.~(\ref{rratef}), it is shown that the pair-production rate in the Sauter field 
becomes smaller as the confining size of the 
field becomes smaller. 
In Ref.~\cite{krx2007}, we present detailed calculations and discussions of pair-productions rate 
in various cases of nonuniform electric fields: the Sauter and Coulomb fields, as well as fields $E(z)\sim z$ and $E(z)\not=0, z>0$.

\section{Plasma oscillations of electron-positron pairs in electric fields}\label{osci}


We have discussed the
Sauter-Heisenberg-Euler-Schwinger process for electron-positron pair production. 
However, we neglect the following dynamics:

\begin{enumerate}
\item the back reaction of pair production on the external electric field; 
\item the screening effect of pairs on the external electric field strengths;
\item the motion of pairs and their interactions.
\end{enumerate}

When these dynamics are considered, a phenomenon of electron-positron oscillation, 
{\it plasma oscillation}, takes place. We quantitatively discuss
this phenomenon by using the relativistic Boltzmann-Vlasov equations \cite{RVX03d}
\begin{align}
\partial_{t}f_{e}+e\mathbf{E\cdot\nabla}_{\mathbf{p}}f_{e}  &  =\mathcal{S}%
\left(  \mathbf{E},\mathbf{p}\right)  -\mathcal{C}_{e}\left(  t,\mathbf{p}%
\right)  ,\label{pairs}\\
\partial_{t}f_{\gamma}  &  =2\mathcal{C}_{\gamma}\left(  t,\mathbf{k}\right)
, \label{photons}%
\end{align}
where $f_e(f_\gamma)$ is spatially independent distribution function of electrons (photons) in phase space;
and the homogeneous Maxwell equations,
\begin{equation}
\partial_{t}\mathbf{E}=-\mathbf{j}_{p}\left(  \mathbf{E}\right)
-\mathbf{j}_{c}\left(  t\right)  , \label{Maxwell}%
\end{equation}
where $\mathbf{j}_{p}$ is the polarization current and $\mathbf{j}_{c}$ conduction current.
The terms $\mathcal{C}_{e}\left(  t,\mathbf{p}\right)$ and $\mathcal{C}_{\gamma}\left(  t,\mathbf{k}\right)$ 
stand for collisions between electrons, positrons and photons.
$\mathcal{S}\left(  \mathbf{E},\mathbf{p}\right)$ is related to the pair-production rate (\ref{wkbprobability}),
\begin{equation}
\mathcal{S}\left(  \mathbf{E},\mathbf{p}\right)=[1\pm 2f_e]\left(\frac{\Gamma}{V}\right)\delta^3({\bf p}),
\label{11bosonrate}
\end{equation}
where $[1\pm 2f_e]$ accounts for the Bose enhancement(+) and Pauli blocking (-). 

\begin{figure}[ptb]
\def\fsz{\footnotesize}
\def\ssz{\scriptsize}
\def\tsz{\tiny}
\def\dst{\displaystyle}\unitlength1mm
\includegraphics[width=6cm,height=9cm]{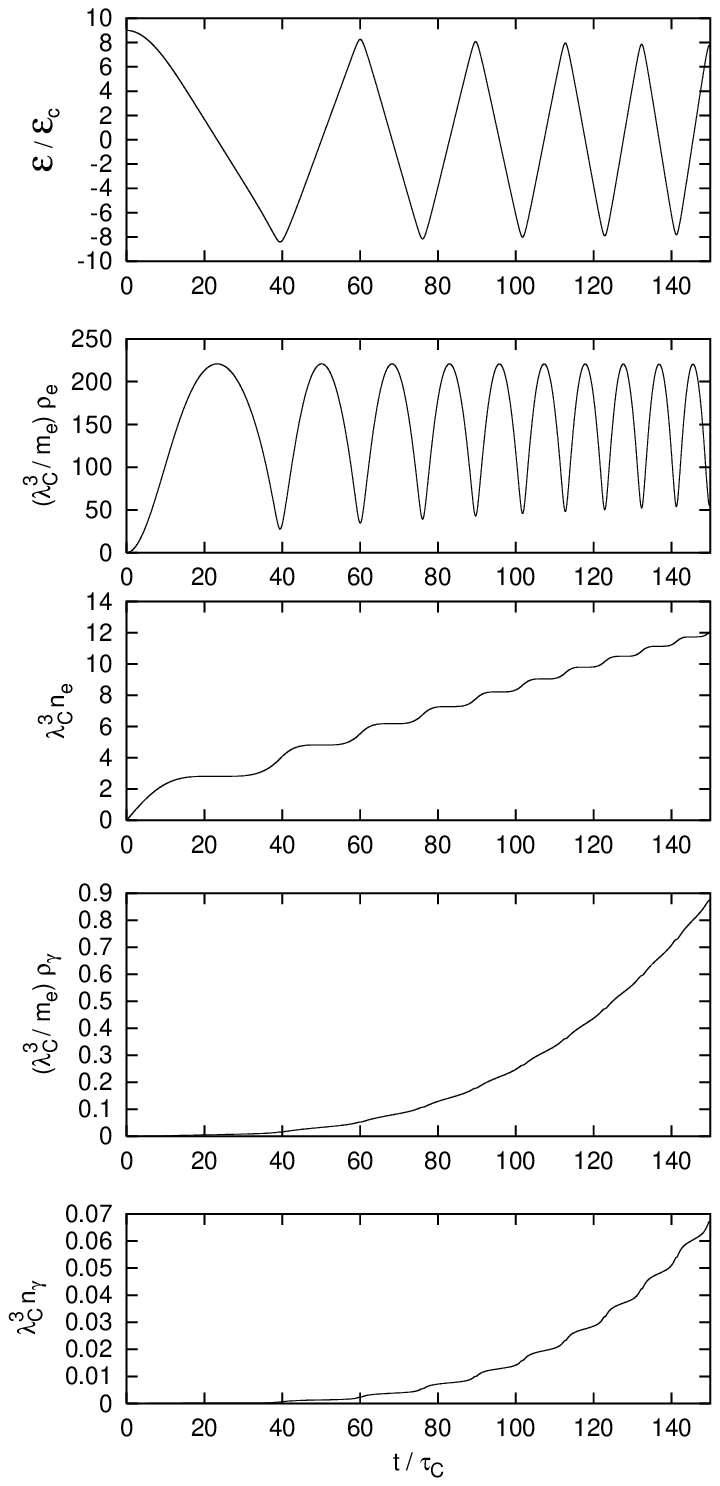}
\includegraphics[width=6cm,height=9cm]{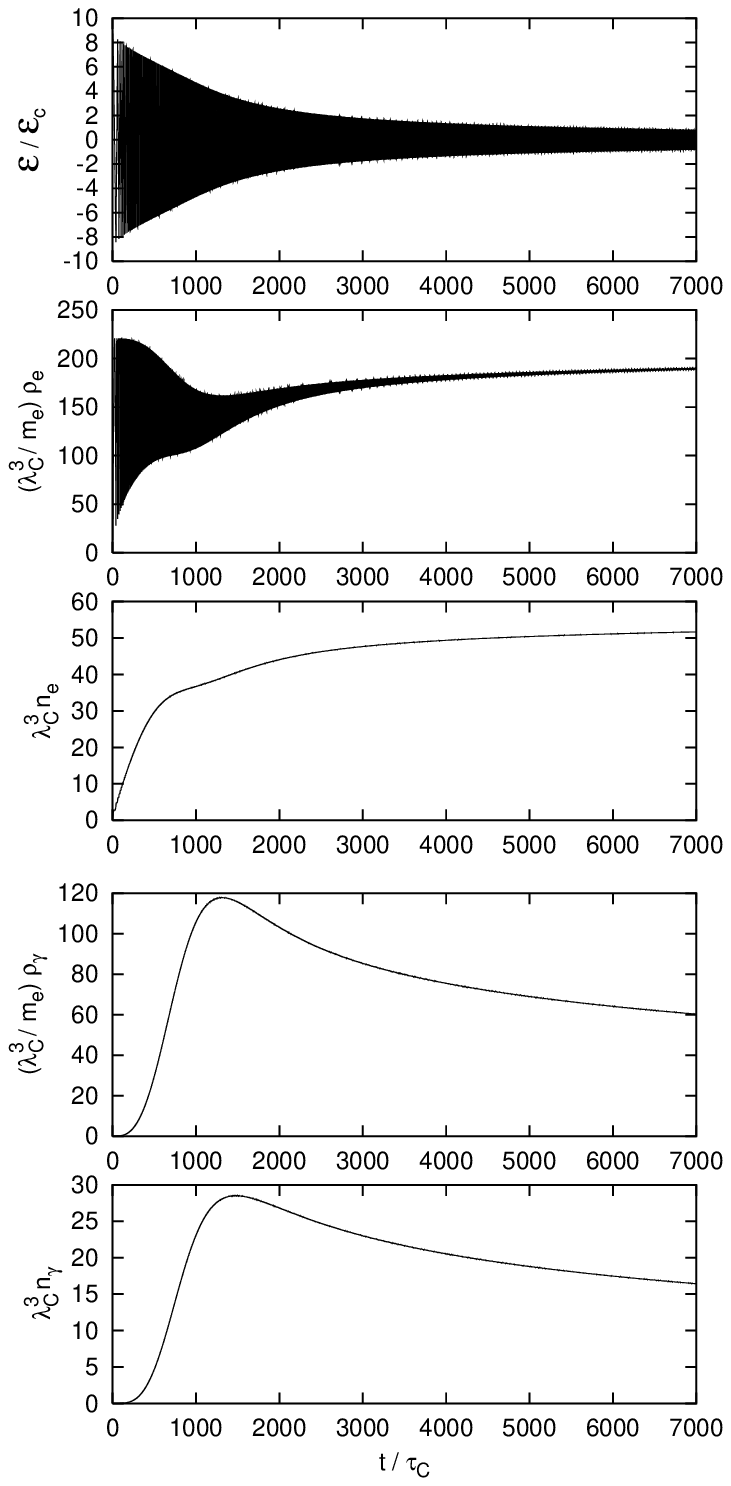}
\caption{In left figure: We plot for 
$t<150\tau_{\mathrm{C}}$, from the top to the bottom panel: a)
electromagnetic field strength; b) electrons energy density; c) electrons
number density; d) photons energy density; e) photons number density as
functions of time. The right figure: We plot for $t<7000\tau_{\mathrm{C}}$ as the same quantities as in left. 
} 
\label{OscillationLT}%
\end{figure}

These Equations (\ref{pairs}-\ref{Maxwell}) are integrated with the following initial
conditions of $|{\bf E}|=E_0=9E_c$ and null densities of electrons, positron and photons.
The results of the numerical integration in units of the Compton time $\tau_C$ and length $\lambda_C$ 
are shown in Fig.~\ref{OscillationLT}: at early times, 
\begin{enumerate}
\item the electric field does not abruptly reach the equilibrium value but
rather oscillates with decreasing amplitude;
\item electrons and positrons oscillates in the electric field direction,
reaching ultra relativistic velocities;
\item the role of the $e^{+}e^{-}\rightleftarrows$ $\gamma\gamma$ scatterings
is marginal in the early time of the evolution.
\end{enumerate}
At late times the system is expected to relax to a plasma configuration of
thermal equipartition with the asymptotic behavior:
\begin{enumerate}
\item the electric field is screened to about the critical value:
$E\simeq E_{\mathrm{c}}$ for $t\sim10^{3}-10^{4}%
\tau_{\mathrm{C}}\gg\tau_{\mathrm{C}}$;
\item the initial electromagnetic energy density is distributed over
electron--positron pairs and photons, indicating energy equipartition;
\item photons and electron--positron pairs number densities are asymptotically
comparable, indicating number equipartition.
\end{enumerate}
Note that we show \cite{rvx2007} that such phenomenon of plasma oscillation occurs not only for strong fields, but also for weak fields,
in addition, a detailed study of thermalization of electrons--positrons--photons
plasma is given in Ref.~\cite{arv2007}. The thermalized plasma 
starts hydrodynamical expansion described by hydrodynamic equations \cite{RVX03c,RSWX00,rswx1999}.

\section{Super critical field on the surface of massive nuclear cores}\label{neutron}

\noindent{\it Electromagnetic properties of massive nuclear cores}\hskip0.3cm
 
Based on numerical \cite{rrx2006} and analytical \cite{rrx2007,rrxmg2007} 
approaches to the relativistic Thomas-Fermi theory, 
we study electron configurations and electromagnetic properties of massive nuclear cores of mass number $A$ and radius $R_c$
\begin{eqnarray}
A &\!\!\!\!\simeq\!\!\!\!& \left(\frac{M_{\rm Planck}}{m_N}\right)^3\!\!\!\sim\!\!\! 10^{57},
\!\!\!\quad\!\!\!
R_c\!\!\!\simeq\!\!\! \frac{\hbar}{m_\pi c}N_p ^{1/3}\!\!\!\sim\!\!\! 10{\rm km},
\label{ar}
\end{eqnarray}
where $M_{\rm Planck}, m_N,m_\pi$ are Planck, nucleon and pion masses, 
with the global neutrality condition: the same proton and electron numbers $N_p=N_e$. 
We show that close to core surface, it exists supercritical electric field $E>E_c$, 
prove that this configuration is stable and energetically favorable against 
the configuration with the local neutrality condition: the same proton and electron densities $n_e(x)=n_p(x)$, 
usually adopted.    

\begin{figure}[ptb]
\def\fsz{\footnotesize}
\def\ssz{\scriptsize}
\def\tsz{\tiny}
\def\dst{\displaystyle}\unitlength1mm
\includegraphics[width=7cm,clip]{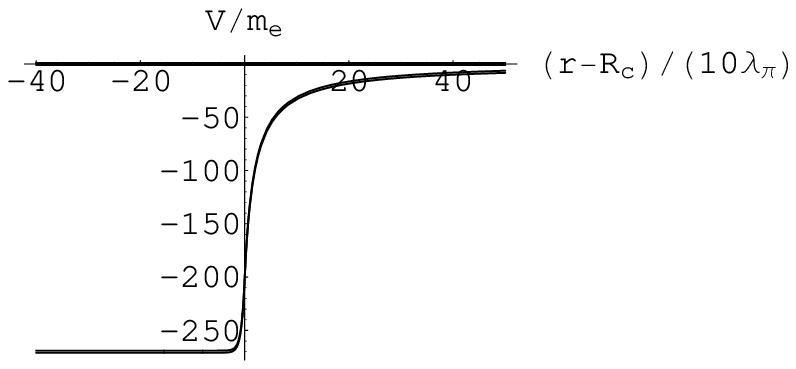}
\includegraphics[width=7cm,clip]{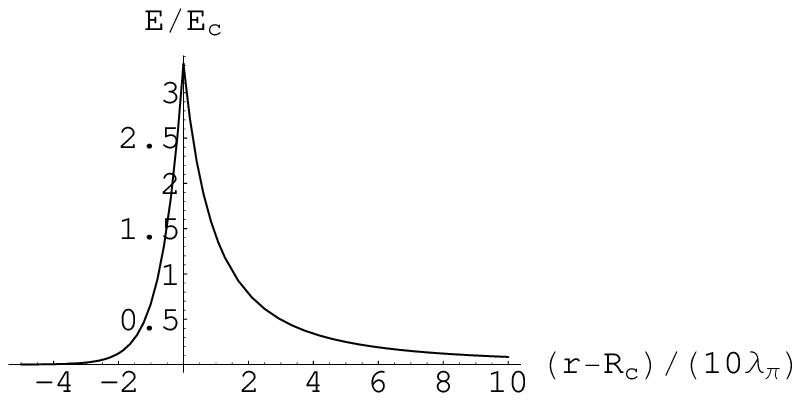}
\caption{Potential energy $-V_{\rm coul}(r)$ (left) and electric field $E(r)\sim -V'_{\rm coul}(r)$ (right)
are plotted as a function of $(r-R_c)/(10\lambda_\pi)$.} 
\label{eandp}%
\end{figure}

The Thomas-Fermi theory for the electrostatic equilibrium of electron distributions 
$n_e(r)$ around extended nuclear cores can be described as follow.
Degenerate electron density $n_e(r)$, Fermi momentum $P_e^F$ 
and Fermi-energy ${\mathcal E}_e(P_e^F)$ are related by
\begin{eqnarray}
n_e (r)&=& \frac {(P_e^F)^3}{3\pi^2\hbar^3},\label{en}\\
\quad {\mathcal E}_e(P_e^F)&\! =\!&[(P_e^Fc)^2+m_e^2c^4]^{1/2}\!-\!m_ec^2\!-\! V_{\rm coul}(r),   
\nonumber
\end{eqnarray}     
where $V_{\rm coul}(r)$ is Coulomb potential energy.  The electrostatic equilibrium of electron distributions is determined by
\begin{eqnarray}
{\mathcal E}_e(P_e^F) = 0,  
\label{eeq}
\end{eqnarray}
which means the balance of electron's kinetic and potential energies in Eq.~(\ref{en}) and degenerate electrons 
occupy energy-levels up to $+m_ec^2$. Eqs.~(\ref{en},\ref{eeq}) give:
\begin{eqnarray}
n_e(r) &=& \frac {1}{3\pi^2(c\hbar)^3}\left[V^2_{\rm coul}(r)+2m_ec^2V_{\rm coul}(r)\right]^{3/2}.  
\label{en1}
\end{eqnarray}
The Gauss law leads the following Poisson equation and boundary conditions,
\begin{eqnarray}
\Delta V_{\rm coul}(r)&=& 4\pi \alpha\left[n_p(r)-n_e(r)\right];\nonumber\\
V_{\rm coul}(\infty)&=&0,\quad V_{\rm coul}(0)={\rm finite}.
\label{eposs}
\end{eqnarray}
Degenerate proton and densities $n_{p,n}(r)$ are constants inside core $r \le R_c$ and vanishes outside the core $r > R_c$. 
\begin{eqnarray}
\!\!\!\!\!\! \!\!\! n_{p,n}(r) &=& \frac {(P_{p,n}^F)^3}{3\pi^2\hbar^3},\label{pn}\\ 
{\mathcal E}_{p,n}(P_e^F)&\! =\!&[(P_{p,n}^Fc)^2\!+\! m_{p,n}^2c^4]^{1/2}\! -\! m_{p,n}c^2\!  +\!  V_{\rm coul}(r)\delta_p,  
\nonumber
\end{eqnarray}
where $P_{p,n}^F, {\mathcal E}_{p,n}(P_e^F)$ are Fermi momenta, energies of protons and neutrons, and
$\delta_p$ indicates $V_{\rm coul}(r)$ for protons only. 
Neutrinos assumed to escape from massive cores, the energetic equation for the $\beta-$equilibrium of neutrons, protons and electrons is
\begin{eqnarray}
{\mathcal E}_n(P_n^F) = {\mathcal E}_p(P_p^F)+{\mathcal E}_e(P_e^F),  
\label{npeq}
\end{eqnarray}
which gives the relationship between the neutron, proton and mass numbers $N_n,N_p,A=N_n+N_p$. We integrate these 
equations are integrated and show results in Fig.~(\ref{eandp})
The configuration is electrostatic stable, since the mean repulsive energy is much 
smaller than mean gravitational binding $0.1M_Nc^2$ for protons in the surface layer. 

\section{Geometry of gravitationally collapsing cores}

\noindent{\it The Tolman-Oppenheimer-Snyder solution.}\hskip0.3cm 

Oppenheimer and Snyder first found a solution of the Einstein equations
describing the gravitational collapse of spherically symmetric star of mass
greater than $\sim0.7M_{\odot}$. In this section we briefly review their
pioneering work as presented in Ref. \cite{OS39}.

In a sperically symmetric space--time they can be found coordinates
$(t,r,\theta,\phi)$ such that the line element takes the form
\begin{equation}
ds^{2}=e^{\nu}dt^{2}-e^{\lambda}dr^{2}-r^{2}d\Omega^{2},\label{E0}%
\end{equation}
$d\Omega^{2}=d\theta^{2}+\sin^{2}\theta d\phi^{2}$, $\nu=\nu(t,r)$,
$\lambda=\lambda(t,r)$. However the gravitational collapse problem is better
solved in a system of coordinates $(\tau,R,\theta,\phi)$ which are comoving
with the matter inside the star. In comoving coordinates the line element
takes the form
\[
ds^{2}=d\tau^{2}-e^{\sigma}dR^{2}-e^{\omega}d\Omega^{2},
\]
$\overline{\omega}=\overline{\omega}(\tau,R)$, $\omega=\omega(\tau,R)$.
Einstein equations read
\begin{align}
8\pi T_{1}^{1}  & =e^{-\omega}-e^{-\sigma}\tfrac{\omega^{\prime2}}{4}%
+\ddot{\omega}+\tfrac{3}{4}\dot{\omega}^{2}\label{E1}\\
8\pi T_{2}^{2}  & =8\pi T_{3}^{3}=-\tfrac{e^{-\sigma}}{4}\left(
2\omega^{\prime\prime}+\omega^{\prime2}-\sigma^{\prime}\omega^{\prime}\right)
\nonumber\\
& +\tfrac{1}{4}(2\ddot{\sigma}+\dot{\sigma}^{2}+2\ddot{\omega}+\dot{\omega
}^{2}+\dot{\sigma}\dot{\omega})\label{E2}\\
8\pi T_{4}^{4}  & =e^{-\omega}-e^{-\sigma}\left(  \omega^{\prime\prime}%
+\tfrac{3}{4}\omega^{\prime2}-\tfrac{\sigma^{\prime}\omega^{\prime}}%
{2}\right)  +\tfrac{\dot{\omega}^{2}}{4}+\tfrac{\dot{\sigma}\dot{\omega}}%
{2}\label{E3}\\
8\pi e^{\sigma}T_{4}^{1}  & =-8\pi T_{1}^{4}=\tfrac{1}{2}\omega^{\prime}%
(\dot{\omega}-\dot{\sigma})+\dot{\omega}^{\prime}.\label{E4}%
\end{align}
Where $T_{\mu\nu}$ is the energy--momentum tensor of the stellar matter, a dot
denotes a derivative with respect to $\tau$ and a prime denotes a derivative
with respect to $R$. Oppenheimer and Snyder were only able to integrate Eqs.
(\ref{E1})--(\ref{E4}) in the case when the pressure $p$ of the stellar matter
vanishes and no energy is radiated outwards. In the following we thus $p=0$.
In this hypothesis
\[
T_{1}^{1}=T_{2}^{2}=T_{3}^{3}=T_{4}^{1}=T_{1}^{4}=0,\quad T_{4}^{4}=\rho
\]
where $\rho$ is the comoving density of the star. Eq.~(\ref{E4}) was first
integrated by Tolman in Ref. \cite{T34}. The solution is
\begin{equation}
e^{\sigma}=e^{\omega}\omega^{\prime2}/4f^{2}(R),\label{E5}%
\end{equation}
where $f=f(R)$ is an arbitrary function. In Ref. \cite{OS39} was studied the
case $f(R)=1$. The hypothesis $f(R)=1$ will be relaxed in the case
of a shell of dust. Using Eq.~(\ref{E5}) into Eq.~(\ref{E1}) with $f(R)=1$
gives%
\begin{equation}
\ddot{\omega}+\tfrac{3}{4}\dot{\omega}^{2}=0,\label{E6}%
\end{equation}
which can be integrated to
\begin{equation}
e^{\omega}=(F\tau+G)^{4/3},\label{E7}%
\end{equation}
where $F=F(R)$ and $G=G(R)$ are arbitrary function. Using Eq.~(\ref{E5}) into
Eq.~(\ref{E2}) gives Eq.~(\ref{E6}) again. From Eqs.~(\ref{E3}), (\ref{E5})
and (\ref{E7}) the density $\rho$ can be found as
\begin{equation}
8\pi\rho=\tfrac{4}{3}\left(  \tau+\tfrac{G}{F}\right)  ^{-1}\left(
\tau+\tfrac{G^{\prime}}{F^{\prime}}\right)  ^{-1}.\label{E8}%
\end{equation}
There is still the gauge freedom of choosing $R$ so to have
\[
G=R^{3/2}.
\]
Moreover, it can be freely chosen the initial density profile, i.e., the
density at the initial time $\tau=0$, $\rho_{0}=\rho_{0}(R)$. Eq.~(\ref{E8})
then becomes
\[
FF^{\prime}=9\pi R^{2}\rho_{0}(R)
\]
whose solution contains only one arbitrary integration constant. It is thus
seen the the choice of Oppenheimer and Snyder of putting $f(R)=1$ allows one
to assign only a one--parameter family of functions for the initial values
$\dot{\rho}_{0}=\dot{\rho}_{0}(R)$ of $\dot{\rho}$. However in general one
should be able to assign the initial values of $\dot{\rho}$ arbitarrily. This
will be done in the following section in the case of a shell of dust.

Choosing, for instance,
\[
\rho_{0}=\left\{
\begin{array}
[c]{cc}%
\mathrm{const}>0 & \text{if }R<R_{b}\\
0 & \text{if }R\geq R_{b}%
\end{array}
\right.  ,
\]
$R_{b}$ being the comoving radius of the boundary of the star, gives
\[
F=\left\{
\begin{array}
[c]{cc}%
-\tfrac{3}{2}r_{+}^{1/2}\left(  \tfrac{R}{R_{b}}\right)  ^{3/2} & \text{if
}R<R_{b}\\
-\tfrac{3}{2}r_{+}^{1/2} & \text{if }R\geq R_{b}%
\end{array}
\right.
\]
where $r_{+}=2M$ is the Shwarzschild radius of the star.

We are finally in the position of performing a coordinate transformation from
the comoving coordinates $(\tau,R,\theta,\phi)$ to new coordinates
$(t,r,\theta,\phi)$ in which the line elements looks like (\ref{E0}). The
requirement that the line element be the Schwarzschild one outside the star
fix the form of such a coordinate transformation to be
\begin{align*}
r  & =(F\tau+G)^{2/3}\\
t  & =\left\{
\begin{array}
[c]{cc}%
\tfrac{2}{3}r_{+}^{-1/2}(R_{b}^{3/2}-r_{+}^{3/2}y^{3/2})-2r_{+}y^{1/2}%
+r_{+}\log\tfrac{y^{1/2}+1}{y^{1/2}-1} & 
\\
\tfrac{2}{3r_{+}^{1/2}}(R^{3/2}-r^{3/2})-2(rr_{+})^{1/2}+r_{+}\log
\tfrac{r^{1/2}+r_{+}^{1/2}}{r^{1/2}-r_{+}^{1/2}} & 
\end{array}
\right.  
\end{align*}
where the first line for $\text{if }R<R_{b}$, the second line for
$\text{if }R\geq R_{b}$ and 
\[
y=\tfrac{1}{2}\left[  \left(  \tfrac{R}{R_{b}}\right)  ^{2}-1\right]
+\tfrac{R_{b}r}{r_{+}R}.
\]

\noindent{\it Gravitational collapse of charged and uncharged
shells.}\hskip0.3cm 

It is well known that the role of exact solutions has been fundamental in the
development of general relativity. In this section, we present here exact
solutions for a charged shell of matter collapsing into an Electromagnetic Black Hole (EMBH). Such
solutions were found in Ref.~\cite{CRV02} and are new with respect to the
Tolman-Oppenheimer-Snyder class. For simplicity we consider the case of zero
angular momentum and spherical symmetry. This problem is relevant for its own
sake as an addition to the existing family of interesting exact solutions and
also represents some progress in understanding the role of the formation of
the horizon and of the irreducible mass discussed in Ref.~
\cite{RV02}. It is also essential in improving the treatment of the vacuum
polarization processes occurring during the formation of an EMBH discussed in
Ref.~\cite{RVX03c}. As we already mentioned, both of these
issues are becoming relevant to explaining gamma ray bursts, see e.g.
\cite{RBCFX01a,RBCFX01b,RBCFX01c,RBCFX01d} and references therein.

W. Israel and V. de La Cruz \cite{I66,IdlC67} showed that the problem of a
collapsing charged shell can be reduced to a set of ordinary differential
equations. We reconsider here the following relativistic system: a spherical
shell of electrically charged dust which is moving radially in the
Reissner-Nordstr\"{o}m background of an already formed nonrotating EMBH of
mass $M_{1}$ and charge $Q_{1}$, with $Q_{1}\leq M_{1}$. The Einstein-Maxwell
equations with a charged spherical dust as source are
\begin{eqnarray}
G_{\mu\nu}&=&8\pi\left[  T_{\mu\nu}^{\left(  \mathrm{d}\right)  }+T_{\mu\nu
}^{\left(  \mathrm{em}\right)  }\right]  ,\quad\nabla_{\mu}F^{\nu\mu}=4\pi
j^{\nu},\nonumber\\
\nabla_{\lbrack\mu}F_{\nu\rho]}&=&0,
\end{eqnarray}
where
\begin{eqnarray}
T_{\mu\nu}^{\left(  \mathrm{em}\right)  }&=&\tfrac{1}{4\pi}\left(  F_{\mu}%
{}^{\rho}F_{\rho\nu}-\tfrac{1}{4}g_{\mu\nu}F^{\rho\sigma}F_{\rho\sigma
}\right)  ,\nonumber\\
T_{\mu\nu}^{\left(  \mathrm{d}\right)  }&=&\varepsilon u_{\mu}u_{\nu},\quad
 j^{\mu}=\sigma u^{\mu}.
\end{eqnarray}
Here $T_{\mu\nu}^{\left(  \mathrm{d}\right)  }$, $T_{\mu\nu}^{\left(
\mathrm{em}\right)  }$ and $j^{\mu}$ are respectively the energy-momentum
tensor of the dust, the energy-momentum tensor of the electromagnetic field
$F_{\mu\nu}$ and the charge $4-$current. The mass and charge density in the
comoving frame are given by $\varepsilon$, $\sigma$ and $u^{a}$ is the
$4$-velocity of the dust. In spherical-polar coordinates the line element is
\begin{equation}
ds^{2}\equiv g_{\mu\nu}dx^{\mu}dx^{\nu}=-e^{\nu\left(  r,t\right)  }%
dt^{2}+e^{\lambda\left(  r,t\right)  }dr^{2}+r^{2}d\Omega^{2},
\end{equation}
where $d\Omega^{2}=d\theta^{2}+\sin^{2}\theta d\phi^{2}$.

We describe the shell by using the $4$-dimensional Dirac distribution
$\delta^{\left(  4\right)  }$ normalized as
\begin{equation}
\int\delta^{\left(  4\right)  }\left(  x,x^{\prime}\right)  \sqrt{-g}d^{4}x=1
\end{equation}
where $g=\det\left\Vert g_{\mu\nu}\right\Vert $. We then have
\begin{align}
\varepsilon\left(  x\right)   &  =M_{0}\int\delta^{\left(  4\right)  }\left(
x,x_{0}\right)  r^{2}d\tau d\Omega,\label{eq1a}\\
\sigma\left(  x\right)   &  =Q_{0}\int\delta^{\left(  4\right)  }\left(
x,x_{0}\right)  r^{2}d\tau d\Omega. \label{eq2a}%
\end{align}
$M_{0}$ and $Q_{0}$ respectively are the rest mass and the charge of the shell
and $\tau$ is the proper time along the world surface $S:$ $x_{0}=x_{0}\left(
\tau,\Omega\right)  $ of the shell. $S$ divides the space-time into two
regions: an internal one $\mathcal{M}_{-}$ and an external one $\mathcal{M}%
_{+}$. As we will see in the next section for the description of the collapse
we can choose either $\mathcal{M}_{-}$ or $\mathcal{M}_{+}$. The two
descriptions, clearly equivalent, will be relevant for the physical
interpretation of the solutions.

Introducing the orthonormal tetrad
\begin{eqnarray}
{\boldsymbol{\omega}}_{\pm}^{\left(  0\right)  }&=&f_{\pm}^{1/2}dt,\quad
{\boldsymbol{\omega}}_{\pm}^{\left(  1\right)  }=f_{\pm}^{-1/2}dr,\quad
{\boldsymbol{\omega}}^{\left(  2\right)  }=rd\theta,\nonumber\\
\quad{\boldsymbol{\omega}%
}^{\left(  3\right)  }&=&r\sin\theta d\phi,
\end{eqnarray}
we obtain the tetrad components of the electric field%
\begin{equation}
{\boldsymbol{\mathcal{E}}}=\mathcal{E\ }{\boldsymbol{\omega}}^{\left(
1\right)  }=\left\{
\begin{array}
[c]{l}%
\frac{Q}{r^{2}}\ {\boldsymbol{\omega}}_{+}^{\left(  1\right)  }\quad
\text{outside the shell}\\
\frac{Q_{1}}{r^{2}}\ {\boldsymbol{\omega}}_{-}^{\left(  1\right)  }%
\quad\text{inside the shell}%
\end{array}
\right.   \label{E31}%
\end{equation}
where $Q=Q_{0}+Q_{1}$ is the total charge of the system. From the $G_{tt}$
Einstein equation we get
\begin{equation}
ds^{2}=\left\{
\begin{array}
[c]{l}%
-f_{+}dt_{+}^{2}+f_{+}^{-1}dr^{2}+r^{2}d\Omega^{2}\qquad\text{outside the
shell}\\
-f_{-}dt_{-}^{2}+f_{-}^{-1}dr^{2}+r^{2}d\Omega^{2}\qquad\text{inside the
shell}%
\end{array}
\right.  , \label{E01}%
\end{equation}
where $f_{+}=1-\tfrac{2M}{r}+\tfrac{Q^{2}}{r^{2}}$, $f_{-}=1-\tfrac{2M_{1}}%
{r}+\tfrac{Q_{1}^{2}}{r^{2}}$ and $t_{-}$ and $t_{+}$ are the
Schwarzschild-like time coordinates in $\mathcal{M}_{-}$ and $\mathcal{M}_{+}$
respectively. Here $M$ is the total mass-energy of the system formed by the
shell and the EMBH, measured by an observer at rest at infinity.

Indicating by $r_{0}$ the Schwarzschild-like radial coordinate of the shell
and by $t_{0\pm}$ its time coordinate, from the $G_{tr}$ Einstein equation we
have
\begin{equation}
\tfrac{M_{0}}{2}\left[  f_{+}\left(  r_{0}\right)  \tfrac{dt_{0+}}{d\tau
}+f_{-}\left(  r_{0}\right)  \tfrac{dt_{0-}}{d\tau}\right]  =M-M_{1}%
-\tfrac{Q_{0}^{2}}{2r_{0}}-\tfrac{Q_{1}Q_{0}}{r_{0}}. \label{eq3a}%
\end{equation}
The remaining Einstein equations are identically satisfied. From (\ref{eq3a})
and the normalization condition $u_{\mu}u^{\mu}=-1$ we find
\begin{align}
\left(  \tfrac{dr_{0}}{d\tau}\right)  ^{2}  &  =\tfrac{1}{M_{0}^{2}}\left(
M-M_{1}+\tfrac{M_{0}^{2}}{2r_{0}}-\tfrac{Q_{0}^{2}}{2r_{0}}-\tfrac{Q_{1}Q_{0}%
}{r_{0}}\right)  ^{2}-f_{-}\left(  r_{0}\right) \nonumber\\
&  =\tfrac{1}{M_{0}^{2}}\left(  M-M_{1}-\tfrac{M_{0}^{2}}{2r_{0}}-\tfrac
{Q_{0}^{2}}{2r_{0}}-\tfrac{Q_{1}Q_{0}}{r_{0}}\right)  ^{2}-f_{+}\left(
r_{0}\right)  ,\label{EQUY}\\
\tfrac{dt_{0\pm}}{d\tau}  &  =\tfrac{1}{M_{0}f_{\pm}\left(  r_{0}\right)
}\left(  M-M_{1}\mp\tfrac{M_{0}^{2}}{2r_{0}}-\tfrac{Q_{0}^{2}}{2r_{0}}%
-\tfrac{Q_{1}Q_{0}}{r_{0}}\right)  . \label{EQUYa}%
\end{align}

We now define, as usual, $r_{\pm}\equiv M\pm\sqrt{M^{2}-Q^{2}}$: when $Q<M$,
$r_{\pm}$ are real and they correspond to the horizons of the new black hole
formed by the gravitational collapse of the shell. We similarly introduce the
horizons $r_{\pm}^{1}=M_{1}\pm\sqrt{M_{1}^{2}-Q_{1}^{2}}$ of the already
formed EMBH. From (\ref{eq3a}) we have that the inequality
\begin{equation}
M-M_{1}-\tfrac{Q_{0}^{2}}{2r_{0}}-\tfrac{Q_{1}Q_{0}}{r_{0}}>0
\label{Constraint}%
\end{equation}
holds for $r_{0}>r_{+}$ if $Q<M$ and for $r_{0}>r_{+}^{1}$ if $Q>M$ since in
these cases the left hand side of (\ref{eq3a}) is clearly positive. Eqs.~(\ref{EQUY}) and (\ref{EQUYa}) (together with (\ref{E01}), (\ref{E31}))
completely describe a 5-parameter ($M$, $Q$, $M_{1}$, $Q_{1}$, $M_{0}$) family
of solutions of the Einstein-Maxwell equations.

For astrophysical applications \cite{RVX03c} the trajectory of the shell
$r_{0}=r_{0}\left(  t_{0+}\right)  $ is obtained as a function of the time
coordinate $t_{0+}$ relative to the space-time region $\mathcal{M}_{+}$. In
the following we drop the $+$ index from $t_{0+}$. From (\ref{EQUY}) and
(\ref{EQUYa}) we have
\begin{equation}
\tfrac{dr_{0}}{dt_{0}}=\tfrac{dr_{0}}{d\tau}\tfrac{d\tau}{dt_{0}}=\pm\tfrac
{F}{\Omega}\sqrt{\Omega^{2}-F}, \label{EQUAISRDLC}%
\end{equation}
where
\begin{eqnarray}
F\equiv f_{+}\left(  r_{0}\right)  &=&1-\tfrac{2M}{r_{0}}+\tfrac{Q^{2}}%
{r_{0}^{2}},\quad\Omega\equiv\Gamma-\tfrac{M_{0}^{2}+Q^{2}-Q_{1}^{2}}%
{2M_{0}r_{0}},\nonumber\\
\Gamma &\equiv&\tfrac{M-M_{1}}{M_{0}}.
\end{eqnarray}
Since we are interested in an imploding shell, only the minus sign case in
(\ref{EQUAISRDLC}) will be studied. We can give the following physical
interpretation of $\Gamma$. If $M-M_{1}\geq M_{0}$, $\Gamma$ coincides with
the Lorentz $\gamma$ factor of the imploding shell at infinity; from
(\ref{EQUAISRDLC}) it satisfies
\begin{equation}
\Gamma=\tfrac{1}{\sqrt{1-\left(  \frac{dr_{0}}{dt_{0}}\right)  _{r_{0}=\infty
}^{2}}}\geq1.
\end{equation}
When $M-M_{1}<M_{0}$ then there is a \emph{turning point} $r_{0}^{\ast}$,
defined by $\left.  \tfrac{dr_{0}}{dt_{0}}\right\vert _{r_{0}=r_{0}^{\ast}}%
=0$. In this case $\Gamma$ coincides with the \textquotedblleft effective
potential\textquotedblright\ at $r_{0}^{\ast}$ :
\begin{equation}
\Gamma=\sqrt{f_{-}\left(  r_{0}^{\ast}\right)  }+M_{0}^{-1}\left(
-\tfrac{M_{0}^{2}}{2r_{0}^{\ast}}+\tfrac{Q_{0}^{2}}{2r_{0}^{\ast}}%
+\tfrac{Q_{1}Q_{0}}{r_{0}^{\ast}}\right)  \leq1.
\end{equation}

The solution of the differential equation (\ref{EQUAISRDLC}) is given by:
\begin{equation}
\int dt_{0}=-\int\tfrac{\Omega}{F\sqrt{\Omega^{2}-F}}dr_{0}. \label{GRYD}%
\end{equation}
The functional form of the integral (\ref{GRYD}) crucially depends on the
degree of the polynomial $P\left(  r_{0}\right)  =r_{0}^{2}\left(  \Omega
^{2}-F\right)  $, which is generically two, but in special cases has lower
values. We therefore distinguish the following cases:

\begin{enumerate}
\item {\boldmath $M=M_{0}+M_{1}$}; {\boldmath $Q_{1}=M_{1}$}; {\boldmath$Q=M$}:
$P\left(  r_{0}\right)  $ is equal to $0$, we simply have
\begin{equation}
r_{0}(t_{0})=\mathrm{{const}.}%
\end{equation}

\item {\boldmath$M=M_{0}+M_{1}$}; {\boldmath$M^{2}-Q^{2}=M_{1}^{2}-Q_{1}^{2}$%
}; {\boldmath$Q\neq M$}: $P\left(  r_{0}\right)  $ is a constant, we have
\begin{eqnarray}
t_{0}&=&\mathrm{const}+\tfrac{1}{2\sqrt{M^{2}-Q^{2}}}\Big[  \left(
r_{0}+2\right)  r_{0}+r_{+}^{2}\log\left(  \tfrac{r_{0}-r_{+}}{M}\right)\nonumber\\
&+&r_{-}^{2}\log\left(  \tfrac{r_{0}-r_{-}}{M}\right)  \Big]  . \label{CASO1}%
\end{eqnarray}

\item {\boldmath$M=M_{0}+M_{1}$}; {\boldmath$M^{2}-Q^{2}\neq M_{1}^{2}%
-Q_{1}^{2}$}: $P\left(  r_{0}\right)  $ is a first order polynomial and
\begin{align}
t_{0}  &  =\mathrm{const}+2r_{0}\sqrt{\Omega^{2}-F}\left[  \tfrac{M_{0}r_{0}%
}{3\left(  M^{2}-Q^{2}-M_{1}^{2}+Q_{1}^{2}\right)  }\right. \nonumber\\
&  \left.  +\tfrac{\left(  M_{0}^{2}+Q^{2}-Q_{1}^{2}\right)  ^{2}%
-9MM_{0}\left(  M_{0}^{2}+Q^{2}-Q_{1}^{2}\right)  +12M^{2}M_{0}^{2}%
+2Q^{2}M_{0}^{2}}{3\left(  M^{2}-Q^{2}-M_{1}^{2}+Q_{1}^{2}\right)  ^{2}%
}\right] \nonumber\\
&  -\tfrac{1}{\sqrt{M^{2}-Q^{2}}}\Big[  r_{+}^{2}\mathrm{arctanh}\left(
\tfrac{r_{0}}{r_{+}}\tfrac{\sqrt{\Omega^{2}-F}}{\Omega_{+}}\right)  \nonumber\\
&-r_{-}%
^{2}\mathrm{arctanh}\left(  \tfrac{r_{0}}{r_{-}}\tfrac{\sqrt{\Omega^{2}-F}%
}{\Omega_{-}}\right)  \Big]  , \label{CASO2}%
\end{align}

where $\Omega_{\pm}\equiv\Omega\left(  r_{\pm}\right)  $.

\item {\boldmath$M\neq M_{0}+M_{1}$}: $P\left(  r_{0}\right)  $ is a second
order polynomial and
\begin{align}
t_{0}  &  =\mathrm{const}-\tfrac{1}{2\sqrt{M^{2}-Q^{2}}}\left\{
\tfrac{2\Gamma\sqrt{M^{2}-Q^{2}}}{\Gamma^{2}-1}r_{0}\sqrt{\Omega^{2}-F}\right.
\nonumber\\
&  +r_{+}^{2}\log\left[  \tfrac{r_{0}\sqrt{\Omega^{2}-F}}{r_{0}-r_{+}}%
+\tfrac{r_{0}^{2}\left(  \Omega^{2}-F\right)  +r_{+}^{2}\Omega_{+}^{2}-\left(
\Gamma^{2}-1\right)  \left(  r_{0}-r_{+}\right)  ^{2}}{2\left(  r_{0}%
-r_{+}\right)  r_{0}\sqrt{\Omega^{2}-F}}\right] \nonumber\\
&  -r_{-}^{2}\log\left[  \tfrac{r_{0}\sqrt{\Omega^{2}-F}}{r_{0}-r_{-}}%
+\tfrac{r_{0}^{2}\left(  \Omega^{2}-F\right)  +r_{-}^{2}\Omega_{-}^{2}-\left(
\Gamma^{2}-1\right)  \left(  r_{0}-r_{-}\right)  ^{2}}{2\left(  r_{0}%
-r_{-}\right)  r_{0}\sqrt{\Omega^{2}-F}}\right] \nonumber\\
&  -\tfrac{\left[  2MM_{0}\left(  2\Gamma^{3}-3\Gamma\right)  +M_{0}^{2}%
+Q^{2}-Q_{1}^{2}\right]  \sqrt{M^{2}-Q^{2}}}{M_{0}\left(  \Gamma^{2}-1\right)
^{3/2}}\log\left[  \tfrac{r_{0}}{M}\sqrt{\Omega^{2}-F}\right. \nonumber\\
&  \left.  \left.  +\tfrac{2M_{0}\left(  \Gamma^{2}-1\right)  r_{0}-\left(
M_{0}^{2}+Q^{2}-Q_{1}^{2}\right)  \Gamma+2M_{0}M}{2M_{0}M\sqrt{\Gamma^{2}-1}%
}\right]  \right\}  . \label{CASO3}%
\end{align}

\end{enumerate}

In the case of a shell falling in a flat background ($M_{1}=Q_{1}=0$) it is of
particular interest to study the \emph{turning points} $r_{0}^{\ast}$ of the
shell trajectory. In this case equation (\ref{EQUY}) reduces to
\begin{equation}
\left(  \tfrac{dr_{0}}{d\tau}\right)  ^{2}=\tfrac{1}{M_{0}^{2}}\left(
M+\tfrac{M_{0}^{2}}{2r_{0}}-\tfrac{Q^{2}}{2r_{0}}\right)  ^{2}-1.
\label{EQUY2}%
\end{equation}
Case $(2)$ has no counterpart in this new regime and Eq.~(\ref{Constraint})
constrains the possible solutions to only the following cases:

\begin{enumerate}
\item {\boldmath$M=M_{0}$}; {\boldmath$Q=M_{0}$. }$r_{0}=r_{0}\left(
0\right)  $ constantly.

\item {\boldmath$M=M_{0}$}; {\boldmath$Q<M_{0}$. }There are no turning points,
the shell starts at rest at infinity and collapses until a
Reissner-Nordstr\"{o}m black-hole is formed with horizons at $r_{0}=r_{\pm
}\equiv M\pm\sqrt{M^{2}-Q^{2}}$ and the singularity in $r_{0}=0$.

\item {\boldmath$M\neq M_{0}$. }There is one turning point $r_{0}^{\ast}$.

\begin{enumerate}
\item {\boldmath$M<M_{0}$}, then necessarily is {\boldmath$Q<M_{0}$}.
Positivity of rhs of (\ref{EQUY2}) requires $r_{0}\leq r_{0}^{\ast}$, where
$r_{0}^{\ast}=\frac{1}{2}\frac{Q^{2}-M_{0}^{2}}{M-M_{0}}$ is the unique
turning point. Then the shell starts from $r_{0}^{\ast}$ and collapses until
the singularity at $r_{0}=0$ is reached.

\item {\boldmath$M>M_{0}$}. The shell has finite radial velocity at infinity.

\begin{enumerate}
\item {\boldmath$Q\leq M_{0}$}. The dynamics are qualitatively analogous to
case (2).

\item {\boldmath$Q>M_{0}$}. Positivity of the rhs of (\ref{EQUY2}) and
(\ref{Constraint}) requires that $r_{0}\geq$ $r_{0}^{\ast}$, where
$r_{0}^{\ast}=\frac{1}{2}\frac{Q^{2}-M_{0}^{2}}{M-M_{0}}$. The shell starts
from infinity and bounces at $r_{0}=r_{0}^{\ast}$, reversing its motion.
\end{enumerate}
\end{enumerate}
\end{enumerate}

In this regime the analytic forms of the solutions are given by Eqs.~(\ref{CASO2}) and (\ref{CASO3}), simply setting $M_{1}=Q_{1}=0$.

\begin{figure}[th]
\includegraphics[width=7cm,height=9cm]{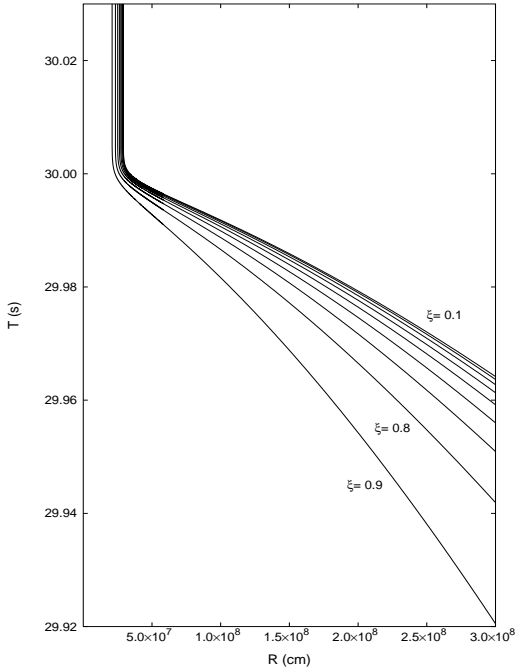}
\caption{Collapse curves in the plane $(T,R)$ for $M=20M_{\odot}$ and for
different values of the parameter $\xi$. The asymptotic behavior is the clear
manifestation of general relativistic effects as the horizon of the EMBH is
approached}%
\label{fig1}%
\end{figure}

Of course, it is of particular interest for the issue of vacuum polarization
the time varying electric field $\mathcal{E}_{r_{0}}=\tfrac{Q}{r_{0}^{2}}$ on
the external surface of the shell. In order to study the variability of
$\mathcal{E}_{r_{0}}$ with time it is useful to consider in the tridimensional
space of parameters $(r_{0},t_{0},\mathcal{E}_{r_{0}})$ the parametric curve
$\mathcal{C}:\left(  r_{0}=\lambda,\quad t_{0}=t_{0}(\lambda),\quad
\mathcal{E}_{r_{0}}=\tfrac{Q}{\lambda^{2}}\right)  $. In astrophysical
applications \cite{RVX03c} we are specially interested in the family of
solutions such that $\frac{dr_{0}}{dt_{0}}$ is 0 when $r_{0}=\infty$ which
implies that $\Gamma=1$. In Fig.~\ref{fig1} we plot the collapse curves in the
plane $(t_{0},r_{0})$ for different values of the parameter $\xi\equiv\frac
{Q}{M}$, $0<\xi<1$. The initial data are chosen so that the integration
constant in Eq.~(\ref{CASO2}) is equal to 0. In all the cases we can
follow the details of the approach to the horizon which is reached in an
infinite Schwarzschild time coordinate.

\begin{figure}[th]
\includegraphics[width=7cm,height=9cm]{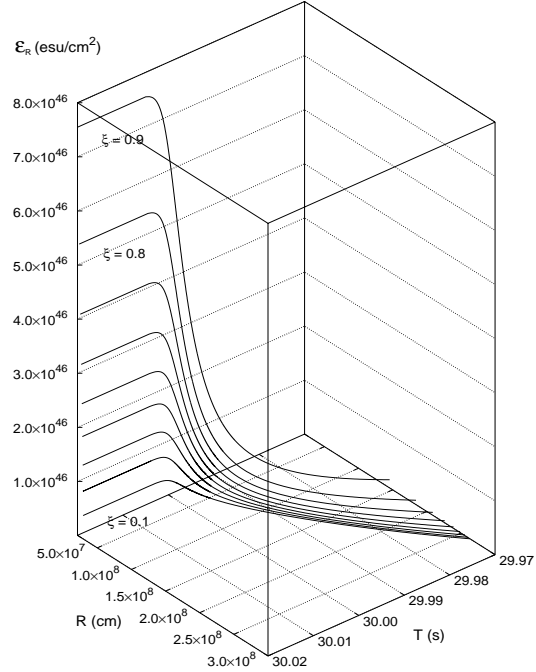}
\caption{Electric field behaviour at the surface of the shell for
$M=20M_{\odot}$ and for different values of the parameter $\xi$. The
asymptotic behavior is the clear manifestation of general relativistic effects
as the horizon of the EMBH is approached}%
\label{fig2}%
\end{figure}

In Fig.~\ref{fig2} we plot the parametric curves $\mathcal{C}$ in the space
$(r_{0},t_{0},\mathcal{E}_{r_{0}})$ for different values of $\xi$. Again we
can follow the exact asymptotic behavior of the curves $\mathcal{C}$,
$\mathcal{E}_{r_{0}}$ reaching the asymptotic value $\frac{Q}{r_{+}^{2}}$. The
detailed knowledge of this asymptotic behavior is of great relevance for the
observational properties of the EMBH formation, see e.g. 
\cite{RVX03c}, \cite{RV02}.

\subsubsection{Irreducible mass of an EMBH and energy extraction processes}

The main objective of this section is to clarify the interpretation of the
mass-energy formula \cite{dr2} for an EMBH. For simplicity we study the case
of a nonrotating EMBH using the results presented in the previous section. As
we saw there, the collapse of a nonrotating charged shell can be described by
exact analytic solutions of the Einstein-Maxwell equations. Consider to two
complementary regions in which the world surface of the shell divides the
space-time: $\mathcal{M}_{-}$ and $\mathcal{M}_{+}$. They are static
space-times; we denote their time-like Killing vectors by $\xi_{-}^{\mu}$ and
$\xi_{+}^{\mu}$ respectively. $\mathcal{M}_{+}$ is foliated by the family
$\left\{  \Sigma_{t}^{+}:t_{+}=t\right\}  $ of space-like hypersurfaces of
constant $t_{+}$.

The splitting of the space-time into the regions $\mathcal{M}_{-}$ and
$\mathcal{M}_{+}$ allows two physically equivalent descriptions of the
collapse and the use of one or the other depends on the question one is
studying. The use of $\mathcal{M}_{-}$ proves helpful for the identification
of the physical constituents of the irreducible mass while $\mathcal{M}_{+}$
is needed to describe the energy extraction process from EMBH. The equation of
motion for the shell, Eq.~\ref{EQUY}, reduces in this case to
\begin{equation}
\left(  M_{0}\tfrac{dr_{0}}{d\tau}\right)  ^{2}=\left(  M+\tfrac{M_{0}^{2}%
}{2r_{0}}-\tfrac{Q^{2}}{2r_{0}}\right)  ^{2}-M_{0}^{2} \label{EQA}%
\end{equation}
in $\mathcal{M}_{-}$ and
\begin{equation}
\left(  M_{0}\tfrac{dr_{0}}{d\tau}\right)  ^{2}=\left(  M-\tfrac{M_{0}^{2}%
}{2r_{0}}-\tfrac{Q^{2}}{2r_{0}}\right)  ^{2}-M_{0}^{2}f_{+} \label{EQAb}%
\end{equation}
in $\mathcal{M}_{+}$. The constraint \ref{Constraint} becomes
\begin{equation}
M-\tfrac{Q^{2}}{2r_{0}}>0. \label{EQO}%
\end{equation}
Since $\mathcal{M}_{-}$ is a flat space-time we can interpret $-\tfrac
{M_{0}^{2}}{2r_{0}}$ in (\ref{EQA}) as the gravitational binding energy of the
system. $\tfrac{Q^{2}}{2r_{0}}$ is its electromagnetic energy. Then Eqs.~(\ref{EQA}), (\ref{EQAb}) differ by the gravitational and electromagnetic
self-energy terms from the corresponding equations of motion of a test particle.

Introducing the total radial momentum $P\equiv M_{0}u^{r}=M_{0}\tfrac{dr_{0}%
}{d\tau}$ of the shell, we can express the kinetic energy of the shell as
measured by static observers in $\mathcal{M}_{-}$ as $T\equiv-M_{0}u_{\mu}%
\xi_{-}^{\mu}-M_{0}=\sqrt{P^{2}+M_{0}^{2}}-M_{0}$. Then from Eq.~(\ref{EQA}) we have
\begin{equation}
M=-\tfrac{M_{0}^{2}}{2r_{0}}+\tfrac{Q^{2}}{2r_{0}}+\sqrt{P^{2}+M_{0}^{2}%
}=M_{0}+T-\tfrac{M_{0}^{2}}{2r_{0}}+\tfrac{Q^{2}}{2r_{0}}. \label{EQC}%
\end{equation}
where we choose the positive root solution due to the constraint (\ref{EQO}).
Eq.~(\ref{EQC}) is the \emph{mass formula} of the shell, which depends on the
time-dependent radial coordinate $r_{0}$ and kinetic energy $T$. If $M\geq Q$,
an EMBH is formed and we have
\begin{equation}
M=M_{0}+T_{+}-\tfrac{M_{0}^{2}}{2r_{+}}+\tfrac{Q^{2}}{2r_{+}}\,, \label{EQL}%
\end{equation}
where $T_{+}\equiv T\left(  r_{+}\right)  $ and $r_{+}=M+\sqrt{M^{2}-Q^{2}}$
is the radius of the external horizon of that
\begin{equation}
M=M_{\mathrm{ir}}+\tfrac{Q^{2}}{2r_{+}}, \label{irrmass}%
\end{equation}
so it follows that {}%
\begin{equation}
M_{\mathrm{ir}}=M_{0}-\tfrac{M_{0}^{2}}{2r_{+}}+T_{+}, \label{EQM}%
\end{equation}
namely that $M_{\mathrm{ir}}$ is the sum of only three contributions: the rest
mass $M_{0}$, the gravitational potential energy and the kinetic energy of the
rest mass evaluated at the horizon. $M_{\mathrm{ir}}$ is independent of the
electromagnetic energy, a fact noticed by Bekenstein \cite{B71}. We have taken
one further step here by identifying the independent physical contributions to
$M_{\mathrm{ir}}$. This will have important consequences for the energetics of
black hole formation (see \cite{RV02}).

Next we consider the physical interpretation of the electromagnetic term
$\tfrac{Q^{2}}{2r_{0}}$, which can be obtained by evaluating the Killing
integral
\begin{eqnarray}
\int_{\Sigma_{t}^{+}}\xi_{+}^{\mu}T_{\mu\nu}^{\mathrm{(em)}}d\Sigma^{\nu}%
&=&\int_{r_{0}}^{\infty}r^{2}dr\int_{0}^{1}d\cos\theta\int_{0}^{2\pi}%
d\phi\ T^{\mathrm{(em)}}{}{}_{0}{}^{0}\nonumber\\
&=&\tfrac{Q^{2}}{2r_{0}}\,, \label{EQR}%
\end{eqnarray}
where $\Sigma_{t}^{+}$ is the space-like hypersurface in $\mathcal{M}_{+}$
described by the equation $t_{+}=t=\mathrm{const}$, with $d\Sigma^{\nu}$ as
its surface element vector. The quantity in Eq.~(\ref{EQR}) differs from the
purely electromagnetic energy
\begin{equation}
\int_{\Sigma_{t}^{+}}n_{+}^{\mu}T_{\mu\nu}^{\mathrm{(em)}}d\Sigma^{\nu}%
=\tfrac{1}{2}\int_{r_{0}}^{\infty}dr\sqrt{g_{rr}}\tfrac{Q^{2}}{r^{2}},
\end{equation}
where $n_{+}^{\mu}=f_{+}^{-1/2}\xi_{+}^{\mu}$ is the unit normal to the
integration hypersurface and $g_{rr}=f_{+}$. This is similar to the analogous
situation for the total energy of a static spherical star of energy density
$\epsilon$ within a radius $r_{0}$, $m\left(  r_{0}\right)  =4\pi\int
_{0}^{r_{0}}dr\ r^{2}\epsilon$, which differs from the pure matter energy
$m_{\mathrm{p}}\left(  r_{0}\right)  =4\pi\int_{0}^{r_{0}}dr\sqrt{g_{rr}}%
r^{2}\epsilon$ by the gravitational energy (see \cite{Gravitation}). Therefore the
term $\tfrac{Q^{2}}{2r_{0}}$ in the mass formula (\ref{EQC}) is the
\emph{total} energy of the electromagnetic field and includes its own
gravitational binding energy. This energy is stored throughout the region
$\mathcal{M}_{+}$, extending from $r_{0}$ to infinity.

We now turn to the problem of extracting the electromagnetic energy from an
EMBH (see \cite{dr2}). We can distinguish between two conceptually physically
different processes, depending on whether the electric field strength
$\mathcal{E}=\frac{Q}{r^{2}}$ is smaller or greater than the critical value
$\mathcal{E}_{\mathrm{c}}$. The maximum value $\mathcal{E}_{+}=\tfrac{Q}%
{r_{+}^{2}}$ of the electric field around an EMBH is reached at the horizon.
We then have the following:

\begin{enumerate}
\item For $\mathcal{E}_{+}<\mathcal{E}_{\mathrm{c}}$ the leading energy
extraction mechanism consists of a sequence of descrete elementary decay
processes of a particle into two oppositely charged particles.
The condition $\mathcal{E}_{+}<\mathcal{E}_{\mathrm{c}}$ implies
\begin{equation}
\xi\equiv\tfrac{Q}{\sqrt{G}M}\lesssim\left\{
\begin{array}
[c]{r}%
\tfrac{GM/c^{2}}{\lambda_{\mathrm{C}}}\left(  \tfrac{e}{\sqrt{G}m_{e}}\right)
^{-1}\sim10^{-6}\tfrac{M}{M_{\odot}}\nonumber\\
1%
\end{array}
\right.  , \label{critical3}%
\end{equation}
where the first line is for $\text{if }\tfrac{M}{M_{\odot}}
\leq10^{6}$, the second line for $\text{if }\tfrac{M}{M_{\odot}}>10^{6}$ and $\lambda_{\mathrm{C}}$ is the Compton wavelength of the electron.
Denardo and Ruffini \cite{dr5a} and Denardo, Hively and Ruffini \cite{dr5b}
have defined as the \emph{effective ergosphere} the region around an EMBH
where the energy extraction processes occur. This region extends from the
horizon $r_{+}$ up to a radius
\begin{equation}
r_{\mathrm{Eerg}}=\tfrac{GM}{c^{2}}\left[  1+\sqrt{1-\xi^{2}\left(
1-\tfrac{e^{2}}{G{m_{e}^{2}}}\right)  }\right]  \simeq\tfrac{e}{m_{e}}%
\tfrac{Q}{c^{2}}\,. \label{EffErg}%
\end{equation}
The energy extraction occurs in a finite number $N_{\mathrm{PD}}$ of such
discrete elementary processes, each one corresponding to a decrease of the EMBH
charge. We have
\begin{equation}
N_{\mathrm{PD}}\simeq\tfrac{Q}{e}\,.
\end{equation}
Since the total extracted energy is (see Eq.~(\ref{irrmass})) $E^{\mathrm{tot}%
}=\tfrac{Q^{2}}{2r_{+}}$, we obtain for the mean energy per accelerated
particle $\left\langle E\right\rangle _{\mathrm{PD}}=\tfrac{E^{\mathrm{tot}}%
}{N_{\mathrm{PD}}}$
\begin{equation}
\left\langle E\right\rangle _{\mathrm{PD}}=\tfrac{Qe}{2r_{+}}=\tfrac{1}%
{2}\tfrac{\xi}{1+\sqrt{1-\xi^{2}}}\tfrac{e}{\sqrt{G}m_{e}}\ m_{e}c^{2}%
\simeq\tfrac{1}{2}\xi\tfrac{e}{\sqrt{G}m_{e}}\ m_{e}c^{2},
\end{equation}
which gives
\begin{equation}
\left\langle E\right\rangle _{\mathrm{PD}}\lesssim\left\{
\begin{array}
[c]{r}%
\tfrac{M}{M_{\odot}}10^{21}eV\quad\text{if }\tfrac{M}{M_{\odot}}\leq10^{6}\\
10^{27}eV\quad\quad\text{if }\tfrac{M}{M_{\odot}}>10^{6}%
\end{array}
\right.  . \label{UHECR}%
\end{equation}

\noindent{\it The theorem of maximum energy extraction from gravitational collapse.}\hskip0.3cm 

One of the crucial aspects of the energy extraction process from an EMBH is
its back reaction on the irreducible mass expressed in \cite{dr2}. Although
the energy extraction processes can occur in the entire effective ergosphere
defined by Eq.~(\ref{EffErg}), only the limiting processes occurring on the
horizon with zero kinetic energy can reach the maximum efficiency while
approaching the condition of total reversibility (see Fig.~2 in \cite{dr2}
for details). The farther from the horizon that a decay occurs, the more it
increases the irreducible mass and loses efficiency. Only in the complete
reversibility limit \cite{dr2} can the energy extraction process from an
extreme EMBH reach the upper value of $50\%$ of the total EMBH energy.

\item As we already discussed, for $\mathcal{E}_{+}\geq
\mathcal{E}_{\mathrm{c}}$ the leading extraction process is the
\emph{collective} process based on the electron-positron plasma generated by
the vacuum polarization. The condition $\mathcal{E}_{+}\geq\mathcal{E}%
_{\mathrm{c}}$ implies
\begin{equation}
\tfrac{GM/c^{2}}{\lambda_{\mathrm{C}}}\left(  \tfrac{e}{\sqrt{G}m_{e}}\right)
^{-1}\simeq2\cdot10^{-6}\tfrac{M}{M_{\odot}}\leq\xi\leq1\,.
\end{equation}
This vacuum polarization process can occur only for an EMBH with mass smaller
than $2\cdot10^{6}M_{\odot}$. The electron-positron pairs are now produced in
the dyadosphere of the EMBH. We have
\begin{equation}
r_{\mathrm{dya}}\ll r_{\mathrm{Eerg}}. \label{dya1}%
\end{equation}
The number of particles created \cite{prx98} is then
\begin{eqnarray}
N_{\mathrm{dya}}&=&\tfrac{1}{3}\left(  \tfrac{r_{\mathrm{dya}}}{\lambda
_{\mathrm{C}}}\right)  \left(  1-\tfrac{r_{+}}{r_{\mathrm{dya}}}\right)
\left[  4+\tfrac{r_{+}}{r_{\mathrm{dya}}}+\left(  \tfrac{r_{+}}%
{r_{\mathrm{dya}}}\right)  ^{2}\right]  \tfrac{Q}{e}\nonumber\\
&\simeq&\tfrac{4}{3}\left(
\tfrac{r_{\mathrm{dya}}}{\lambda_{\mathrm{C}}}\right)  \tfrac{Q}{e}\,.
\label{numdya}%
\end{eqnarray}
The total energy stored in the dyadosphere is \cite{prx98}
\begin{equation}
E_{\mathrm{dya}}^{\mathrm{tot}}=\left(  1-\tfrac{r_{+}}{r_{\mathrm{dya}}%
}\right)  \left[  1-\left(  \tfrac{r_{+}}{r_{\mathrm{dya}}}\right)
^{4}\right]  \tfrac{Q^{2}}{2r_{+}}\simeq\tfrac{Q^{2}}{2r_{+}}\,.
\label{enedya}%
\end{equation}
The mean energy per particle produced in the dyadosphere $\left\langle
E\right\rangle _{\mathrm{dya}}=\tfrac{E_{\mathrm{dya}}^{\mathrm{tot}}%
}{N_{\mathrm{dya}}}$ is then
\begin{equation}
\left\langle E\right\rangle _{\mathrm{dya}}=\tfrac{3}{2}\tfrac{1-\left(
\tfrac{r_{+}}{r_{\mathrm{dya}}}\right)  ^{4}}{4+\tfrac{r_{+}}{r_{\mathrm{dya}%
}}+\left(  \tfrac{r_{+}}{r_{\mathrm{dya}}}\right)  ^{2}}\left(  \tfrac
{\lambda_{\mathrm{C}}}{r_{\mathrm{dya}}}\right)  \tfrac{Qe}{r_{+}}\simeq
\tfrac{3}{8}\left(  \tfrac{\lambda_{\mathrm{C}}}{r_{\mathrm{dya}}}\right)
\tfrac{Qe}{r_{+}}\,, \label{meanenedya}%
\end{equation}
which can be also rewritten as
\begin{equation}
\left\langle E\right\rangle _{\mathrm{dya}}\simeq\tfrac{3}{8}\left(
\tfrac{r_{\mathrm{dya}}}{r_{+}}\right)  \ m_{e}c^{2}\sim\sqrt{\tfrac{\xi
}{M/M_{\odot}}}10^{5}keV\,. \label{GRB}%
\end{equation}
We stress again that the vacuum polarization around an EMBH has been observed
to reach the maximum efficiency limit of $50\%$ of the total mass-energy of an
extreme EMBH (see e.g. \cite{prx98}). The conceptual justification of this
result needs, however, the dynamical analysis of the vacuum polarization
process during the gravitational collapse and the implementation of the
screening of the $e^{+}e^{-}$ neutral plasma generated in this process. This
analysis conceptually validates the reversibility of the process and is given
in the next chapter.
\end{enumerate}

Let us now compare and contrast these two processes. We have
\begin{eqnarray}
r_{\mathrm{Eerg}}&\simeq&\left(  \tfrac{r_{\mathrm{dya}}}{\lambda_{\mathrm{C}}%
}\right)  r_{\mathrm{dya}},\quad N_{\mathrm{dya}}\simeq\left(  \tfrac
{r_{\mathrm{dya}}}{\lambda_{\mathrm{C}}}\right)  N_{\mathrm{PD}}%
,\nonumber\\
\left\langle E\right\rangle _{\mathrm{dya}}&\simeq &\left(  \tfrac
{\lambda_{\mathrm{C}}}{r_{\mathrm{dya}}}\right)  \left\langle E\right\rangle
_{\mathrm{PD}}.
\end{eqnarray}
Moreover we see (Eqs.~(\ref{UHECR}), (\ref{GRB})) that $\left\langle
E\right\rangle _{\mathrm{PD}}$ is in the range of energies of UHECR (see
\cite{NW00} and references therein), while for $\xi\sim0.1$ and $M\sim
10M_{\odot}$, $\left\langle E\right\rangle _{\mathrm{dya}}$ is in the gamma
ray range. In other words, the discrete particle decay process involves a
small number of particles with ultra high energies ($\sim10^{21}eV$), while
vacuum polarization involves a much larger number of particles with lower mean
energies ($\sim10MeV$).

The new conceptual understanding of the mass formula has important
consequences for the energetics of a black hole. The expression for the
irreducible mass in terms of its different physical constituents (Eq.~(\ref{EQM})) leads to a reinterpretation of the energy extraction process
during the formation of a black hole as expressed in \cite{RV02}. It will
certainly be interesting to reach an understanding of the new expression for
the irreducible mass in terms of its thermodynamical analogues.

\noindent{\it The theorem of the maximum energy extraction from gravitational
collapse.}\hskip0.3cm 

In this section we turn to the following aim: pointing out how formula
\ref{EQM} for the $M_{\mathrm{irr}}$ leads to a deeper physical understanding
of the role of the gravitational interaction in the maximum energy extraction
process of an EMBH. This formula can also be of assistance in clarifying some
long lasting epistemological issue on the role of general relativity, quantum
theory and thermodynamics.

It is well known that if a spherically symmetric mass distribution without any
electromagnetic structure undergoes free gravitational collapse, its total
mass-energy $M$ is conserved according to the Birkhoff theorem: the increase
in the kinetic energy of implosion is balanced by the increase in the
gravitational energy of the system. If one considers the possibility that part
of the kinetic energy of implosion is extracted then the situation is very
different: configurations of smaller mass-energy and greater density can be
attained without violating Birkhoff theorem.

We illustrate our considerations with two examples. Concerning the first
example, it is well known from the work of Landau \cite{L32} that at the
endpoint of thermonuclear evolution, the gravitational collapse of a
spherically symmetric star can be stopped by the Fermi pressure of the
degenerate electron gas (white dwarf). A configuration of equilibrium can be
found all the way up to the critical number of particles
\begin{equation}
N_{\mathrm{crit}}=0.775\tfrac{m_{Pl}^{3}}{m_{0}^{3}},
\end{equation}
where the factor $0.775$ comes from the coefficient $\tfrac{3.098}{\mu^{2}}$
of the solution of the Lane-Emden equation with polytropic index $n=3$, and
$m_{Pl}=\sqrt{\tfrac{\hbar c}{G}}$ is the Planck mass, $m_{0}$ is the nucleon
mass and $\mu$ the average number of electrons per nucleon. As the kinetic
energy of implosion is carried away by radiation the star settles down to a
configuration of mass
\begin{equation}
M=N_{\mathrm{crit}}m_{0}-U, \label{BE}%
\end{equation}
where the gravitational binding energy $U$ can be as high as $5.72\times
10^{-4}N_{\mathrm{crit}}m_{0}$.

Similarly Gamov \cite{G51} has shown that a gravitational collapse process to
still higher densities can be stopped by the Fermi pressure of the neutrons
(neutron star) and Oppenheimer \cite{OV39} has shown that, if the effects of
strong interactions are neglected, a configuration of equilibrium exists also
in this case all the way up to a critical number of particles
\begin{equation}
N_{\mathrm{crit}}=0.398\tfrac{m_{Pl}^{3}}{m_{0}^{3}},
\end{equation}
where the factor $0.398$ comes now from the integration of the
Tolman-Oppenheimer-Volkoff equation (see, e.g., Harrison et al.~(1965)
\cite{HTWW}). If the kinetic energy of implosion is again carried away by
radiation of photons or neutrinos and antineutrinos the final configuration is
characterized by the formula (\ref{BE}) with $U\lesssim2.48\times
10^{-2}N_{\mathrm{crit}}m_{0}$. These considerations and the existence of such
large values of the gravitational binding energy have been at the heart of the
explanation of astrophysical phenomena such as red-giant stars and supernovae:
the corresponding measurements of the masses of neutron stars and white dwarfs
have been carried out with unprecedented accuracy in binary systems
\cite{GR75}.

From a theoretical physics point of view it is still an open question how far
such a sequence can go: using causality nonviolating interactions, can one
find a sequence of braking and energy extraction processes by which the
density and the gravitational binding energy can increase indefinitely and the
mass-energy of the collapsed object be reduced at will? This question can also
be formulated in the mass-formula language \cite{dr2} (see also Ref.~\cite{RV02}): given a collapsing core of nucleons with a given rest
mass-energy $M_{0}$, what is the minimum irreducible mass of the black hole
which is formed?

Following the previous two sections, consider a spherical shell of rest mass
$M_{0}$ collapsing in a flat space-time. In the neutral case the irreducible
mass of the final black hole satisfies Eq.~\ref{EQM}. The minimum irreducible
mass $M_{\mathrm{irr}}^{\left(  {\mathrm{min}}\right)  }$ is obtained when the
kinetic energy at the horizon $T_{+}$ is $0$, that is when the entire kinetic
energy $T_{+}$ has been extracted. We then obtain, form Eq.~\ref{EQM}, the
simple result
\begin{equation}
M_{\mathrm{irr}}^{\left(  \mathrm{min}\right)  }=\tfrac{M_{0}}{2}.
\label{Mirrmin}%
\end{equation}
We conclude that in the gravitational collapse of a spherical shell of rest
mass $M_{0}$ at rest at infinity (initial energy $M_{\mathrm{i}}=M_{0}$), an
energy up to $50\%$ of $M_{0}c^{2}$ can in principle be extracted, by braking
processes of the kinetic energy. In this limiting case the shell crosses the
horizon with $T_{+}=0$. The limit $\tfrac{M_{0}}{2}$ in the extractable
kinetic energy can further increase if the collapsing shell is endowed with
kinetic energy at infinity, since all that kinetic energy is in principle extractable.

\begin{figure}[ptb]
\includegraphics[width=7cm,height=9cm
]{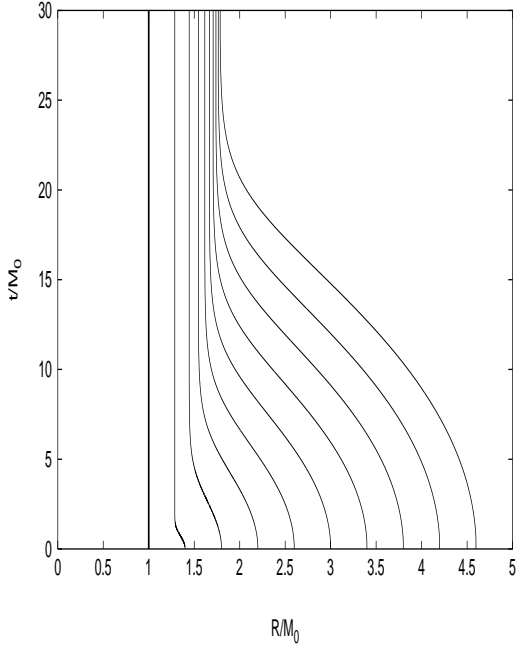}
\caption{Collapse curves for neutral shells with rest mass $M_{0}$ starting at
rest at selected radii $R^{\ast}$ computed by using the exact solutions given
in Ref.~\cite{CRV02}. A different value of
$M_{\mathrm{irr}}$ (and therefore of $r_{+}$) corresponds to each curve. The
time parameter is the Schwarzschild time coordinate $t$ and the asymptotic
behaviour at the respective horizons is evident. The limiting configuration
$M_{\mathrm{irr}}=\tfrac{M_{0}}{2}$ (solid line) corresponds to the case in
which the shell is trapped, at the very beginning of its motion, by the
formation of the horizon.}%
\label{CC}%
\end{figure}

\begin{figure}[ptb]
\includegraphics[height=11.21cm]{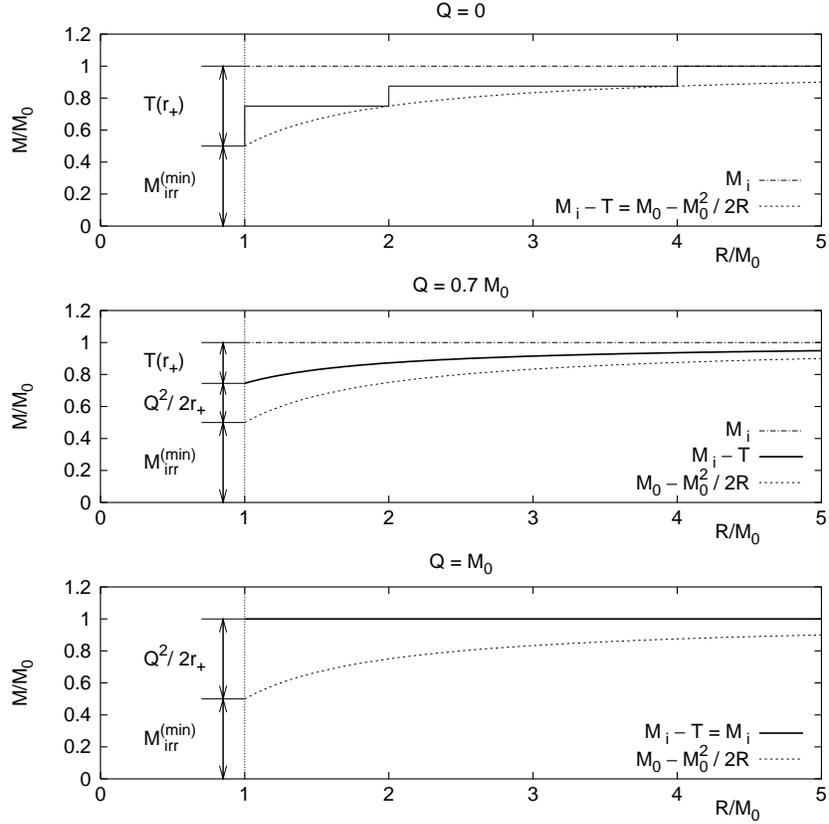}
\caption{Energetics of a shell such that $M_{\mathrm{i}}=M_{0} $, for selected
values of the charge. In the first diagram $Q=0$; the dashed line represents
the total energy for a gravitational collapse without any braking process as a
function of the radius $R$ of the shell; the solid, stepwise line represents a
collapse with suitable braking of the kinetc energy of implosion at selected
radii; the dotted line represents the rest mass energy plus the gravitational
binding energy. In the second and third diagram $Q/M_{0}=0.7$, $Q/M_{0}=1$
respectively; the dashed and the dotted lines have the same meaning as above;
the solid lines represent the total energy minus the kinetic energy. The
region between the solid line and the dotted line corresponds to the stored
electromagnetic energy. The region between the dashed line and the solid line
corresponds to the kinetic energy of collapse. In all the cases the sum of the
kinetic energy and the electromagnetic energy at the horizon is 50\% of
$M_{0}$. Both the electromagnetic and the kinetic energy are extractable. It
is most remarkable that the same underlying process occurs in the three cases:
the role of the electromagnetic interaction is twofold: a) to reduce the
kinetic energy of implosion by the Coulomb repulsion of the shell; b) to store
such an energy in the region around the EMBH. The stored electromagnetic
energy is extractable as shown in Ref.~\cite{RV02}}%
\label{fig3}%
\end{figure}

We have represented in Fig.~\ref{CC} the
world lines of spherical shells of the same rest mass $M_{0}$, starting their
gravitational collapse at rest at selected radii $r_{0}^{\ast}$. These initial
conditions can be implemented by performing suitable braking of the collapsing
shell and concurrent kinetic energy extraction processes at progressively
smaller radii (see also Fig.~\ref{fig3}). The reason for the existence of the
minimum (\ref{Mirrmin}) in the black hole mass is the \textquotedblleft self
closure\textquotedblright\ occurring by the formation of a horizon in the
initial configuration (thick line in Fig.~\ref{CC}).

Is the limit $M_{\mathrm{irr}}\rightarrow\tfrac{M_{0}}{2}$ actually attainable
without violating causality? Let us consider a collapsing shell with charge
$Q$. If $M\geq Q$ an EMBH is formed. As pointed out in the previous section
the irreducible mass of the final EMBH does not depend on the charge $Q$.
Therefore Eqs.~(\ref{EQM}) and (\ref{Mirrmin}) still hold in the charged case.
In Fig.~\ref{fig3} we consider the special case in which the shell is
initially at rest at infinity, i.e. has initial energy $M_{\mathrm{i}}=M_{0}$,
for three different values of the charge $Q$. We plot the initial energy
$M_{i}$, the energy of the system when all the kinetic energy of implosion has
been extracted as well as the sum of the rest mass energy and the
gravitational binding energy $-\tfrac{M_{0}^{2}}{2r_{0}}$ of the system (here
$r_{0}$ is the radius of the shell). In the extreme case $Q=M_{0}$, the shell
is in equilibrium at all radii (see Ref.~\cite{CRV02}) and the kinetic energy
is identically zero. In all three cases, the sum of the extractable kinetic
energy $T$ and the electromagnetic energy $\tfrac{Q^{2}}{2r_{0}}$ reaches
$50\%$ of the rest mass energy at the horizon, according to Eq.~(\ref{Mirrmin}).

What is the role of the electromagnetic field here? If we consider the case of
a charged shell with $Q\simeq M_{0}$, the electromagnetic repulsion implements
the braking process and the extractable energy is entirely stored in the
electromagnetic field surrounding the EMBH (see Ref.~\cite{RV02}). In the
previous section we have outlined two different processes of electromagnetic
energy extraction. We emphasize here that the extraction of $50\%$ of the
mass-energy of an EMBH is not specifically linked to the electromagnetic field
but depends on three factors: a) the increase of the gravitational energy
during the collapse, b) the formation of a horizon, c) the reduction of the
kinetic energy of implosion. Such conditions are naturally met during the
formation of an extreme EMBH but are more general and can indeed occur in a
variety of different situations, e.g. during the formation of a Schwarzschild
black hole by a suitable extraction of the kinetic energy of implosion (see
Fig.~\ref{CC} and Fig.~\ref{fig3}).

Now consider a test particle of mass $m$ in the gravitational field of an
already formed Schwarzschild black hole of mass $M$ and go through such a
sequence of braking and energy extraction processes. Kaplan \cite{K49} found
for the energy $E$ of the particle as a function of the radius $r$
\begin{equation}
E=m\sqrt{1-\tfrac{2M}{r}}. \label{pointtest}%
\end{equation}
It would appear from this formula that the entire energy of a particle could
be extracted in the limit $r\rightarrow2M$. Such $100\%$ efficiency of energy
extraction has often been quoted as evidence for incompatibility between
General Relativity and the second principle of Thermodynamics (see Ref.~\cite{B73} and references therein). J. Bekenstein and S. Hawking have gone as
far as to consider General Relativity not to be a complete theory and to
conclude that in order to avoid inconsistencies with thermodynamics, the
theory should be implemented through a quantum description \cite{B73,hawking}.
Einstein himself often expressed the opposite point of view (see, e.g., Ref.~\cite{D02} and references therein).

The analytic treatment presented 
can clarify this
fundamental issue. It allows to express the energy increase $E$ of a black
hole of mass $M_{1}$ through the accretion of a shell of mass $M_{0}$ starting
its motion at rest at a radius $r_{0}$ in the following formula which
generalizes Eq.~(\ref{pointtest}):
\begin{equation}
E\equiv M-M_{1}=-\tfrac{M_{0}^{2}}{2r_{0}}+M_{0}\sqrt{1-\tfrac{2M_{1}}{r_{0}}%
},
\end{equation}
where $M=M_{1}+E$ is clearly the mass-energy of the final black hole. This
formula differs from the Kaplan formula (\ref{pointtest}) in three respects:
a) it takes into account the increase of the horizon area due to the accretion
of the shell; b) it shows the role of the gravitational self energy of the
imploding shell; c) it expresses the combined effects of a) and b) in an exact
closed formula.

The minimum value $E_{\mathrm{\min}}$ of $E$ is attained for the minimum value
of the radius $r_{0}=2M$: the horizon of the final black hole. This
corresponds to the maximum efficiency of the energy extraction. We have
\begin{eqnarray}
E_{\min}&=&-\tfrac{M_{0}^{2}}{4M}+M_{0}\sqrt{1-\tfrac{M_{1}}{M}}=-\tfrac
{M_{0}^{2}}{4(M_{1}+E_{\min})}\nonumber\\
&+& M_{0}\sqrt{1-\tfrac{M_{1}}{M_{1}+E_{\min}}},
\end{eqnarray}
or solving the quadratic equation and choosing the positive solution for
physical reasons
\begin{equation}
E_{\min}=\tfrac{1}{2}\left(  \sqrt{M_{1}^{2}+M_{0}^{2}}-M_{1}\right)  .
\end{equation}
The corresponding efficiency of energy extraction is
\begin{equation}
\eta_{\max}=\tfrac{M_{0}-E_{\min}}{M_{0}}=1-\tfrac{1}{2}\tfrac{M_{1}}{M_{0}%
}\left(  \sqrt{1+\tfrac{M_{0}^{2}}{M_{1}^{2}}}-1\right)  , \label{efficiency}%
\end{equation}
which is strictly \emph{smaller than} 100\% for \emph{any} given $M_{0}\neq0$.
It is interesting that this analytic formula, in the limit $M_{1}\ll M_{0}$,
properly reproduces the result of equation (\ref{Mirrmin}), corresponding to
an efficiency of $50\%$. In the opposite limit $M_{1}\gg M_{0}$ we have
\begin{equation}
\eta_{\max}\simeq1-\tfrac{1}{4}\tfrac{M_{0}}{M_{1}}.
\end{equation}
Only for $M_{0}\rightarrow0$, Eq.~(\ref{efficiency}) corresponds to an
efficiency of 100\% and correctly represents the limiting reversible
transformations. It seems that the difficulties of reconciling General
Relativity and Thermodynamics are ascribable not to an incompleteness of
General Relativity but to the use of the Kaplan formula in a regime in which
it is not valid.

\section{Dyadosphere formed in gravitational collapses}\label{gr}

\noindent{\it Electric field amplified in gravitational collapse.}\hskip0.3cm 

\begin{figure}[ptb]
\includegraphics[width=12cm,height=6cm]{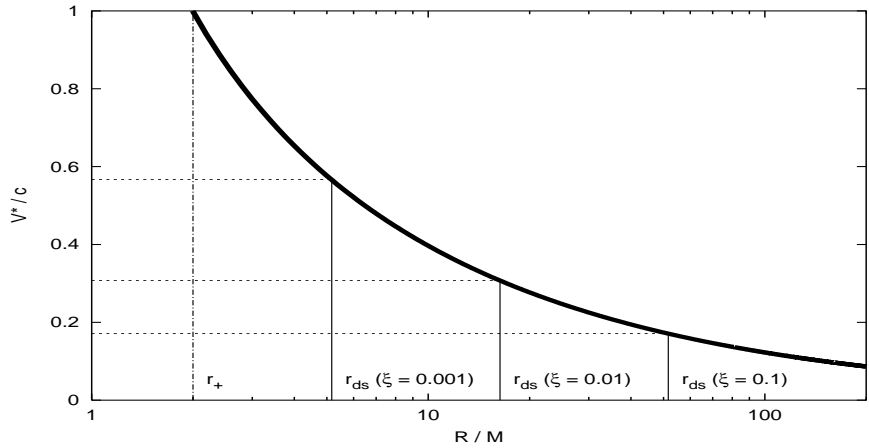}
\caption{Collapse velocity of a charged stellar core of mass $M_{0}%
=20M_{\odot}$ as measured by static observers as a function of the radial
coordinate of the core surface.  Dyadosphere radii for different charge to mass
ratios ($\xi=10^{-3},10^{-2},10^{-1}$) are indicated in the plot together with
the corresponding velocity.}%
\label{Vstar}%
\end{figure}

Initiating with supercritical electric fields on the core surface, we study pair production together with 
gravitational collapse. 
We use the exact solution of
Einstein--Maxwell equations describing the gravitational collapse of a thin
charged shell.
Recall that the region of space--time external to the core is
Reissner--Nordstr\"{o}m with line element
\begin{equation}
ds^{2}=-\alpha^2dt^{2}+\alpha^{-2}dr^{2}+r^{2}d\Omega^{2} \label{st}%
\end{equation}
in Schwarzschild like coordinate $\left(  t,r,\theta,\phi\right)  ,$ where
$\alpha^2=1-2M/r+Q^{2}/r^{2}$; $M$ is the total energy of the
core as measured at infinity and $Q$ is its total charge. Let us label with $r_{0}$ and $t_{0}$ the radial and
time--like coordinate of the core surface, and the equation of motion of the core is \cite{RV02,CRV02,RV03}:%
\begin{eqnarray}
\tfrac{dr_{0}}{dt_{0}}&=&-\tfrac{\alpha^2\left(  r_{0}\right)  }{\Omega\left(
r_{0}\right)  }\sqrt{\Omega^{2}\left(  r_{0}\right)  -\alpha^2\left(  r_{0}\right)  },\nonumber\\
\Omega\left(  r_{0}\right)  &=& \tfrac{M}{M_{0}}-\tfrac{M_{0}^{2}+Q^{2}}%
{2M_{0}r_{0}};
\label{Motion2}%
\end{eqnarray}
$M_{0}$ being the rest mass of the shell. The analytical solutions of
Eq.~(\ref{Motion2}) were found $
t_{0}=t_{0}\left(  r_{0}\right)  ,
$
and the core collapse speed $V^{\ast}(r_0)$ as a function of $r_0$ is plotted in Fig.~\ref{Vstar}, where we indicate
$V_{\mathrm{ds}}^{\ast}\equiv V^{\ast}\vert _{r_{0}=r_{\mathrm{ds}}}$ as the velocity of the core 
at the Dyadosphere radius $r_{\mathrm{ds}}$.

We now turn to the pair creation and plasma oscillation taking place in the classical electric and gravitational fields
during the gravitational collapse of a charged overcritical stellar core. As already show in Fig.~\ref{OscillationLT}, (i) 
the electric field oscillates with lower and lower amplitude around $0 $;
(ii) electrons and positrons oscillates back and forth in the radial
direction with ultra relativistic velocity, as result the oscillating charges are confined 
in a thin shell whose radial dimension is given by the elongation $\Delta l=\left|  l-l_{0}\right|  $ of
the oscillations, where $l_{0}$ is the radial coordinate of the center of
oscillation and
\begin{equation}
\Delta l=\int_{0}^{\Delta t}\tfrac{\pi_{e\parallel}}{\rho_{e}}dt,
\end{equation}
where $\pi_{e\parallel}/\rho_{e}\equiv v$ is the radial mean velocity of
charges.
\begin{figure}[ptb]
\def\fsz{\footnotesize}
\def\ssz{\scriptsize}
\def\tsz{\tiny}
\def\dst{\displaystyle}\unitlength1mm
\includegraphics[width=7cm,height=7cm]{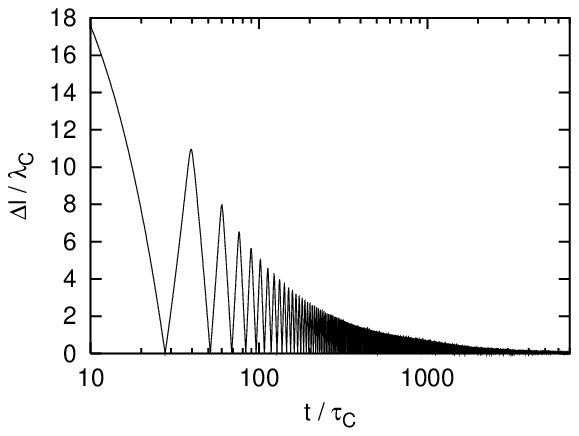}
\includegraphics[width=7cm,height=7cm]{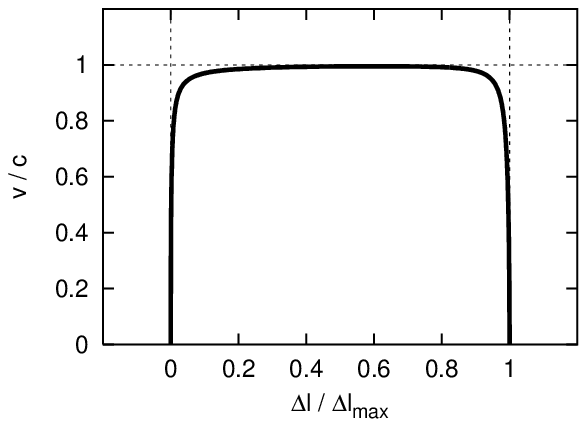}
\caption{In left figure: Electrons elongation as function of time in the case
$r=r_{\mathrm{ds}}/3$. The oscillations are damped in a time of the order of
$10^{3}-10^{4}\tau_{\mathrm{C}}$. The right figure: Electrons mean velocity as a function of the elongation during the
first half oscillation. The plot summarize the oscillatory behaviour: as the
electrons move, the mean velocity grows up from $0$ to the speed of light and
then falls down at $0$ again.
} 
\label{Elong&VvsD}%
\end{figure}
In Fig.~\ref{Elong&VvsD}, we plot the elongation $\Delta l$ as a function of time and
electron mean velocity $v$ as a function of the
elongation during the first half period $\Delta t$ of oscillation. This shows
precisely the characteristic time $\Delta t$ and size $\Delta l$ of charge confinement due to plasma oscillation.

In the time
$\Delta t$ the charge oscillations prevent a macroscopic current from flowing
through the surface of the core. Namely in the time $\Delta t$ the core moves
inwards of
\begin{equation}
\Delta r^{\ast}=V^{\ast}\Delta t\gg\Delta l. \label{Dr*}%
\end{equation}
Since the plasma charges are confined within a region of thickness $\Delta l$,
due to Eq.~(\ref{Dr*}) no charge ``reaches'' the surface of the core which can
neutralize it and the initial charge of the core remains untouched. For
example in the case
$M=20M_{\odot},
\xi=0.1$, and $
r=\tfrac{1}{3}r_{\mathrm{ds}},
$ we have
\begin{equation}
\Delta l\lesssim30\lambda_{\mathrm{C}},\quad
\Delta t\sim10^{3}\tau_{\mathrm{C}},\quad
V^{\ast}\sim0.3c,
\end{equation}
and $\Delta r^{\ast}\gg\Delta l$.
We conclude that the core is not discharged or, in other words, the electric
charge of the core is stable against vacuum polarization and electric field $E=Q/r_0^2$ is amplified during the
gravitational collapse. As a consequence, an enormous amount ($N\sim
Qr_{\mathrm{ds}}/e\lambda_{\mathrm{C}})$ (\ref{n}) as claimed in Ref.~\cite{prx98}) of
pairs is left behind the collapsing core and Dyadosphere \cite{prx98} is formed.

\noindent{\it Plasma expansion during gravitational collapse.}\hskip0.3cm 

The $e^{+}e^{-}$ pairs generated by the vacuum polarization process around the
core are entangled in the electromagnetic field \cite{RVX03b}, and 
thermalize in an electron--positron--photon plasma on a time scale
$\sim 10^{4}\tau_C$ \cite{RVX03d} (see Fig.~\ref{OscillationLT}). 
As soon as the thermalization has occurred, the hydrodynamic expansion of this electrically neutral
plasma starts \cite{RSWX00,rswx2000}. While the temporal evolution of the
$e^{+}e^{-}\gamma$ plasma takes place, the gravitationally collapsing core
moves inwards, giving rise to a further amplified supercritical field, which
in turn generates a larger amount of $e^{+}e^{-}$ pairs leading to a yet
higher temperature in the newly formed $e^{+}e^{-}\gamma$ plasma.
We report progress in this theoretically
challenging process which is marked by distinctive and precise quantum and
general relativistic effects. As presented in Ref.~\cite{RVX03c}: 
we do not consider an already
formed EMBH, but we follow the dynamical phase of the formation
of Dyadosphere and of the asymptotic approach to the horizon by examining the
time varying process at the surface of the gravitationally collapsing core. 

It is
worthy to remark that the time--scale of hydrodynamic evolution ($t\sim0.1s$)
is, in any case, much larger than both the time scale needed for
\textquotedblleft all pairs to be created\textquotedblright\ ($\sim10^{3}%
\tau_{\mathrm{C}})$, and the thermalization
time--scale ($\sim10^{4}\tau_{\mathrm{C}}$, see Fig.~\ref{OscillationLT}) and therefore 
it is consistent to consider pair production, plus
thermalization, and hydrodynamic expansion as separate regimes of the system.
We assume the initial condition that the Dyadosphere starts
to be formed at the instant of gravitational collapse $t_{\mathrm{ds}}=t_{0}\left(  r_{\mathrm{ds}%
}\right)  =0$, and $r_{\mathrm{ds}}=R_c$ the radius of massive nuclear core.
Having formulated the core collapse in General Relativity in Eq.~(\ref{Motion2}%
), we discretize the gravitational collapse of a spherically symmetric core by
considering a set of events ($N-$events) along the world line of a point of fixed angular
position on the collapsing core surface. Between each of these events we
consider a spherical shell of plasma of constant coordinate thickness
$\Delta r$ so that:

\begin{enumerate}
\item $\Delta r$ is assumed to be a constant which is small with respect to
the core radius;

\item $\Delta r$ is assumed to be large with respect to the mean free path of
the particles so that the statistical description of the $e^{+}e^{-}\gamma$
plasma can be used;

\item There is no overlap among the slabs and their union describes the
entirety of the process.
\end{enumerate}

We check that the final results are independent of the special value of the
chosen $\Delta r$ and $N$.

In each slab the processes of $e^{+}e^{-}$-pair production, oscillation with electric field and 
thermalization with photons 
are considered. While the average of the electric field
$\mathcal{E}$ over one oscillation is $0$, the average of $\mathcal{E}^{2}$ is
of the order of $\mathcal{E}_{\mathrm{c}}^{2}$, therefore the energy density
in the pairs and photons, as a function of $r_{0}$, is given by 
\begin{equation}
\epsilon_{0}\left(  r_{0}\right)  =\tfrac{1}{8\pi}\left[  \mathcal{E}%
^{2}\left(  r_{0}\right)  -\mathcal{E}_{\mathrm{c}}^{2}\right]  =\tfrac
{\mathcal{E}_{\mathrm{c}}^{2}}{8\pi}\left[  \left(  \tfrac{r_{\mathrm{ds}}%
}{r_{0}}\right)  ^{4}-1\right]  .\label{eps0}%
\end{equation}
For the number densities of $e^{+}e^{-}$ pairs and photons at thermal
equilibrium we have $n_{e^{+}e^{-}}\simeq n_{\gamma}$; correspondingly the
equilibrium temperature $T_{0}$, which is clearly a function of $r_{0}$ and is
different for each slab, is such that \cite{RSWX00}
\begin{equation}
\epsilon\left(  T_{0}\right)  \equiv\epsilon_{\gamma}\left(  T_{0}\right)
+\epsilon_{e^{+}}\left(  T_{0}\right)  +\epsilon_{e^{-}}\left(  T_{0}\right)
=\epsilon_{0},\label{eq0}%
\end{equation}
with $\epsilon$ and $n$ given by Fermi (Bose) integrals (with zero chemical
potential):
\begin{eqnarray}
\epsilon_{e^{+}e^{-}}\left(  T_{0}\right)   &  =&\tfrac{2}{\pi^{2}\hbar^{3}%
}\int_{m_{e}}^{\infty}\tfrac{\left(  E^{2}-m_{e}^{2}\right)  ^{1/2}}%
{\exp\left(  E/kT_{0}\right)  +1}E^{2}dE,\nonumber\\
\epsilon_{\gamma}\left(
T_{0}\right)  &=&\tfrac{\pi^{2}}{15\hbar^{3}}\left(  T_{0}\right)
^{4},\label{Integrals1}\\
n_{e^{+}e^{-}}\left(  T_{0}\right)   &  =&\tfrac{1}{\pi^{2}\hbar^{3}}%
\int_{m_{e}}^{\infty}\tfrac{\left(  E^{2}-m_{e}^{2}\right)  ^{1/2}}%
{\exp\left(  E/kT_{0}\right)  +1}EdE,\nonumber\\ n_{\gamma}\left(  T_{0}\right)
&=&\tfrac{2\zeta\left(  3\right)  }{\hbar^{3}}\left(  T_{0}\right)
^{3}.\label{Integrals2}%
\end{eqnarray}
From the conditions set by
Eqs.~(\ref{eq0}), (\ref{Integrals1}), (\ref{Integrals2}), we can now turn to
the dynamical evolution of the $e^{+}e^{-}\gamma$ plasma in each slab. We use
the covariant conservation of energy momentum and the rate equation for the
number of pairs in the Reissner--Nordstr\"{o}m geometry external to the
core:
\begin{align}
\nabla_{a}T^{ab}  & =0,\quad\label{Tab}\\
\nabla_{a}\left(  n_{e^{+}e^{-}}u^{a}\right)    & =\overline{\sigma v}\left[
n_{e^{+}e^{-}}^{2}\left(  T\right)  -n_{e^{+}e^{-}}^{2}\right]  ,\label{rate}%
\end{align}
where $T^{ab}=\left(  \epsilon+p\right)  u^{a}u^{b}+pg^{ab}$ is the
energy--momentum tensor of the plasma with proper energy density $\epsilon$
and proper pressure $p$, $u^{a}$ is the fluid $4-$velocity, $n_{e^{+}e^{-}}$
is the number of pairs, $n_{e^{+}e^{-}}\left(  T\right)  $ is the equilibrium
number of pairs and $\overline{\sigma v}$ is the mean of the product of the
$e^{+}e^{-}$ annihilation cross-section and the thermal velocity of pairs.
In each slab the plasma remains
at thermal equilibrium in the initial phase of the expansion and the right
hand side of the rate Eq.~(\ref{rate}) is effectively $0$, see Fig.~24 (second
panel) of \cite{Brasile} for details.

If we denote by $\xi^{a}$ the static Killing vector field normalized at unity
at spacial infinity and by $\left\{  \Sigma_{t}\right\}  _{t}$ the family of
space-like hypersurfaces orthogonal to $\xi^{a}$ ($t$ being the Killing time)
in the Reissner--Nordstr\"{o}m geometry, from Eqs.~(\ref{rate}), the following
integral conservation laws can be derived
\begin{equation}
\int_{\Sigma_{t}}\xi_{a}T^{ab}d\Sigma_{b}=E,\quad\int_{\Sigma_{t}}%
n_{e^{+}e^{-}}u^{b}d\Sigma_{b}=N_{e^{+}e^{-}}, \label{Ne}%
\end{equation}
where $d\Sigma_{b}=\alpha^{-2}\xi_{b}r^{2}\sin\theta drd\theta d\phi$ is the
vector surface element, $E$ the total energy and $N_{e^{+}e^{-}}$ the total
number of pairs which remain constant in each slab. We then have
\begin{equation}
\left[  \left(  \epsilon+p\right)  \gamma^{2}-p\right]  r^{2}=\mathfrak{E}%
,\quad n_{e^{+}e^{-}}\gamma\alpha^{-1}r^{2}=\mathfrak{N}_{e^{+}e^{-}},
\label{ne}%
\end{equation}
where $\mathfrak{E}$ and $\mathfrak{N}_{e^{+}e^{-}}$ are constants and
\begin{equation}
\gamma\equiv\alpha^{-1}u^{a}\xi_{a}=\left[  1-\alpha^{-4}\left(  \tfrac
{dr}{dt}\right)  ^{2}\right]  ^{-1/2}%
\end{equation}
is the Lorentz $\gamma$ factor of the slab as measured by static observers. We
can rewrite Eqs.~(\ref{Ne}) for each slab as
\begin{align}
\left(  \tfrac{dr}{dt}\right)  ^{2}  &  =\alpha^{4}f_{r_{0}},\label{eq17}\\
\left(  \tfrac{r}{r_{0}}\right)  ^{2}  &  =\left(  \tfrac{\epsilon+p}%
{\epsilon_{0}}\right)  \left(  \tfrac{n_{e^{+}e^{-}0}}{n_{e^{+}e^{-}}}\right)
^{2}\left(  \tfrac{\alpha}{\alpha_{0}}\right)  ^{2}-\tfrac{p}{\epsilon_{0}%
}\left(  \tfrac{r}{r_{0}}\right)  ^{4},\label{eq18}\\
f_{r_{0}}  &  =1-\left(  \tfrac{n_{e^{+}e^{-}}}{n_{e^{+}e^{-}0}}\right)
^{2}\left(  \tfrac{\alpha_{0}}{\alpha}\right)  ^{2}\left(  \tfrac{r}{r_{0}%
}\right)  ^{4} \label{eq19}%
\end{align}
where pedex $_{0}$ refers to quantities evaluated at selected initial times
$t_{0}>0$, having assumed $r\left(  t_{0}\right)  =r_{0}$, $\left.
dr/dt\right|  _{t=t_{0}}=0$, $T\left(  t_{0}\right)  =T_{0}$.

Eq.~(\ref{eq17}) is only meaningful when $f_{r_{0}}\left(  r\right)  \geq0$.
From the structural analysis of such equation it is clearly identifiable a
critical radius $r_{0}$ such that:

\begin{itemize}
\item for any slab initially located at $r_{0}>\bar{R}$ we have $f_{r_{0}%
}\left(  r\right)  \geq0$ for any value of $r\geq r_{0}$ and $f_{r_{0}}\left(
r\right)  <0$ for $r\lesssim r_{0}$; therefore a slab initially located at a
radial coordinate $r_{0}>\bar{R}$ moves outwards,

\item for any slab initially located at $r_{0}<\bar{R}$ we have $f_{r_{0}%
}\left(  r\right)  \geq0$ for any value of $r_{+}<r\leq r_{0}$ and $f_{r_{0}%
}\left(  r\right)  <0$ for $r\gtrsim r_{0}$; therefore a slab initially
located at a radial coordinate $r_{0}<\bar{R}$ moves inwards and is trapped by
the gravitational field of the collapsing core.
\end{itemize}

\begin{figure}[ptb]
\def\fsz{\footnotesize}
\def\ssz{\scriptsize}
\def\tsz{\tiny}
\def\dst{\displaystyle}\unitlength1mm
\includegraphics[width=7cm,height=9cm]{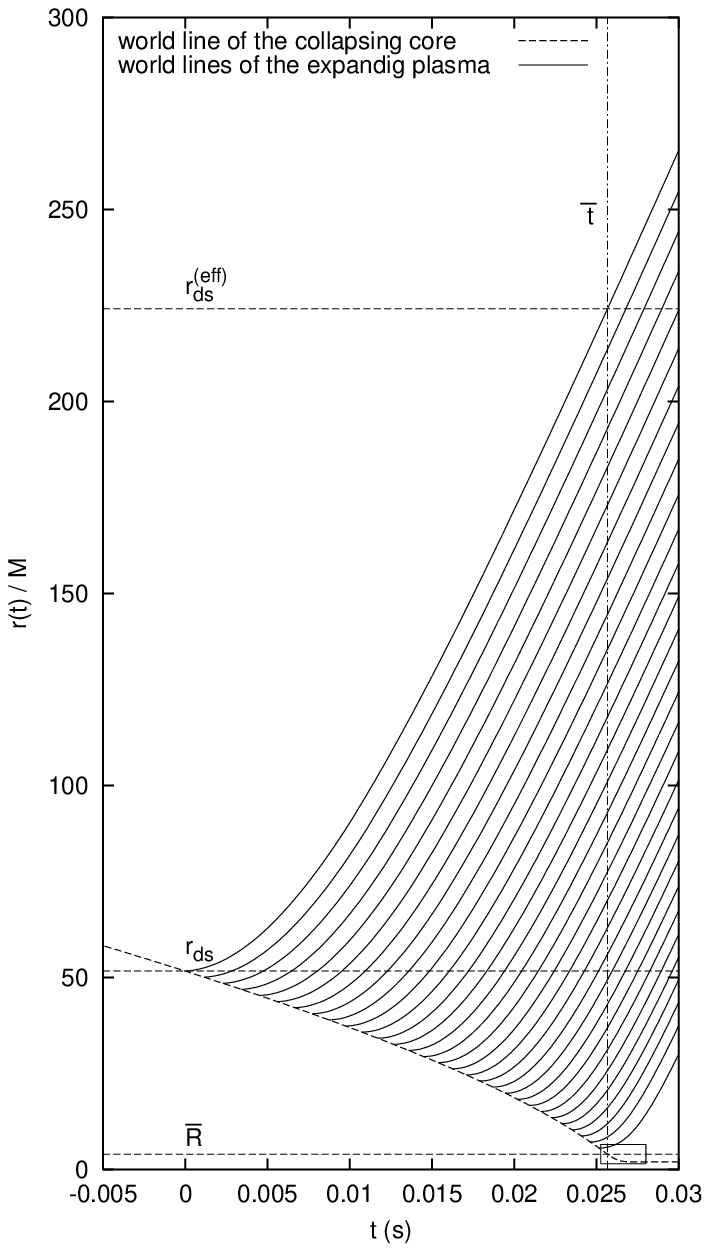}
\includegraphics[width=7cm,height=9cm]{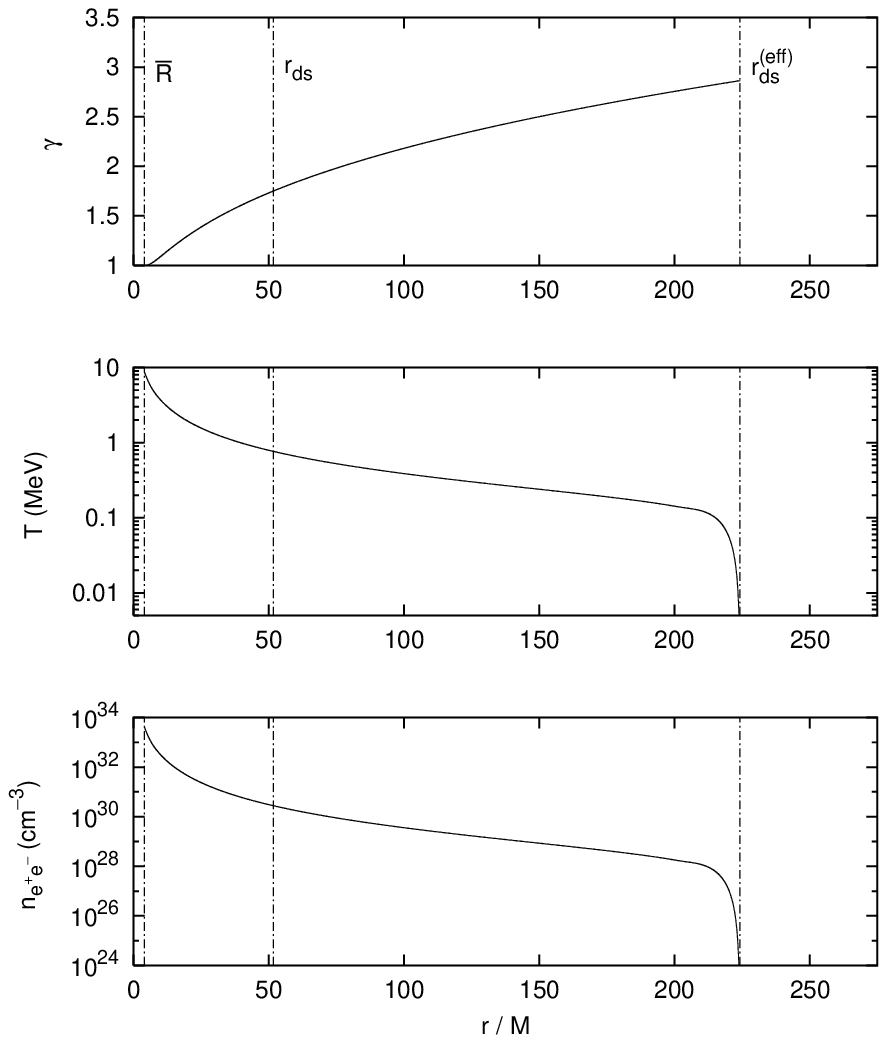}
\caption{In left figure: World line of the collapsing charged core (dashed line) as derived
from Eq.~(\ref{Motion2}); world
lines of slabs of plasma for selected radii $r_{0}$ in the interval $\left(
\bar{R},r_{\mathrm{ds}}\right)  $. At time $\bar{t}$ the expanding plasma
extends over a region which is almost one order of magnitude larger than the
Dyadosphere. The small rectangle in the right bottom is enlarged in
Fig.~\ref{trap}. The right figure: Physical parameters in the effective Dyadosphere: Lorentz $\gamma$
factor, proper temperature and proper $e^{+}e^{-}$ number density as functions
at time $\bar{t}$.
} 
\label{exp&ED}%
\end{figure}

\begin{figure}[th]
\includegraphics[width=12cm,height=6cm]{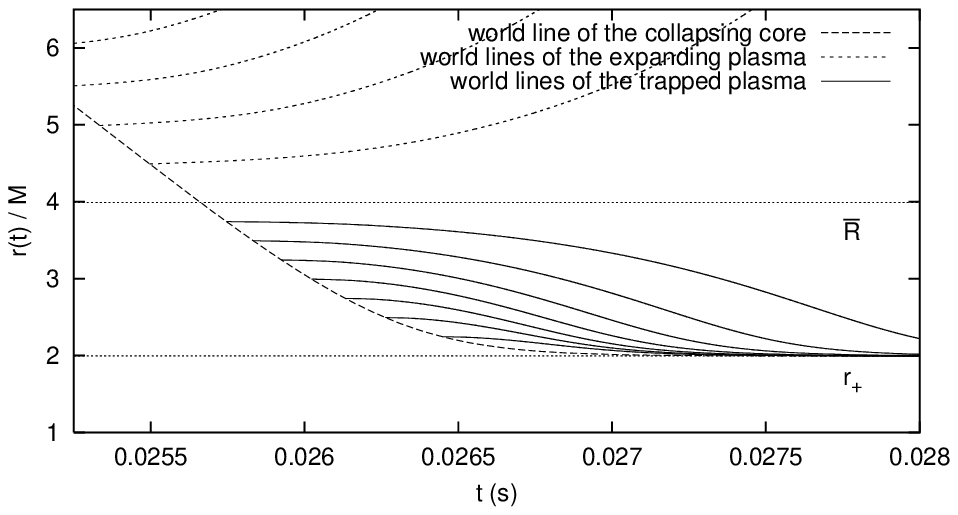}
\caption{Enlargement of the small rectangle in the right bottom of left figure in 
Fig.~\ref{exp&ED}. World--lines of slabs of plasma for selected radii $r_{0}$ in
the interval $\left(  r_{+,}\bar{R}\right)  $.}%
\label{trap}%
\end{figure}

We define the surface $r=\bar{R}$, the \emph{Dyadosphere trapping surface
}(DTS). The radius $\bar{R}$ of DTS is generally evaluated by the condition
$\left.  \tfrac{df_{\bar{R}}}{dr}\right\vert _{r=\bar{R}}=0$. $\bar{R}$ is so
close to the horizon value $r_{+}$ that the initial temperature $T_{0}$
satisfies $kT_{0}\gg m_{e}c^{2}$ and we can obtain for $\bar{R}$ an analytical
expression. Namely the ultra relativistic approximation of all Fermi integrals,
Eqs.~(\ref{Integrals1}) and (\ref{Integrals2}), is justified and we have
$n_{e^{+}e^{-}}\left(  T\right)  \propto T^{3}$ and therefore $f_{r_{0}}%
\simeq1-\left(  T/T_{0}\right)  ^{6}\left(  \alpha_{0}/\alpha\right)
^{2}\left(  r/r_{0}\right)  ^{4}$ ($r\leq\bar{R}$). The defining equation of
$\bar{R}$, together with (\ref{eq19}), then gives
\begin{equation}
\bar{R}=2M\left[  1+\left(  1-3Q^{2}/4M^{2}\right)  ^{1/2}\right]
>r_{+}.\label{Rbar}%
\end{equation}
In the case of an EMBH with $M=20M_{\odot}$, $Q=0.1M$, we compute:
\begin{itemize}
\item the fraction of energy trapped in DTS:
\begin{equation}
\bar{E}=\int_{r_{+}<r<\bar{R}}\alpha\epsilon_{0}d\Sigma\simeq0.53\int
_{r_{+}<r<r_{\mathrm{ds}}}\alpha\epsilon_{0}d\Sigma;
\end{equation}

\item the world--lines of slabs of plasma for selected $r_{0}$ in the interval
$\left(  \bar{R},r_{\mathrm{ds}}\right)  $ (see left figure in Fig.~\ref{exp&ED});

\item the world--lines of slabs of plasma for selected $r_{0}$ in the interval
$\left(  r_{+},\bar{R}\right)  $ (see Fig.~\ref{trap}).
\end{itemize}
At time $\bar{t}\equiv t_{0}\left(  \bar{R}\right)  $ when the DTS is formed,
the plasma extends over a region of space which is almost one order of
magnitude larger than the Dyadosphere and which we define as the
\emph{effective Dyadosphere}. The values of the Lorentz $\gamma$ factor, the
temperature and $e^{+}e^{-}$ number density in the effective Dyadosphere are
given in the right figure in Fig.~\ref{exp&ED}.

\section{Hydrodynamic expansion after gravitational collapse.}

\noindent{\it Plasma expansion after gravitational collapse.}\hskip0.3cm
 
\begin{figure}[ptb]
\def\fsz{\footnotesize}
\def\ssz{\scriptsize}
\def\tsz{\tiny}
\def\dst{\displaystyle}\unitlength1mm
\includegraphics[width=7cm,height=7cm]{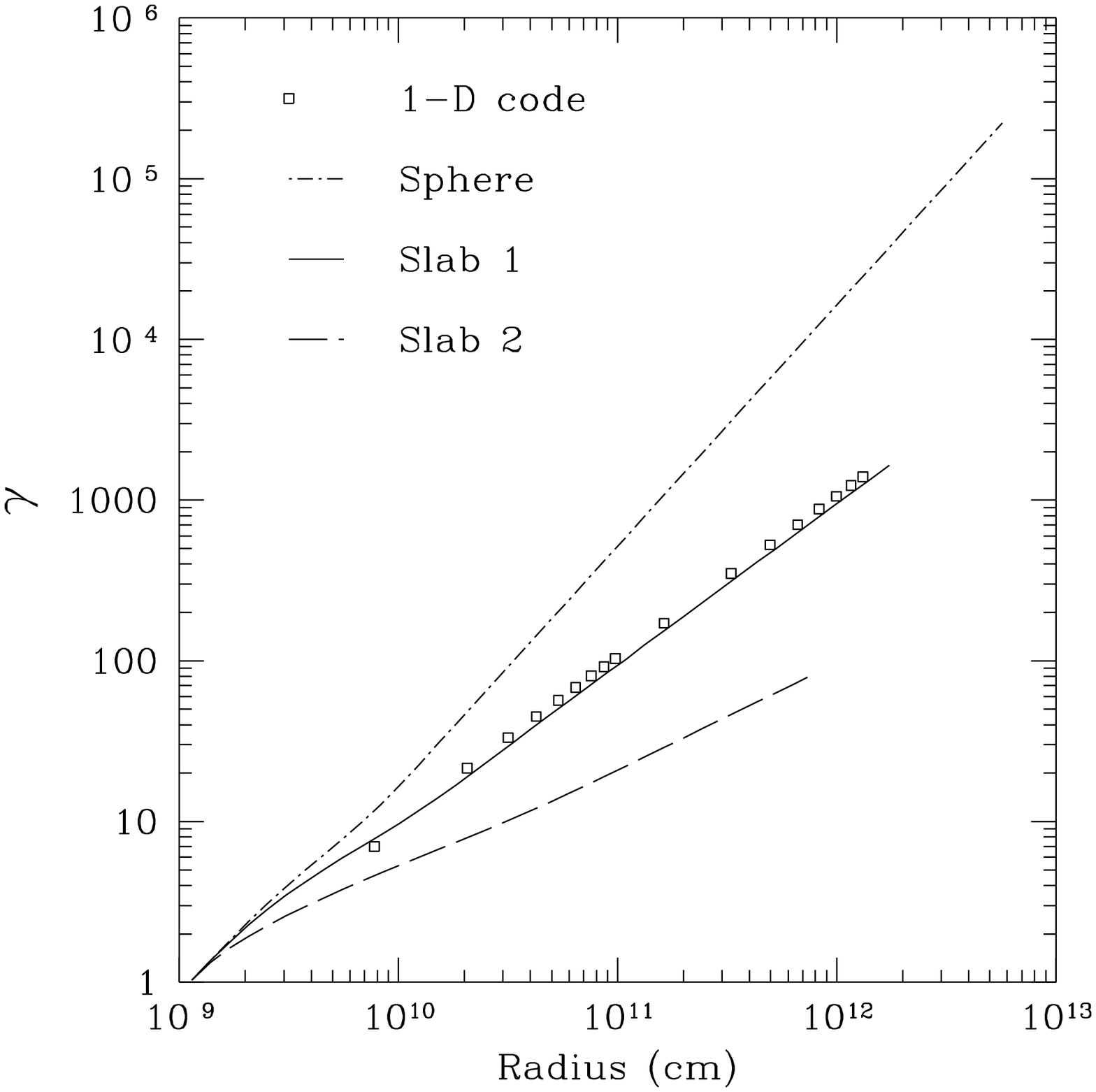}
\includegraphics[width=7cm,height=7cm]{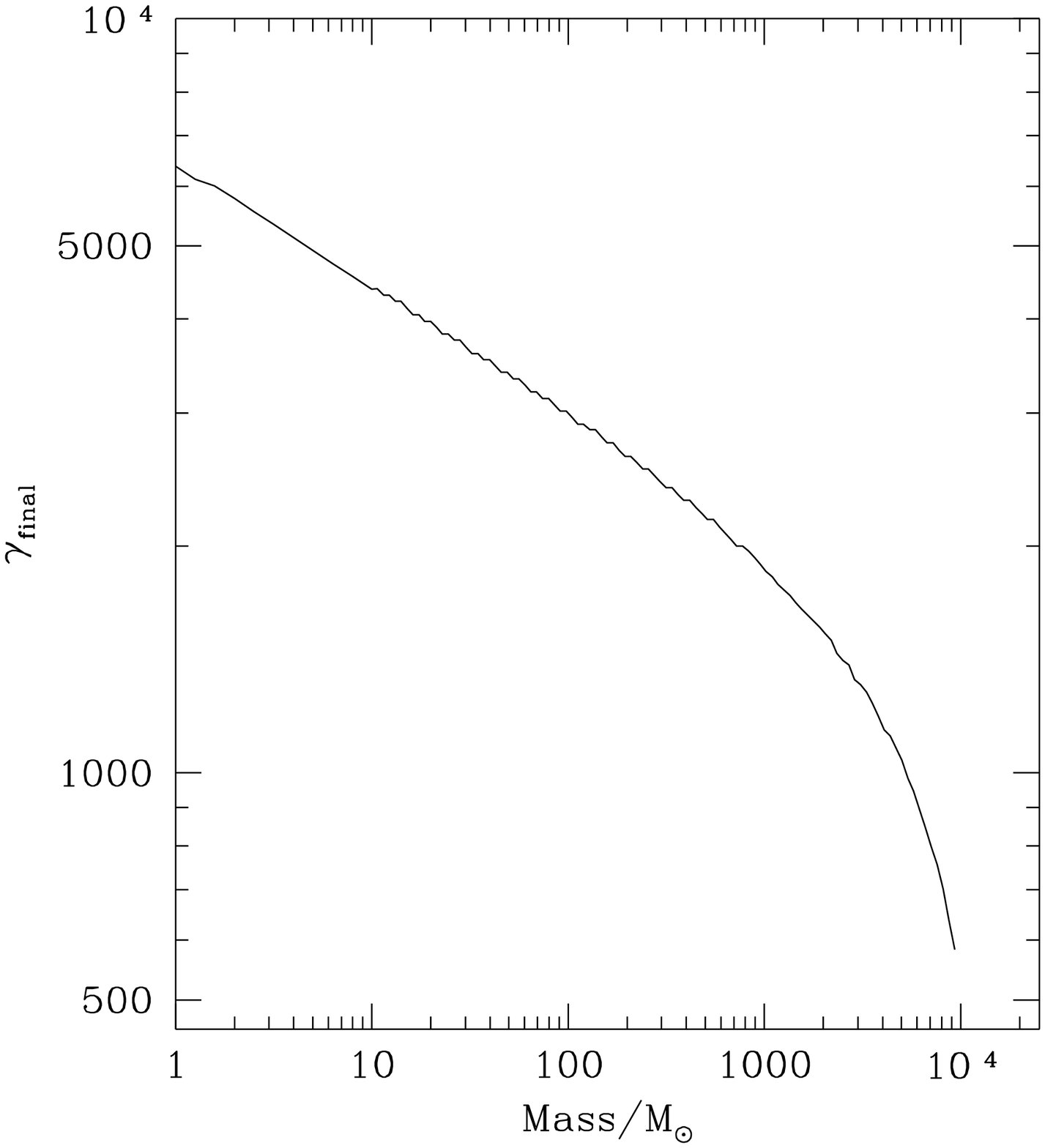}
\caption{In left figure: Lorentz gamma factor $\gamma$ as a function of radius.
Three models for the expansion pattern of the PEM-pulse are
compared with the results of the one dimensional hydrodynamic code for
a $1000 M_\odot$ black hole with charge to mass ratio $\xi=0.1$. The 1-D
code has an expansion pattern that strongly resembles that of a shell
with constant coordinate thickness. The right figure: In the expansion model of a shell with
constant coordinate thickness, the decoupling gamma-factor
$\gamma_{\rm final}$ as a function of EMBH masses is plotted with charge to mass ratio
$\xi=0.1$.
} 
\label{figshells&dgamma}%
\end{figure}

After gravitational collapse, the adiabatically hydrodynamic expansion of $e^+e^-\gamma$ plasma continues, 
obeying hydrodynamic and rate equations (\ref{Tab},\ref{rate}) for 
conservations of energy-momentum, entropy and particle numbers, until it becomes optically
thin:
\begin{equation} 
\int_R dr(n_{e^\pm})\sigma_T\simeq O(1),
\label{thin}
\end{equation}
where $\sigma_T =0.665\times 10^{-24}
{\rm cm^2}$ is the Thomson cross-section and the integration is over the radial interval of the plasma in the
comoving frame. At this point the energy is virtually entirely in the
form of free-streaming photons.
The calculations were independently performed by numerical simulation 
in Lawrence Livermore National Laboratory using the one dimensional (1-D) hydrodynamic code, and analytical approach in ICRA, University of Rome \cite{RSWX00,rswx2000} using approximate hydrodynamic and rate equations (\ref{Tab},\ref{rate}) neglecting gravitational effects,
\begin{eqnarray}
\tfrac{\epsilon_{0}}{\epsilon}  &\!\!\! =\!\!\!&\left(  \tfrac{\gamma\mathcal{V}}%
{\gamma_{0}\mathcal{V}_{0}}\right)  ^{\Gamma},\!\!\!\quad\!\!\!
\tfrac{\gamma}{\gamma_{0}}   =\sqrt{\tfrac{\epsilon_{0}\mathcal{V}_{0}%
}{\epsilon\mathcal{V}}},\!\!\!\quad\!\!\! \Rightarrow  \tfrac{d(\epsilon\gamma^2\mathcal{V})}{dt}=0; \label{Eq3}
\end{eqnarray}
\begin{eqnarray}
\tfrac{\partial}{\partial t}N_{e^{+}e^{-}}  &\!\!\!\!\!\! =\!\!\!\!\!\!&-N_{e^{+}e^{-}}\tfrac
{1}{\mathcal{V}}\tfrac{\partial\mathcal{V}}{\partial t}\!\!\!+\!\!\!\tfrac{\overline{\sigma
v}}{\gamma^{2}}\left[  N_{e^{+}e^{-}}^{2}\left(  T\right)
-N_{e^{+}e^{-}}^{2}\right]  ,\label{Eq3+}
\end{eqnarray}
where the thermal index $\Gamma=1+p/\epsilon$, $\mathcal{V}$ is the volume of a single slab,
$N_{e^{+}e^{-}}=\gamma n_{e^{+}e^{-}}$ is the pair number density as measured
in the laboratory frame by an observer at rest with the black hole, and
$N_{e^{+}e^{-}}\left(  T\right)  $ is the equilibrium laboratory pair number density.   
In Fig.~\ref{figshells&dgamma} we show the Lorentz gamma factor as a function of radius 
and the decoupling gamma factor at the transparency point 
(\ref{thin}) as a function of EMBH masses. We find the expansion pattern of a shell of constant 
coordinate thickness with the astrophysically unprecedented large Lorentz factors, named by the Pair-ElectroMagnetic Pulse (PEM-pulse).
\begin{figure}[ptb]
\def\fsz{\footnotesize}
\def\ssz{\scriptsize}
\def\tsz{\tiny}
\def\dst{\displaystyle}\unitlength1mm
\includegraphics[width=7cm,height=7cm]{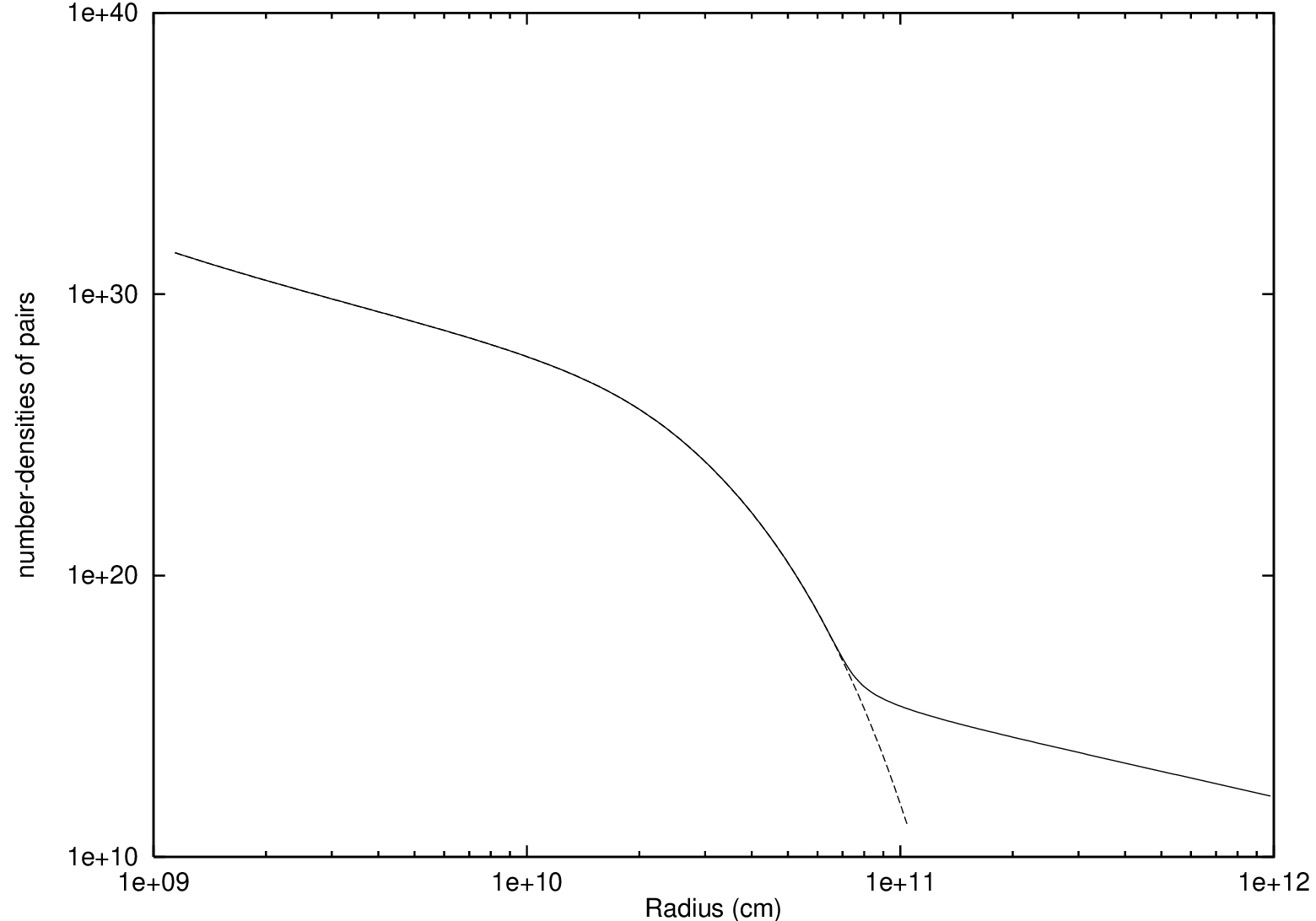}
\includegraphics[width=7cm,height=7cm]{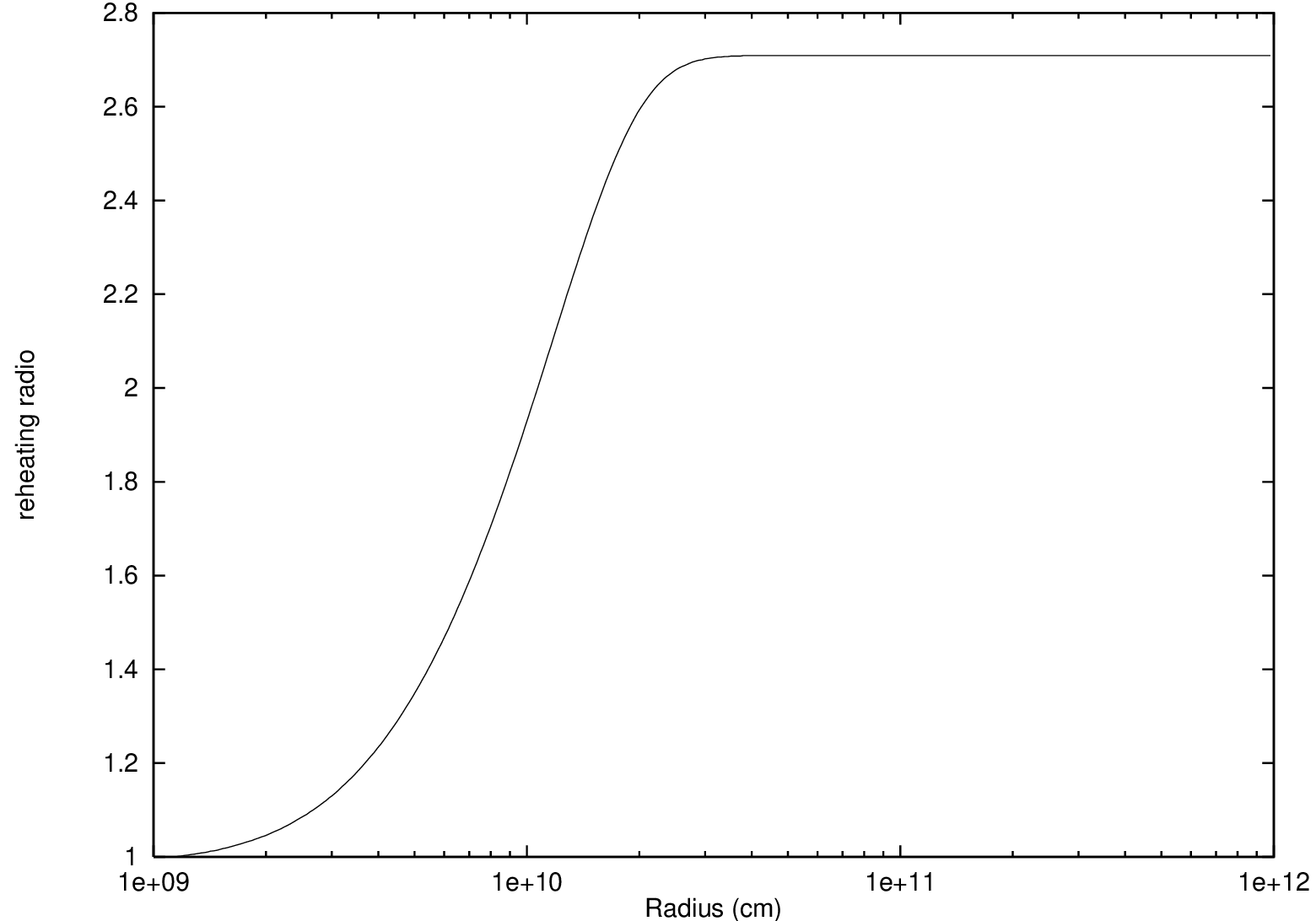}
\caption{In left figure: The number-densities ($N_{e^\pm}$/cm$^{-3}$) 
of pairs $n_{e^\pm}$ (solid line)
as obtained from the rate equation (\ref{Eq3+}), and $n_{e^\pm}(T)$ 
(dashed line) as computed by Fermi-integrals with zero chemical potential (\ref{Integrals2}), provided the temperature $T$ determined by 
the equilibrium condition (\ref{eq0}), are plotted for
a $1000 M_\odot$ black hole with charge to mass ratio $\xi=0.1$. For $T< m_ec^2$, the two curves strongly diverge.
The right figure: The reheating ratio $T^3V/T^3_\circ V_\circ$ defined by Eq.~(\ref{reheat})
is plotted as a function of radius for
a $1000 M_\odot$ black hole with charge to mass ratio $\xi=0.1$. The rate equation (\ref{Eq3+})
naturally leads to the value $\frac{11}{4}$ after 
$e^+e^-$-annihilation has occurred.
} 
\label{figarticlepnpnt&figarticleratio}%
\end{figure} 

\noindent{\it Pair-annihilation and reheating.}\hskip0.3cm

\begin{figure}[ptb]
\def\fsz{\footnotesize}
\def\ssz{\scriptsize}
\def\tsz{\tiny}
\def\dst{\displaystyle}\unitlength1mm
\includegraphics[width=7cm,height=7cm]{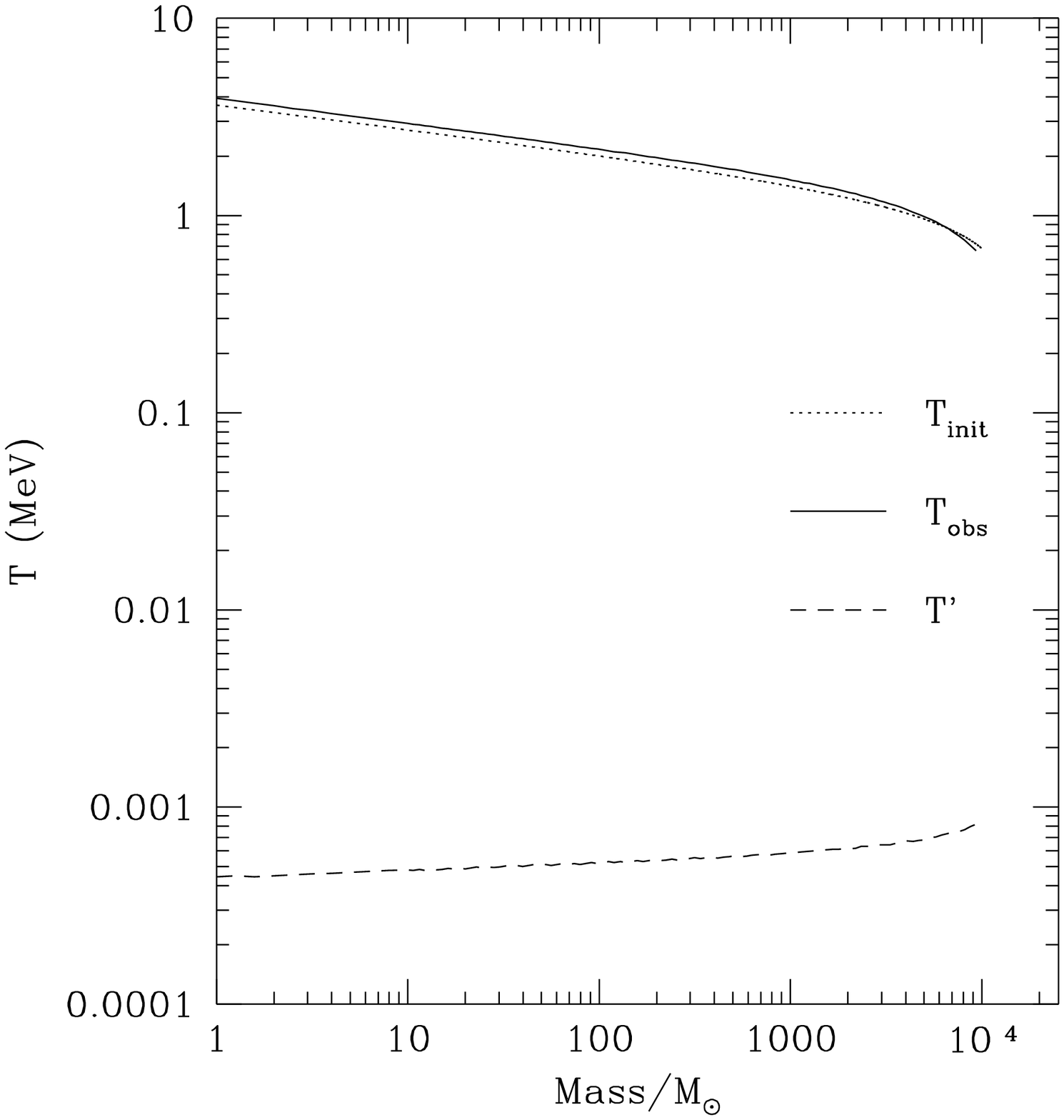}
\includegraphics[width=7cm,height=7cm]{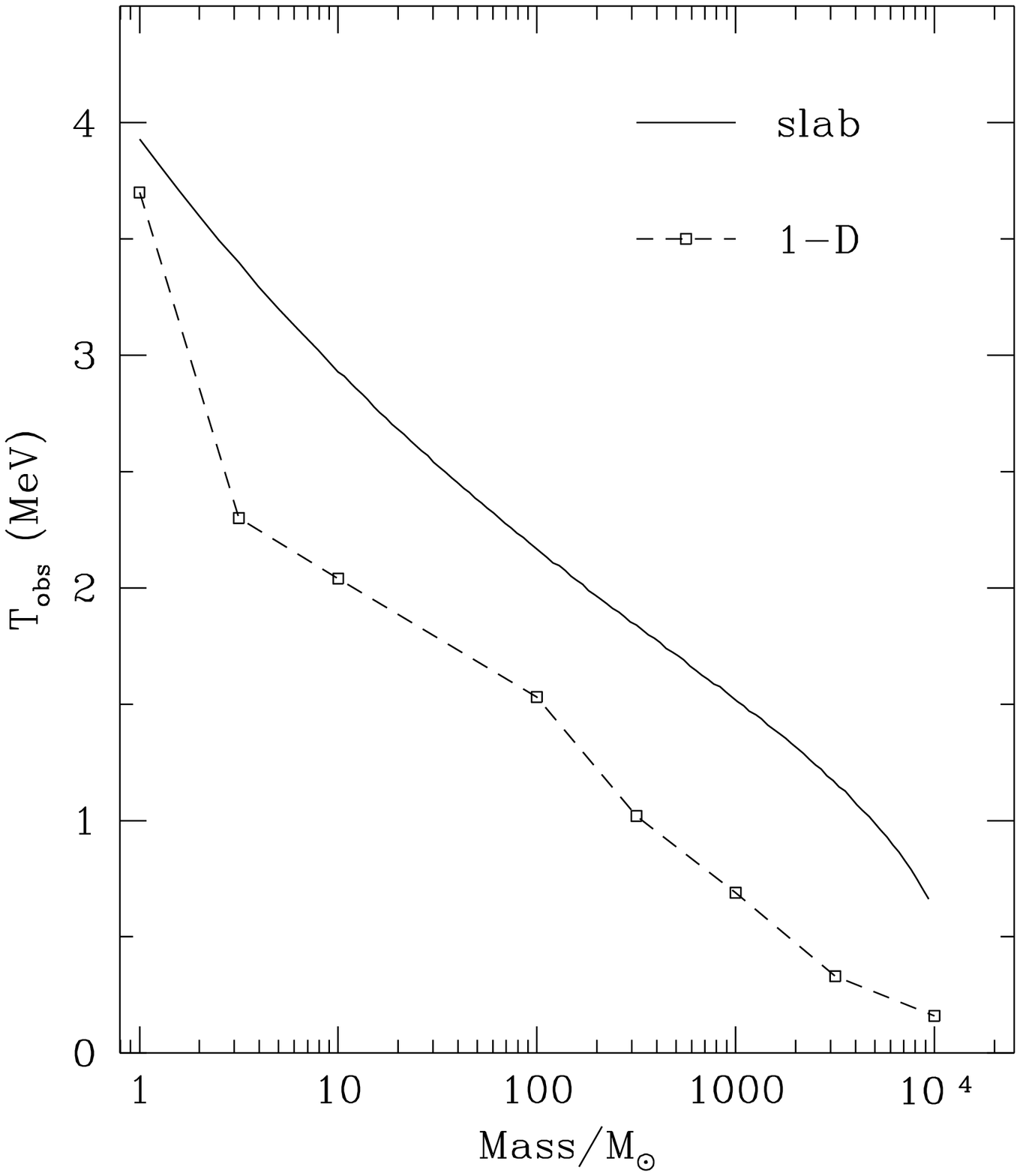}
\caption{In left figure: Temperature of the plasma is shown over a range of masses.  
$T_{\rm init}$ is the average initial temperature of the plasma deposited around the EMBH, 
$T_{\rm obs} = 1.39 \gamma_{\rm final} T'$ is the observed peak temperature of the plasma 
at decoupling while $T'$ is the comoving temperature at decoupling.  
Notice that $\gamma T'\simeq T_{\rm init}$  as expected from Eq.~(\ref{rt}).
The right figure: The peak of the observed number spectrum as a
function of the EMBH mass is plotted with charge to mass ratio $\xi=0.1$.
} 
\label{figallTm&denergy}%
\end{figure}

In Fig.~\ref{figarticlepnpnt&figarticleratio}, we plot the number-densities
of pairs $n_{e^+e^-}$ given by the rate equation (\ref{Eq3+}) and 
$n_{{e^+e^-}}(T)$ computed from a Fermi-integral with zero chemical potential (\ref{Integrals2}) with
the temperature $T$ determined by the equilibrium condition (\ref{eq0}). 
It clearly indicates that the pairs $e^\pm$ fall out of equilibrium as 
the temperature drops below the
threshold of $e^\pm$-pair annihilation. As a consequence of pair 
$e^\pm$-annihilation, the crossover of this reheating process is also shown in 
Fig.~\ref{figarticlepnpnt&figarticleratio}. This can be understood as follows.
From the conservation of entropy it follows that asymptotically we have
\begin{equation}
\frac{(V T^3)_{T<mc^2}}{(V T^3)_{T>mc^2}}  =\frac{11}{4}\ ,
\label{reheat}
\end{equation}
where the comoving volume $V={\mathcal V}/\gamma $. The same considerations when
repeated for the conservation of the total energy 
$\epsilon\gamma V=\epsilon\gamma^2{\mathcal V}$
following from Eq.~(\ref{Eq3}) then lead to
\begin{equation}
      \frac{(V T^4 \gamma)_{T<mc^2}}{(V T^4 \gamma)_{T>mc^2}}  
             =\frac{11}{4}\ .
\end{equation}
The ratio of these last two relations then gives asymptotically 
\begin{equation}
      T_\circ= (T \gamma)_{T>mc^2}= (T \gamma)_{T<mc^2},
\label{rt}
\end{equation}
where $T_\circ$ is the initial average temperature of the Dyadosphere at rest, given in Fig.~\ref{fig: totalenergyanda&afig.4}. 
Eq.~(\ref{Eq3}) also explains the approximate constancy of $T \gamma$ shown 
in Fig.~\ref{figallTm&denergy}. 

\noindent{\it Photon temperature at transparency.}\hskip0.3cm

We are interested in the observed spectrum at the time of decoupling.
To calculate the spectrum, we assume that (i) the plasma fluid of
coupled $e^+e^-$ pairs and photons undergoes adiabatic expansion and
does not emit radiative energy before they decouple \cite{prepa2000}; 
(ii) the $e^+e^-$ pairs and photons are in
equilibrium at the same temperature $T$ when they decouple. Thus the photons that are described by a Planck
distribution in an
emitter's rest-frame with temperature $T'$ will appear Planckian to a
moving observer, but with boosted temperature $T$
\begin{equation}
u_\varepsilon (\theta,v,T') \approx \frac{ \varepsilon^3 }{ \exp(\frac{ \varepsilon }{ T}) - 1 },
\quad
T = \frac{T'}{\gamma (1 - \frac{v}{ c} \cos
\theta)},
\label{boosttem}
\end{equation}
where $\frac{v}{ c} \cos \theta$ is the component of plasma velocity 
directed toward the observer.  
We integrate over angle with respect to the observer, and 
get the observed number spectrum $N_\varepsilon $, per photon energy
$\varepsilon $, per steradian, of a relativistically expanding spherical
shell with radius $R$, thickness $dR$ in cm, velocity $v$, Lorentz
factor $\gamma$ and fluid-frame temperature $T'$ to be (in
photons/eV/$4\pi$)
\begin{align}
N_\varepsilon (v,T',R) \equiv &  \int dV \frac{u_\varepsilon 
 }{ \varepsilon }= (5.23 \times 10^{11}) 4\pi R^2 dR \frac{\varepsilon  T' }{ v
\gamma} \times \nonumber \\
&\times \log \Biggl[ \frac{1 - \exp[- \gamma \varepsilon  (1 + \frac{v}{ c})/T' ] }{ 1 -
\exp[ - \gamma \varepsilon  (1 - \frac{v}{ c})/T' ] } \Biggr],
\label{jay:E:nmax}
\end{align}
which has a maximum at $\varepsilon_{max} \cong 1.39 \gamma T'\ eV$ for
$\gamma \gg 1$.  We may then sum this spectrum over all shells of our PEM-pulse to get the total
spectrum.  Since we had assumed the photons are thermal in the comoving frame, our
spectrum (\ref{jay:E:nmax}) has now an high frequency exponential tail, and the spectrum appear as
not thermal.  
Fig.~\ref{figallTm&denergy} 
illustrates the extreme
relativistic nature of the PEM pulse expansion: at decoupling, the
local comoving plasma temperature is $< 1$ keV, but is boosted by
$\gamma_{\rm final}$ to $\sim 1$ MeV and its relation to the initial energy of the PEM pulse in the Dyadosphere is presented.
In Ref.~ \cite{brx2001} we obtain the observed light curve by decomposing 
the spherical PEM-pulse into concentric shells, considering the light curve in the relative arrival time $t_{a}$ 
of the first light from each shell, then summing over contributions from all shells.

\section{Predications in connection with short Gamma Ray Bursts}\label{shortGRBs}

\begin{figure}[ptb]
\def\fsz{\footnotesize}
\def\ssz{\scriptsize}
\def\tsz{\tiny}
\def\dst{\displaystyle}\unitlength1mm
\includegraphics[width=7cm,height=7cm]{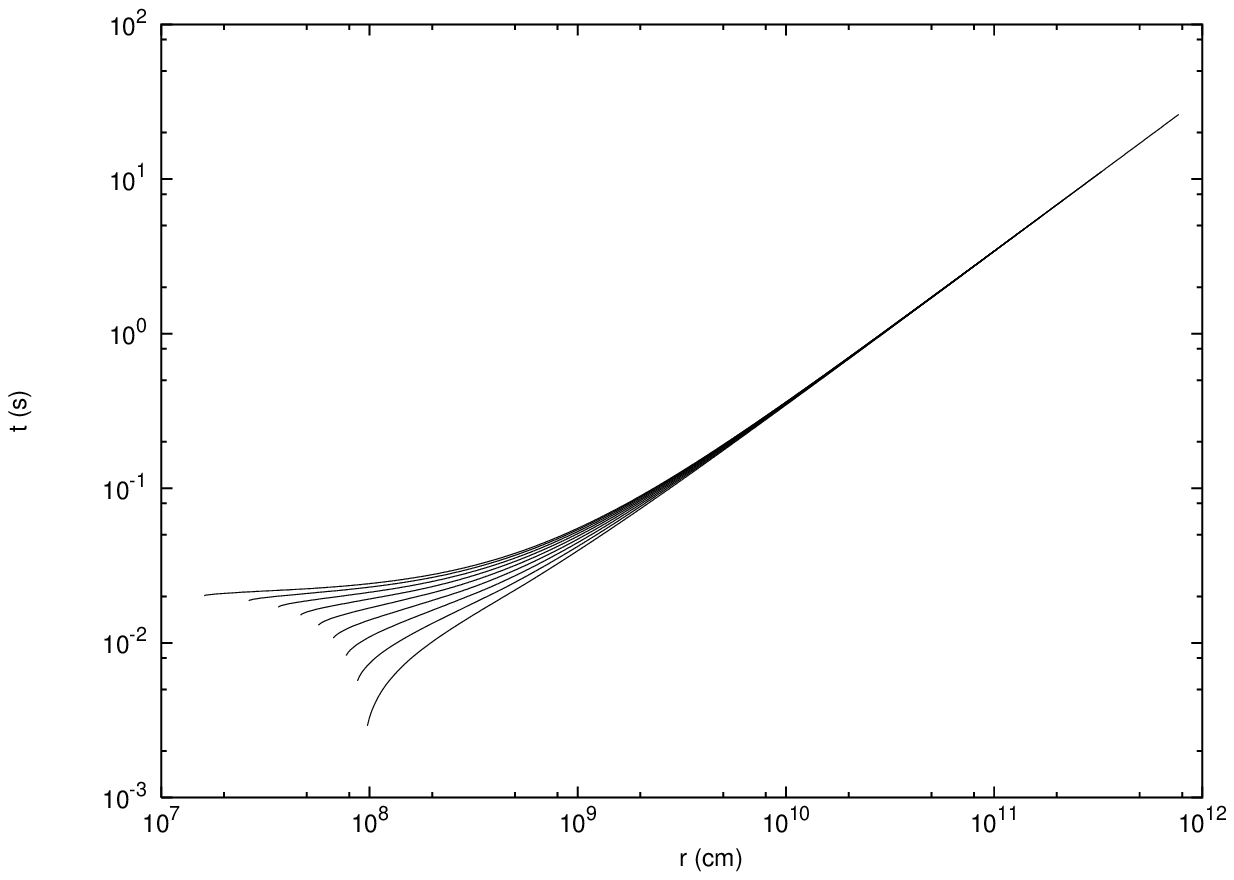}
\includegraphics[width=7cm,height=7cm]{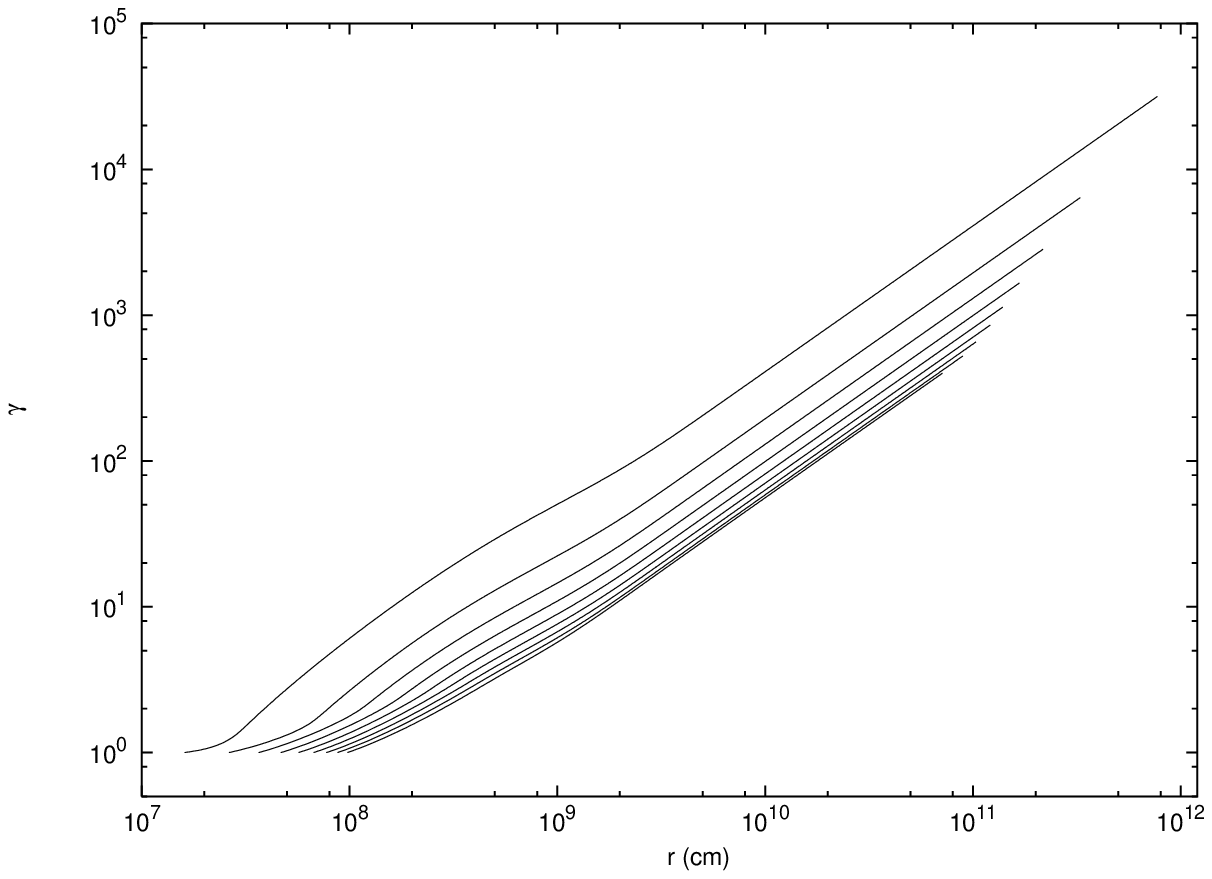}
\caption{Expansion of the plasma created around an overcritical collapsing
stellar core with $M=10M_{\odot}$ and $Q=0.1\sqrt{G}M$. Left diagram: world
lines of the plasma. Right diagram: Lorentz $\gamma$ factor as a function of
the radial coordinate $r$, showing the PEM-pulse reaches ultra relativistic regimes with Lorentz factor
$\gamma\sim10^{2}$--$10^{4}$.
} 
\label{ff1}%
\end{figure}

The formation of Dyadosphere during gravitational collapse and its hydrodynamic expansion after
gravitational collapse continuously connect when gravitational collapse ends. 
Eqs.~(\ref{Eq3},\ref{Eq3+}) must be integrated and matched with the solution found by integrating Eqs.~(\ref{Tab},\ref{rate})
at the transition between the two regimes. 
In this general framework we analyze in Refs.~\cite{rfvx05,frvx2007}, the gravitational
formation and then hydrodynamic evolution of the Dyadosphere.
We recall that {separatrix} was found in
the motion of the plasma at a critical radius $\bar{R}$ (\ref{Rbar}):  the plasma created at
radii smaller than $\bar{R}$ is trapped by the gravitational field of the
collapsing core and implodes toward the black hole, while the one created at
radii larger than $\bar{R}$ expands outward.
The plasma ($r>\bar R$) is divided into $N$ slabs with thickness $\Delta r$, to describe
the adiabatic hydrodynamic expansion of the optically thick
plasma slabs (the PEM-pulse) all the way to the point where the transparency condition is reached.
Eqs.~(\ref{Tab}), (\ref{rate}) and (\ref{Eq3},\ref{Eq3+}) have been integrated and results are presented 
in Fig.~\ref{ff1}. The integration
stops when each slab of plasma reaches the optical transparency condition
given by
\begin{equation}
\int_{0}^{\Delta r}\sigma_{T}n_{e^{+}e^{-}}dr\sim1\,,
\end{equation}
where the integral extends over
the radial thickness $\Delta r$ of the slab. The overall independence of the result of the dynamics on the number
$N$ of the slabs adopted in the discretization process or analogously on the
value of $\Delta r$ has also been checked. We have repeated the integration
for $N=10$, $N=100$ reaching the same result to extremely good accuracy. The
results in Fig.~\ref{ff1} correspond to the case $N=10$.
The evolution of each slab occurs
without any collision or interaction with the other slabs; see the left
diagram in Fig.~\ref{ff1}. The outer layers are colder and less dense than the inner ones and
therefore reach transparency earlier; see the right diagram in Fig.~\ref{ff1}.

\begin{figure}[th]
\includegraphics[width=7cm]{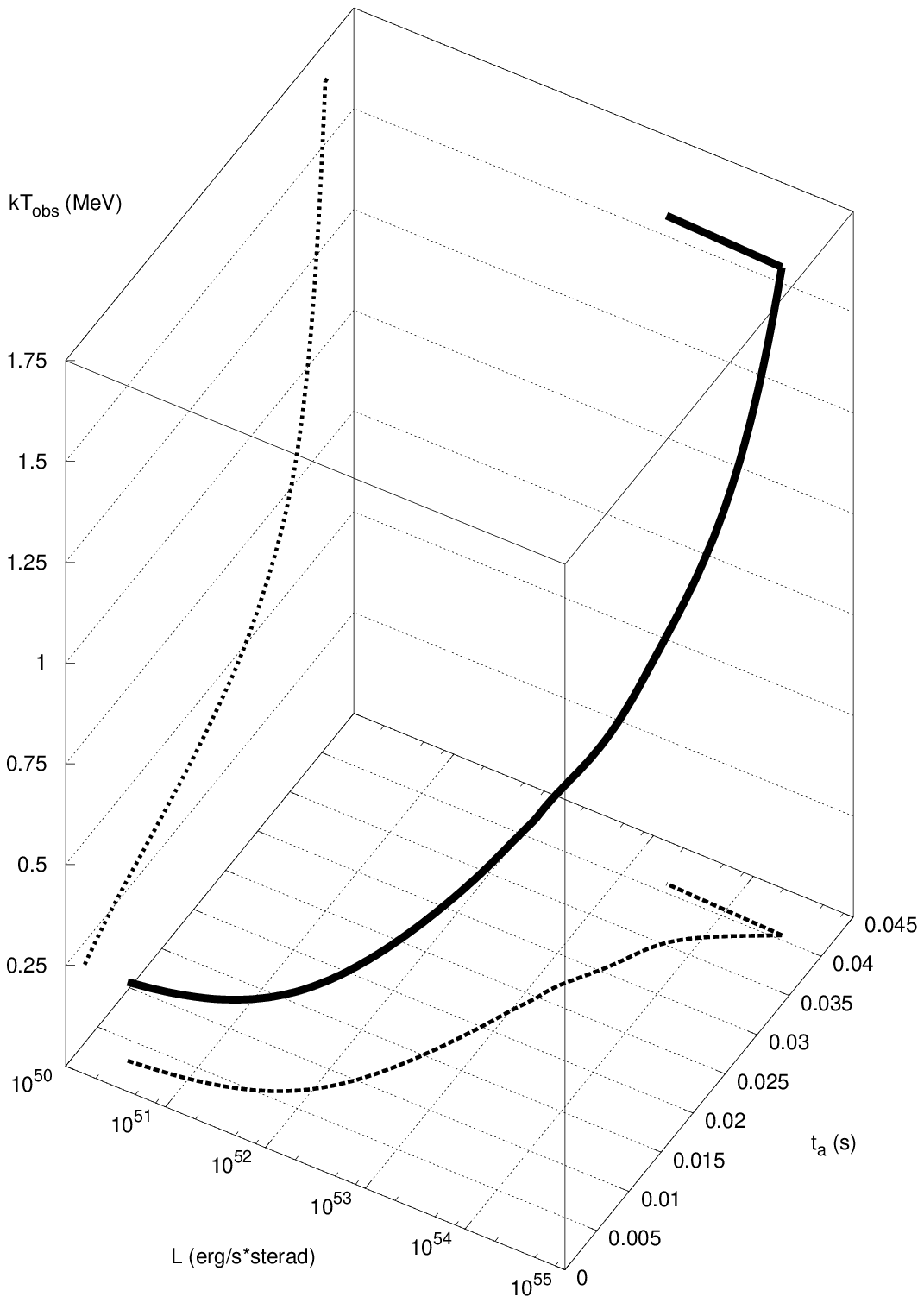}
\caption{Predicted observed luminosity and observed spectral hardness of the
electromagnetic signal from the gravitational collapse of a collapsing core
with $M=10M_{\odot}$, $Q=0.1\sqrt{G}M$ at $z=1$ as functions of the arrival
time $t_{a}$.}%
\label{ff4}%
\end{figure}

\begin{figure}[ptb]
\def\fsz{\footnotesize}
\def\ssz{\scriptsize}
\def\tsz{\tiny}
\def\dst{\displaystyle}\unitlength1mm
\includegraphics[width=7cm,height=7cm]{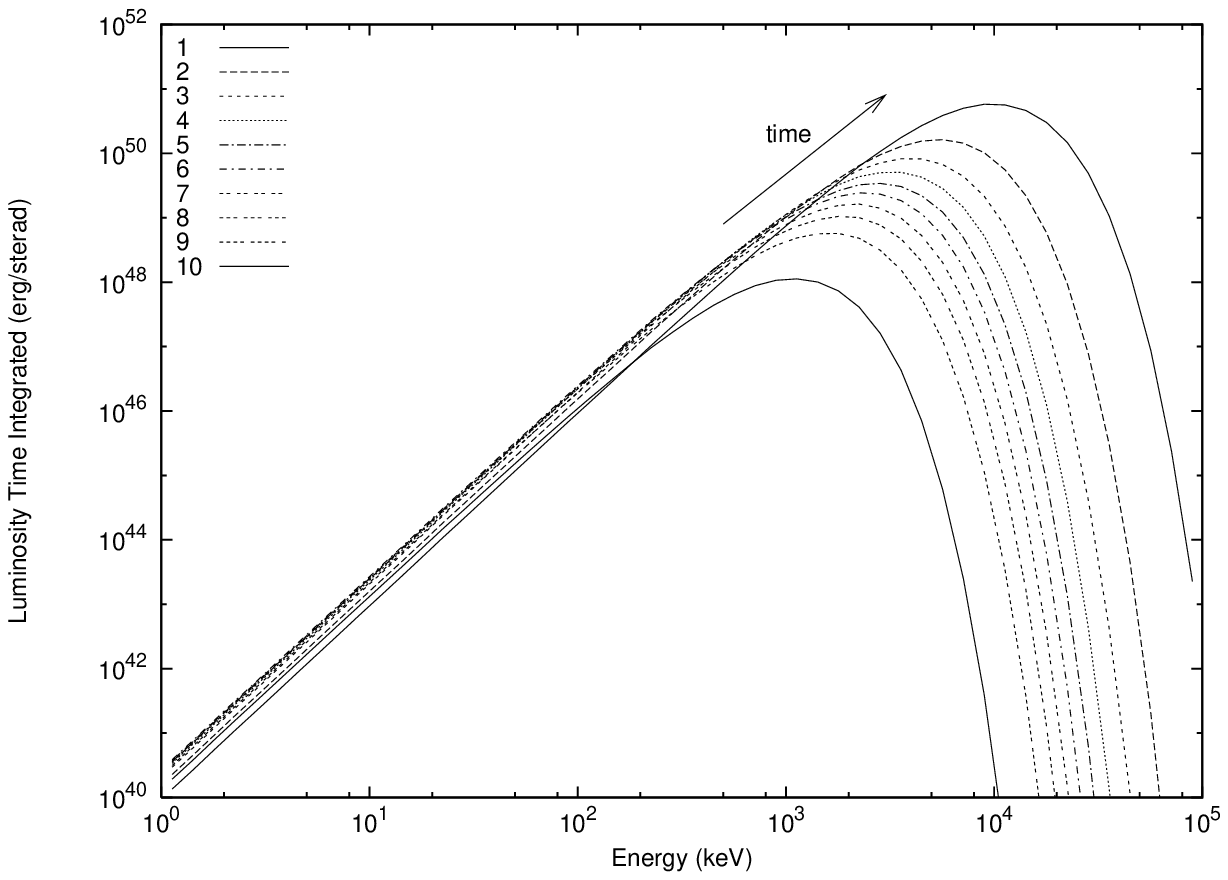}
\includegraphics[width=7cm,height=7cm]{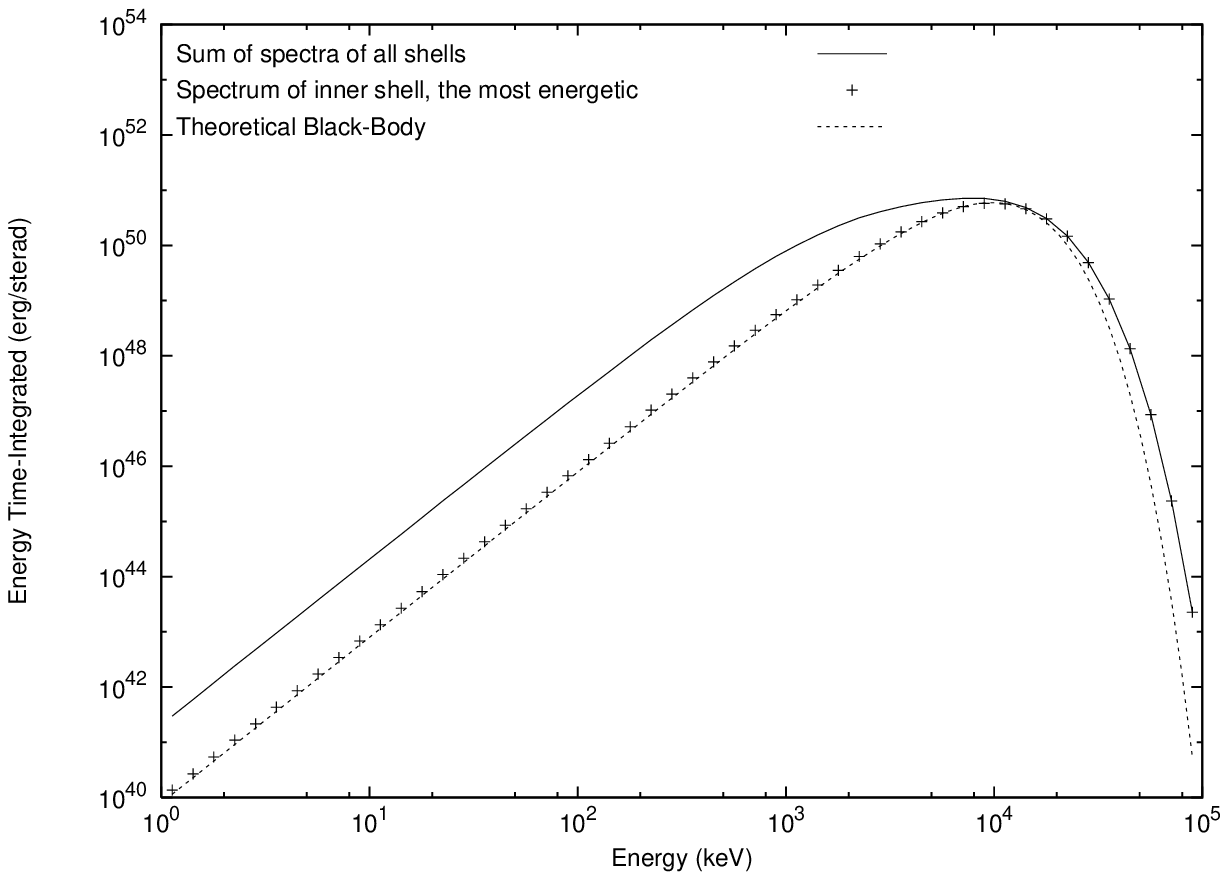}
\caption{Left diagram: Time-integrated spectra  from a plasma created around an overcritical collapsing stellar 
core with  $M=10M_{\odot}$ and $Q=0.1\sqrt{G}M$ with discretization in 10 subshells. As expected, 
the outer subshell corresponds to the minimal energy peak since the electrostatic energy density radially decreases. 
The soft-to-hard evolution of spectral hardness is confirmed by a direct spectral analysis. 
Right diagram: The theoretical prediction of total time integrated spectrum for a short GRB 
is compared with the spectrum of the inner shell, which is the most energetic. As can be seen, the contribution 
of all the other shells is to shift the energy peak to lower energies and to broaden the curve. 
For comparison  the spectrum of a pure black body is also reported.
} 
\label{spectra&spectrumtot}%
\end{figure}

\begin{figure}[th]
\includegraphics[width=9cm,height=6cm]{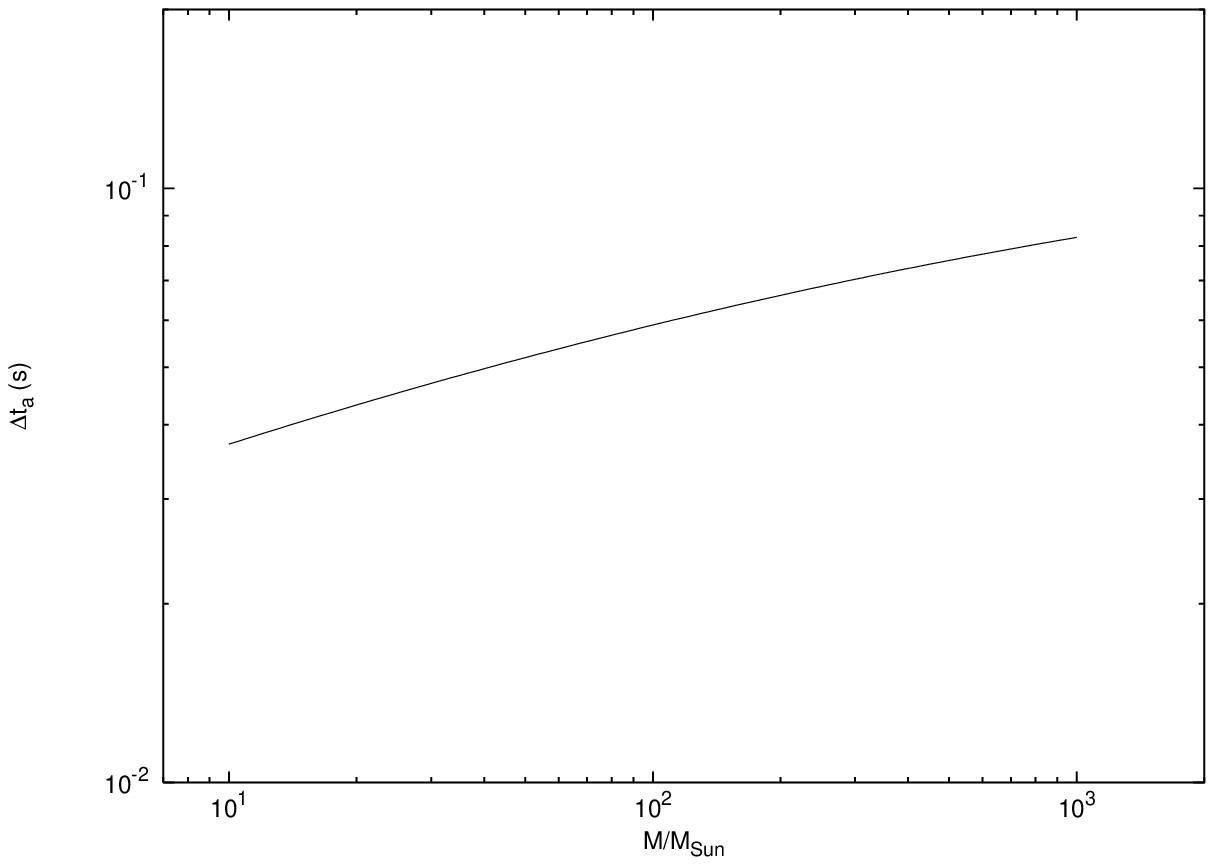}
\caption{Arrival time duration of the electromagnetic signal from the
gravitational collapse of a stellar core with charge $Q=0.1\sqrt{G}M$ as a
function of the mass $M$ of the core.}%
\label{ff5}%
\end{figure}

As the plasma becomes transparent, gamma ray photons are emitted,
and the luminosity and spectrum are calculated, analogously to Eqs.~(\ref{boosttem},\ref{jay:E:nmax}).   
In Fig.~\ref{ff4}, where we plot both the
theoretically predicted luminosity $L$ and the spectral hardness $T_{\rm obs}$ of the signal
reaching a far-away observer as functions of the arrival time $t_{a}$. Since three quantities 
depend in an essential way on the cosmological
redshift factor $z$, see Refs.~\cite{brx2001,r03}, we have adopted a
cosmological redshift $z=1$ for this figure.
The energy of the observed photon is $kT_{\rm obs}=k\gamma T'/\left(
1+z\right)$, where $k$ is the Boltzmann constant, $T'$ is the temperature in
the comoving frame of the pulse and $\gamma$ is the Lorentz factor of the
plasma at the transparency time. The initial zero of
time is chosen as the time when the first photon is observed, then the arrival
time $t_{a}$ of a photon at the detector in spherical coordinates centered on
the black hole is given by \cite{brx2001,r03}:
\begin{equation}
t_{a}=\left(  1+z\right)  \left[  t+\tfrac{r_{0}}{c}-\tfrac{r\left(  t\right)
}{c}\cos\theta\right]
\end{equation}
where $\left(  t,r\left(  t\right)  ,\theta,\phi\right)  $ labels the
laboratory emission event along the world line of the emitting slab and
$r_{0}$ is the initial position of the slab. The projection of the plot in
Fig.~\ref{ff4} onto the $t_{a}$-$L$ plane gives the total luminosity as the
sum of the partial luminosities of the single slabs. The sudden decrease of
the intensity at the time $t_{a}=0.040466$ s corresponds to the creation of the
{\itshape separatrix} introduced in Ref.~\cite{RVX03b}. We find that the
duration of the electromagnetic signal emitted by the relativistically
expanding pulse is given in arrival time by
\begin{equation}
\Delta t_{a}\sim5\times10^{-2}\mathrm{s}\,.\label{ta1}%
\end{equation}
The projection of the plot in Fig.~\ref{ff4} onto the $kT_{\mathrm{obs}}$,
$t_{a}$ plane describes the temporal evolution of the spectral hardness. We
observe a precise soft-to-hard evolution of the spectrum of the gamma ray
signal from $\sim10^{2}$ KeV monotonically increasing to $\sim1$ MeV.
The corresponding spectra are presented in Figs.~\ref{spectra&spectrumtot}, which are consistent with observations of 
short GRBs \cite{frvx2007}.
The three quantities $L, kT_{\rm obs}$ and $t_{a}$ are clearly functions of 
the charge $Q$ and the mass $M$ of the collapsing core. We present in
Fig.~\ref{ff5} the arrival time interval for $M$ ranging from $M\sim10M_{\odot}$ to
$10^{3}M_{\odot}$, keeping $Q=0.1\sqrt{G}M$. The arrival time interval is very
sensitive to the mass of the black hole:
\begin{equation}
\Delta t_{a}\sim10^{-2}-10^{-1}\mathrm{s} \,.\label{ta2}%
\end{equation}
Similarly the spectral hardness of the signal is sensitive to the ratio
$Q/\sqrt{G}M$ \cite{rfvx05}. Moreover the duration, the spectral hardness and
luminosity are all sensitive to the cosmological redshift $z$ (see
Ref.~\cite{rfvx05}). 

Ghirlanda et al. have given evidence for the existence of an
exponential cut off at high energies in the spectra of short GRBs. We are
currently comparing and contrasting these observational results with the
predicted cut off in Fig.~\ref{ff4} which results from the existence of the
separatrix introduced in \cite{RVX03c}. The observational confirmation of the
results presented in Fig.~\ref{ff4} would lead for the first time to the
identification of a process of gravitational collapse and its general
relativistic self-closure as seen from an asymptotic observer.

The characteristic spectra, time variabilities and luminosities
of the electromagnetic signals from collapsing overcritical stellar cores,
here derived from first principles, agrees very closely with the observations
of short GRBs \cite{RBCFX01b}. New space missions must be planned, with
temporal resolution down to fractions of $\mu$s and higher collecting area and
spectral resolution than at present, in order to verify the detailed agreement
between our model and the observations. It is now clear that if these
theoretical predictions will be confirmed, we would have a very powerful tool
for cosmological observations: the independent information about luminosity,
time-scale and spectrum can uniquely determine the mass, the electromagnetic
structure and the distance from the observer of the collapsing core, see,
e.g., Fig.~\ref{ff5} and Ref.~\cite{rfvx05,frvx2007}. In that case short-bursts may
become the best example of standard candles in cosmology.

\end{document}